\documentclass[aps,prb,floats,onecolumn,showpacs,amsmath,amssymb,superscriptaddress,floatfix,12pt]{revtex4-2}

\usepackage{graphicx}
\usepackage{dcolumn}
\usepackage{euscript}
\usepackage[bbgreekl]{mathbbol}
\usepackage{float}
\usepackage[utf8]{inputenc}
\usepackage[T1,T2A]{fontenc}
\usepackage[english]{babel}
\newcommand{\e}{\epsilon}

\newcommand{\s}{\sigma}
\newcommand{\fee}{\varphi}
\renewcommand{\t}{\tau}
\newcommand{\V}{{\EuScript{V}}}
\renewcommand{\d}{\textstyle}

\renewcommand{\L}{\check{\Lambda}}

\newcommand{\tr}{\,\mathrm{tr}}

\newcommand{\bB}{\mathcal{B}}

\renewcommand{\ln}{\mathrm{ln}}

\newcommand{\G}{\, \check{G}}
\newcommand{\Gt}{\, \check{\tilde{G}}}
\newcommand{\A}{ \mathcal{A}}

\newcommand{\GammaF}{\mathbb{\Gamma}}

\begin{document}

\title{Superconducting quantum fluctuations in one dimension}
\author{Andrew G. Semenov}
\affiliation{I.E. Tamm Department of Theoretical Physics, P.N. Lebedev Physical Institute, 119991 Moscow, Russia}
\affiliation{National Research University Higher School of Economics, 101000 Moscow, Russia}
\author{Andrei D. Zaikin}
\affiliation{Institute for Quantum Materials and Technologies, Karlsruhe Institute of Nanotechnology (KIT), 76021, Karlsruhe, Germany}
\affiliation{I.E. Tamm Department of Theoretical Physics, P.N. Lebedev Physical Institute, 119991 Moscow, Russia}
\affiliation{National Research University Higher School of Economics, 101000 Moscow, Russia}

\date{\today}

\begin{abstract}
We review some recent developments in the field of quasi-one-dimensional superconductivity. We demonstrate that low temperature properties of superconducting nanowires are essentially determined by quantum fluctuations. Smooth (Gaussian) fluctuations of the superconducting phase (also associated with plasma modes propagating along the wire) may significantly affect the electron density of states in such nanowires and induce persistent current noise in superconducting nanorings. Further interesting phenomena such as, e.g., non-vanishing resistance and shot noise of the voltage in current-biased superconducting nanowires, are caused by non-Gaussian fluctuations of the order parameter -- quantum phase slips (QPS). Such phenomena may be interpreted in terms of tunneling of fluxons playing the role of effective quantum "particles" dual to Cooper pairs and obeying complicated full counting statistics which reduces to Poissonian one in the low frequency limit. We also demonstrate that QPS effects may be particularly pronounced in thinnest wires and rings where quantum phase slips remain unbound and determine a non-perturbative length scale $L_c$ beyond which the supercurrent gets suppressed by quantum fluctuations. Accordingly, for $T \to 0$ such nanowires should become insulating at scales exceeding $L_c$, whereas at shorter length scales they may still exhibit superconducting properties. We argue that certain non-trivial features associated with quantum fluctuations of the order parameter may be sensitive to specific circuit topology and may be observed in structures like, e.g., a system of capacitively coupled superconducting nanowires.\end{abstract}
\maketitle
\tableofcontents

\section{\label{sec:level1}Introduction}

An important role of fluctuations in reduced dimension is widely known. Of a special interest are fluctuation effects in low dimensional superconductors which properties -- in contrast to bulk structures -- cannot in general be adequately described by means of the standard Bardeen-Cooper-Schriffer (BCS) mean field theory. Fluctuations are most strongly pronounced in ultrathin superconducting wires causing a large number of intriguing physical phenomena. Over last decades these phenomena attracted a lot of attention of numerous researchers worldwide and were discussed in details in a number of recent books and review papers, see, e.g., \cite{book,AGZ,LV,Bezr08,Z10,Bezrbook}.

Can superconductivity survive also in structures of lower dimension or do fluctuations disrupt any supercurrent in such systems? The answer to this question is of both fundamental interest and practical importance due to rapidly progressing miniaturization of superconducting nanocircuits. According to the well-known theorem \cite{MW,H} fluctuations destroy the true long-range order in low dimensional superconductors. With this in mind, one could attempt to conclude that low dimensional conductors cannot exhibit superconducting properties.

This conclusion, however, would be somewhat premature  because any generic superconducting system has a finite size in which case phase coherence can be preserved at least to a certain extent. For instance, two-dimensional structures undergo Berezinskii-Kosterlitz-Thouless (BKT) phase transition \cite{b,kt,BKT} as a result of which the decay of correlations in space changes from exponential at high enough temperatures  to power law at lower $T$. This result implies that at low temperatures long range phase coherence does survive in samples of a finite size and, hence, generic two-dimensional films can and do become superconducting.

Likewise, the general theorem \cite{MW,H} does not yet allow one to make any definite conclusion about the presence or absence of superconductivity in quasi-one-dimensional wires of a finite length employed in any realistic experiment. Moreover, as we will see below, in the presence of quantum fluctuations superconducting properties of such structures may crucially depend on particular experimental configuration which makes the whole situation even more complicated.

\begin{figure}
	\begin{center}
		\includegraphics[width=0.49\linewidth]{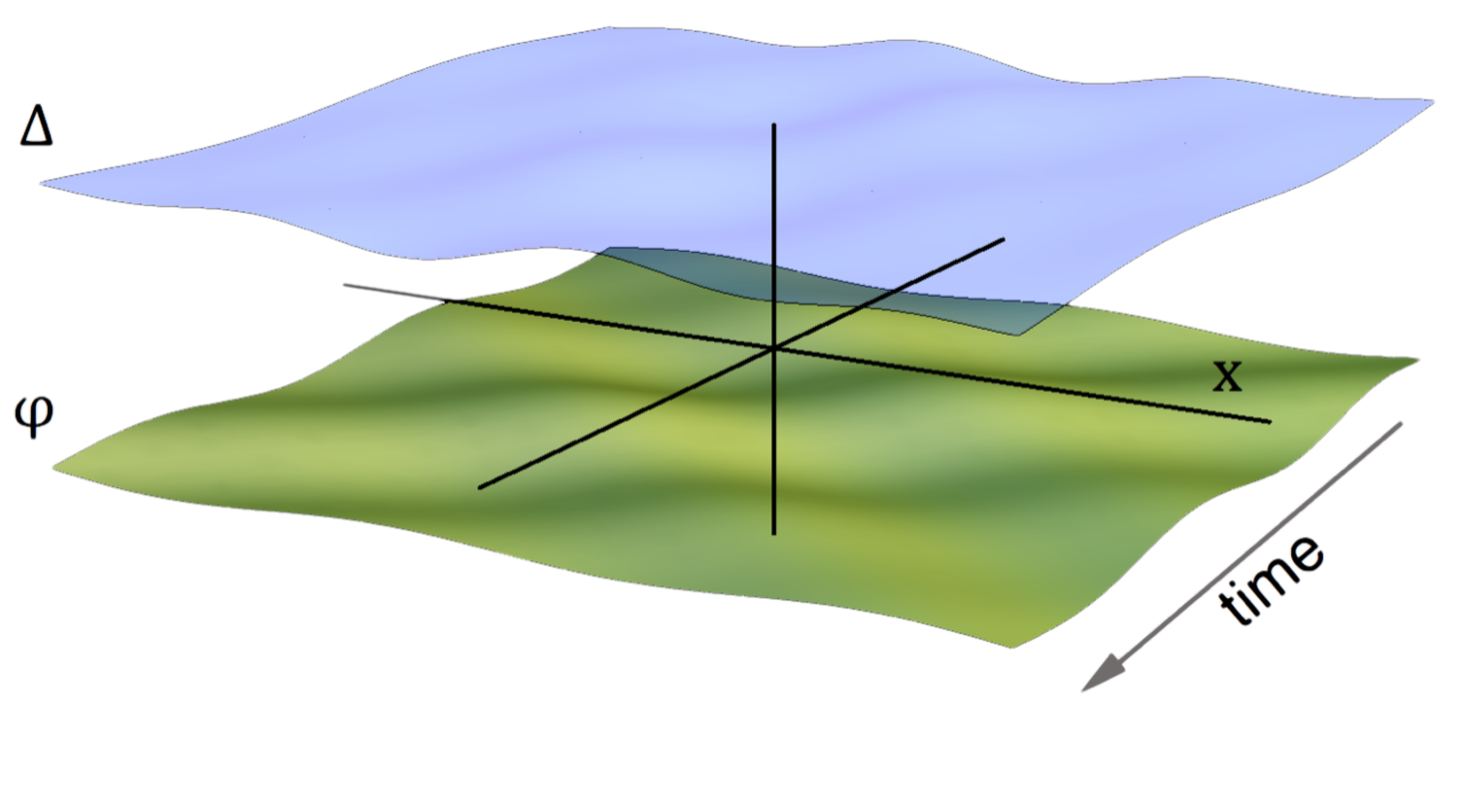}
		\includegraphics[width=0.49\linewidth]{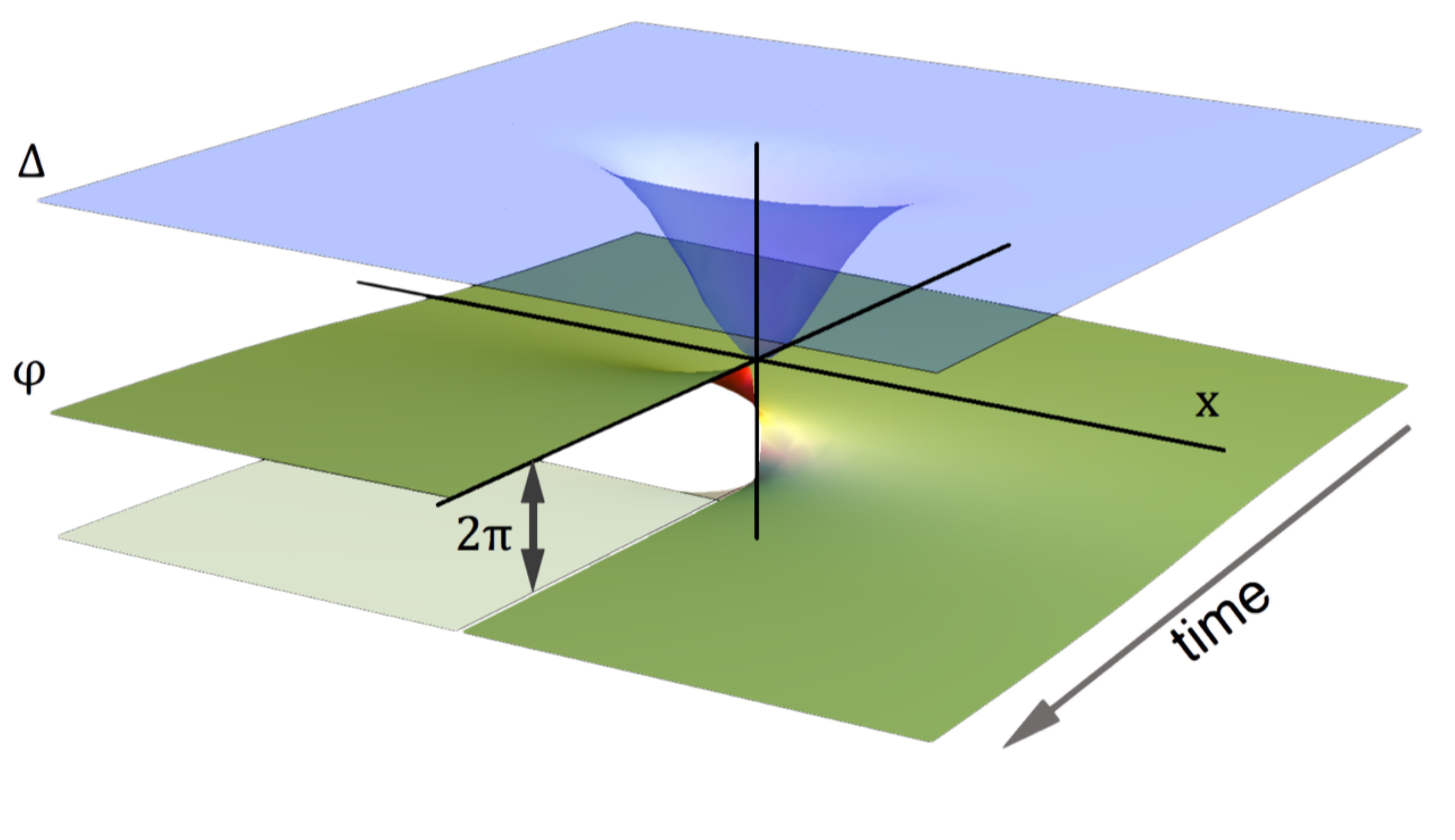}
	\end{center}
\caption{Schematics of small fluctuations of the order parameter (left panel) and the phase slip process (right panel) in superconducting nanowires. During the latter process the absolute value of the order parameter $\Delta$ gets locally and temporarily suppressed due to thermal and/or quantum fluctuations while its phase $\varphi$ suffers a jump by $\pm 2\pi$.}
\label{Fig:1}
\end{figure}

The superconducting state of a quasi-one-dimensional metallic wire can be described by means of a complex order parameter $\Delta(x)=|\Delta(x)|e^{{\rm i}\varphi(x)}$, where $x$ is the coordinate along such a wire. Both thermal and quantum fluctuations cause deviations of the modulus as well as the phase of this order parameter from their equilibrium values. Such fluctuations can be divided into two different types which are (i) small (Gaussian) fluctuations of the order parameter and (ii) non-Gaussian fluctuations, i.e. the so-called {\it phase slips}. Both these types of fluctuations are schematically illustrated in Fig. \ref{Fig:1}.

The effect of Gaussian superconducting fluctuations can be treated in a straightforward manner by expanding the exact gauge invariant expression for the effective action of a superconductor \cite{ZGOZ,OGZB,GZQPS} in the fluctuating part of the order parameter  $\Delta (x)$. In this way one can, e.g., derive the (negative) correction to the mean field (BCS) value of the order parameter $\Delta_{0}$. In particular, at $T\to 0$ one finds $\Delta=\Delta_0-\delta\Delta_0$ with \cite{GZTAPS}
\begin{align}
	\frac{\delta\Delta_{0}}{\Delta_{0}}\sim \frac{1}{g_{\xi}} \sim Gi_{\mathrm{1D}}^{3/2}.\label{estim22}%
\end{align}
Here 
\begin{equation}
	g_\xi =R_q/R_\xi
\label{gxi}
\end{equation}
is dimensionless conductance, $R_q=2\pi/e^2 \simeq 25.8$ K$\Omega$ is the quantum resistance unit, $R_\xi$ is the normal state resistance of the wire segment of length equal to the superconducting coherence length $\xi$ and $Gi_{\mathrm{1D}}$ is the so-called Ginzburg number in one dimension \cite{LV}.

Note that in Eq. (\ref{estim22}) fluctuations of both the phase and the absolute value of the order parameter give contributions of the same order. This estimate demonstrates that at low temperatures the order parameter suppression due to Gaussian fluctuations in superconducting nanowires remains weak as long as $g_{\xi} \gg 1$ and -- in full accordance with  general expectations -- it becomes important only for extremely thin wires with $Gi_{\mathrm{1D}} \sim 1$ and $g_{\xi} \sim 1$.

It is also important to emphasize that even if the absolute value $|\Delta|$ does not fluctuate, the superconducting phase  $\varphi$ can still do so. In the limit $g_\xi \gg 1$ phase fluctuations in superconducting wires can be essentially decoupled from those of $|\Delta|$. Such phase fluctuations are controlled by the dimensionless admittance 
\begin{equation}
	g=R_q/Z_{\rm w},
\label{gaZ}
\end{equation} 
where $Z_{\rm w}=\sqrt{\mathcal{L}_{\rm kin}/C}$ is the wire impedance, $\mathcal{L}_{\rm kin}=1/(\pi\sigma_N\Delta s)$ and $C$ are respectively the kinetic wire inductance (times length) and the geometric wire capacitance (per length), $\sigma_N=2e^2\nu_FD$ is the normal state Drude conductance of the wire, $\nu_F$ is density of states at the Fermi level  and $s$ is the wire cross section. Such phase fluctuations are intimately related to sound-like plasma modes \cite{Mooij} (the so-called Mooij-Sch\"on modes) which can propagate along superconducting wires with the velocity $v=1/\sqrt{\mathcal{L}_{\rm kin}C}$. The dimensionless impedance $g \propto \sqrt{s}$ constitutes  another important parameter which -- along with  $g_\xi \propto s$ -- also accounts for Gaussian superconducting fluctuations in ultrathin nanowires.

Let us now turn to non-Gaussian fluctuations of the order parameter produced by phase slips. A non-trivial fluctuation of that kind corresponds to temporal suppression of $|\Delta(x)|$ down to zero in some point $x=x_0$ inside the wire, as shown in Fig. \ref{Fig:1}(right panel). As soon as the modulus of the order parameter $|\Delta(x_0)|$ vanishes, the phase $\varphi(x_0)$ becomes unrestricted and can jump by the value $2\pi n$, where $n$ is any integer number. After this process the modulus $|\Delta(x_0)|$ gets restored, the phase becomes single valued again and the system returns to its initial state accumulating the net phase shift $2\pi n$. Provided such phase slip events are sufficiently rare, one can restrict $n$ by $n=\pm1$ and totally disregard fluctuations with $|n|\geq2$.

Phase slips may have a strong impact on the behavior of sufficiently thin superconducting wires. For instance, as it was first pointed out by Little \cite{Little}, quasi-one-dimensional wires made of a superconducting material can acquire a finite resistance below the superconducting critical temperature $T_{C}$ of a bulk material due to the mechanism of thermally activated phase slips (TAPS).  This mechanism works as follows.

According to the Josephson relation each phase jump by $\delta \varphi =\pm 2\pi$ implies positive or negative voltage pulse  $\delta V=\dot{\varphi}/2e$. In the absence of any bias current the net average numbers of \textquotedblright positive\textquotedblright\ ($n=+1$) and \textquotedblright negative\textquotedblright\ ($n=-1$) phase slips are equal, thus the net voltage drop across the wire remains zero. Applying the current $I\propto |\Delta | ^{2}\nabla\varphi$ one creates nonzero phase gradient along the wire making \textquotedblright positive\textquotedblright\ phase slips to prevail over
\textquotedblright negative\textquotedblright\ ones. Hence, the net voltage drop $V$ due to TAPS differs from zero, i.e. thermal fluctuations cause non-zero resistance $R=V/I$ of superconducting wires even below $T_{C}$.

A quantitative theory of this TAPS phenomenon was initially worked out by Langer and Ambegaokar \cite{la} who employed the standard Ginzburg-Landau equations and evaluated  the TAPS rate within the exponential accuracy. More accurate analysis of the TAPS rate including the pre-exponential factor was performed by McCumber and Halperin \cite{mh} who employed the so-called time-dependent Ginzburg-Landau (TDGL) equations. More recently it was realized that the TDGL-based approach is not sufficiently accurate to account for the effect of quantum fluctuations and, hence, to correctly determine the pre-exponent in the expression for the TAPS rate. Appropriate modifications of the LAMH theory have been worked out \cite{GZTAPS} employing the general effective action approach \cite{ZGOZ,OGZB,GZQPS}.

This theory predicts that the TAPS creation rate and, hence, resistance of a superconducting wire $R$ below $T_{C}$ are determined by the activation exponent
\begin{equation}
	R(T)\propto\exp(-\delta F/T),\label{TAPS}%
\end{equation}
where $\delta F$ is the free energy difference or, in other words, an effective potential barrier which the system should overcome in order to create a phase slip. The height of this potential barrier is determined by the superconducting condensation energy for a part of the wire where superconductivity is destroyed by thermal fluctuations. At temperatures close to $T_{C}$ Eq. (\ref{TAPS}) yields appreciable resistivity which was indeed detected in experiments  \cite{Webb R(T) in Sn whiskers,Tinkham R(T) in Sn whiskers} performed on small superconducting whiskers with typical diameters in the range of $\sim0.5$ $\mu$m. Close to $T_{C}$ the experimental results fully confirm the activation behaviour of $R(T)$ expected from Eq. (\ref{TAPS}). However, as the temperature is lowered further below $T_{C}$ the number of TAPS inside the wire decreases exponentially and no measurable wire resistance is predicted by Eq. \eqref{TAPS}.

Recent progress in nanolithographic technique allowed to fabricate samples with much smaller diameters down to -- and even below -- 10 nm. In such systems one can consider a possibility for phase slips to occur not only due to thermal, but also due to \textit{quantum} fluctuations of the superconducting order parameter. The physical picture describing such quantum phase slips (QPS) is qualitatively similar to that of TAPS (see Fig. \ref{Fig:1}(right panel)) except the order parameter $|\Delta(x)|$ gets virtually suppressed due the process of quantum tunnelling rather than thermal activation. As a result, the superconducting phase again suffers jumps by $\delta \varphi =\pm 2\pi$.

Following the standard quantum mechanical arguments one can expect that the probability of such tunnelling process should be controlled by the exponent $\sim\exp(-\delta F/\omega_{0})$, i.e. one should just substitute temperature $T$ by some attempt frequency $\omega_{0}$ in the activation exponent (\ref{TAPS}). This is because the order parameter field $\Delta (x)$ now tunnels under the barrier $\delta F $ rather than overcomes it by thermal activation. Since such tunnelling process should obviously persist down to $T=0$, one arrives at a fundamentally important conclusion that such nanowires should demonstrate a non-vanishing resistivity down to lowest temperatures. This effect was predicted and investigated theoretically \cite{ZGOZ,GZQPS} and received its convincing  experimental confirmation \cite{BT,Lau,Zgi08,Leh,liege}.

According to the present theory, quantum phase slip effects are controlled by the QPS amplitude per unit wire length \cite{GZQPS} 
\begin{equation}
	\gamma_{QPS} = b(g_\xi\Delta/\xi)\exp (-ag_\xi),
\label{gaQPS0}
\end{equation}
where, as before, $\Delta$ is the superconducting order parameter, $a \sim 1$ and $b \sim 1$ are numerical prefactors. It follows immediately from Eq. (\ref{gaQPS0}) that -- provided the parameter $g_\xi $ is not too large -- QPS effects in superconducting nanowires (similarly to Gaussian fluctuations)  are pronounced and need to be properly accounted for. Conversely, by choosing the dimensionless conductance $g_\xi $ sufficiently large one can suppress both these types of fluctuations of the superconducting order parameter.

Note that although the dimensionless admittance $g$ \eqref{gaZ} does not enter directly into the QPS amplitude  (\ref{gaQPS0}), it nevertheless plays an important role in the physics of quantum phase slips because -- as we already pointed out -- it accounts for Mooij-Sch\"on plasma modes propagating along the wire. Different quantum phase slips interact by exchanging such plasmons and, hence, the parameter $g$ controls the strength of inter-QPS interactions. By reducing the wire diameter  one also reduces $g$ and eventually arrives at the superconductor-insulator quantum phase transition \cite{ZGOZ} that occurs at $T \to 0$. In other words, quantum fluctuations may drive a superconducting nanowire not only into a resistive but also into an insulating state.

Thus, we conclude that the same dimensionless parameters \eqref{gxi} and \eqref{gaZ} which account for small fluctuations of the order parameter also essentially control the physics of quantum phase slips. These two parameters will play a central role in our further considerations.

The main purpose of this work is to review some recent developments in the field. In doing so, we will merely emphasize fundamental aspects of the phenomena under consideration focusing our attention on recent advances in theory of quantum fluctuations in superconducting nanowires and nanorings. Wherever necessary, we will also briefly indicate relevant experiments and possible applications of the effects in question.

The structure of our review paper is as follows. In Sec. II we will analyze the effect of small (Gaussian) quantum fluctuations of the phase of the order parameter on the electron density of states in ultrathin superconducting wires. The same type of fluctuations causing supercurrent noise in superconducting nanorings will be addressed in Sec. III. Section IV will be devoted to an important issue of phase-charge duality in superconducting nanowires and nanorings in the presence of quantum phase slips. The effect of quantum phase slips on both the supercurrent and its fluctuations in superconducting nanorings will be investigated in Sec. V. Voltage fluctuations in superconducting nanowires, such as, e.g., shot noise associated with quantum phase slips will be described in Sec. VI. In Sec. VII we will outline a theory of full counting statistics for quantum phase slips. Topology controlled quantum phase transitions and superconducting fluctuation effects will be analyzed in Sec. VIII. In Sec. IX we will address several interesting phenomena associated with quantum phase slips in capacitively coupled superconducting nanowires. The paper is concluded by a short summary in Sec. X.

\section{Quantum phase fluctuations and local density of states}

\begin{figure}
	\includegraphics[width=0.99\linewidth]{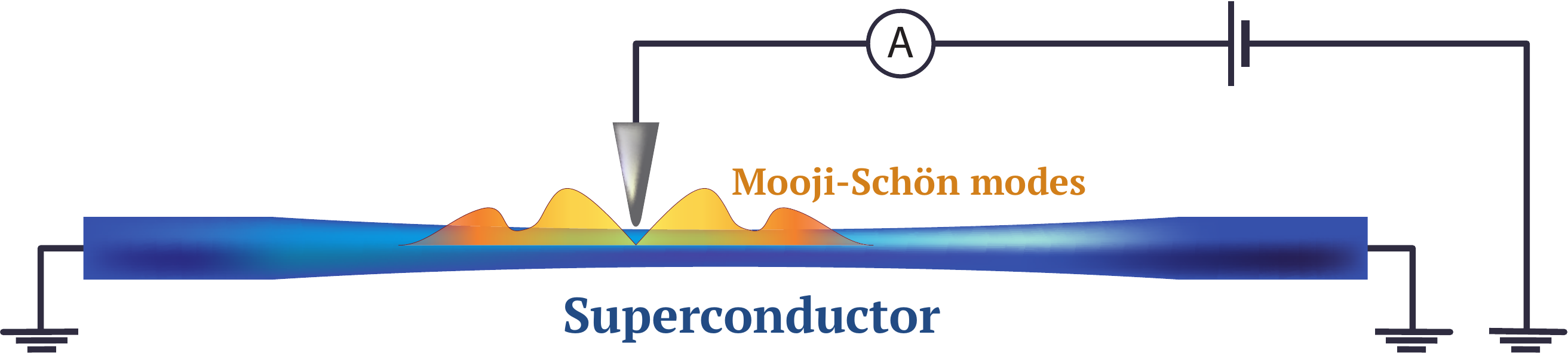}
\caption{A narrow superconducting wire together with a circuit which could be employed for DOS measurements.}
\label{Fig:2}
\end{figure}

Let us consider a long superconducting wire with sufficiently small diameter $\sim \sqrt{s} < \xi$ is attached to two big superconducting reservoirs. This structure is displayed in Fig. \ref{Fig:2}. Superconducting properties of the wire are described by the order parameter field $\Delta (x, t)=|\Delta (x, t)|\exp (i\varphi (x, t))$ which depends both on the coordinate along the wire $x$ and on time $t$. The wire remains in thermodynamic equilibrium at low enough temperature $T \ll |\Delta |$ and its parameters are chosen such that one can safely ignore fluctuations of the absolute value of the order parameter which is set to be independent of both $x$ and $t$, i.e. $|\Delta (x, t)|=\Delta$. As we already described above, this situation can be achieved provided the dimensionless conductance $g_\xi$ remains very large, $g_\xi \gg 1$.

On the other hand, we will allow for fluctuations of the phase variable $\varphi (x, t)$ along the wire and keep the dimensionless impedance $g$ not very large. Our main goal in this section is to demonstrate that such fluctuations can affect and significantly alter the local electron density of states (DOS) of superconducting nanowires. The physical origin of this effect is directly related to the presence of Mooij-Sch\"on plasma modes propagating along the wire and forming an effective environment for electrons inside the system. As it will be demonstrated below, interaction between electrons and such plasma modes can lead to substantial modifications and smearing of local electron DOS inside superconducting nanowires \cite{RSZ}.

\subsection{Green functions in the presence of phase fluctuations} 
In order to proceed, we will note that typically the motion of electrons in metallic nanowires is diffusive implying that the elastic electron mean free path $\ell$ in such wires is much smaller than the superconducting coherence length $\xi$. This electron motion can be described quasiclassically with the aid of the standard approach based on the Keldysh version of the Usadel equations \cite{Usadel,bel}
\begin{equation}
	\left[\partial_t \sigma_3 -i\check{\Delta}+i e\check{V}\hat{1},\check{G}\right]-\frac{D}{2}\hat{\partial}\left[\check{G},\hat{\partial}\check{G}\right]=0
\label{usadel_eq}
\end{equation}
for the quasiclassical electron Green-Keldysh matrix function 
$
\check G=\left(\begin{smallmatrix}\hat G^R & \hat G^K \\
0 & \hat G^A \\
\end{smallmatrix}\right)
$
which also obeys the normalization condition $\check{G}^2=\check{1}$. Both retarded and advanced $2\times2$ Green functions are $2\times2$  matrices in the Nambu space 
$
\hat G^{R,A}=\left(\begin{smallmatrix}
	G^{R,A} & F^{R,A} \\
	\tilde F^{R,A} & - G^{R,A}\\
\end{smallmatrix}\right)
$,
whereas the Keldysh matrix has the form $\hat G^K = \hat G^R \hat h - \hat h \hat G^A$, where $\hat h$ is the matrix distribution function. In Eq. (\ref{usadel_eq}) we defined the covariant spatial derivative $\hat{\partial}(\dots)=\partial_x(\dots)+ie\left[\check{{A}}_x\sigma_3,(\dots)\right]$, $[a,b]=ab-ba$ denotes the commutator, $V$ and $A$ are the scalar and vector potentials of the electromagnetic field, $D=v_F\ell /3$ is the diffusion coefficient, $\t_{1,2,3}$ and $\s_{1,2,3}$ stand for the Pauli matrices respectively in Keldysh and Nambu spaces and $\check{\Delta}$ is the superconducting order parameter matrix.

The electron DOS $\nu(E,x)$ is related to the quasiclassical Green functions in a simple way as
\begin{equation}
	\nu(E,x)=\nu_F \tr\frac{\s_3}{4}\left(G^R(E,x)-G^A(E,x)\right),
\end{equation}
where $\nu_F$ stands for DOS in a normal metal at the Fermi level and
\begin{equation}
	\G(E,x)=\int d(t-t^\prime) {\rm e}^{iE(t-t^\prime)}\G(t,t^\prime,x).
\end{equation}

It will be convenient for us to perform the rotation in the Keldysh space expressing initial field variables, e.g., the phase of the order parameter $\fee_{F,B}$ on the forward and backward branches of the Keldysh time contour in terms of their classical and quantum components $\fee_{+}= \left(\fee_F+\fee_B\right)/2$, $\fee_{-}= \fee_F-\fee_B$. We also define the matrices
\begin{equation}
	\check{\fee}=\begin{pmatrix}
		\fee_{+} & \fee_-/2\\
		\fee_-/2 & \fee_{+}
	\end{pmatrix}
\end{equation}

Employing the gauge transformation
\begin{gather}
	e\check{V}\rightarrow \check{\Phi}\equiv e\check{V}+\frac{\dot{\check{\fee}}}{2},\label{10}\\
	e\check{A}_x\rightarrow \check{\A}\equiv e\check{A}_x-\frac{\partial_x \check{\fee}}{2},\\
	{\Delta}_{\pm}\rightarrow |\Delta|_{\pm},
\label{12}
\end{gather}
we expel the phase of the order parameter from $\Delta (x,t)$ and get
\begin{equation}
	\G(t,t^\prime,x)={\rm e}^{\frac{i}{2}\check{\fee}(t,x)\s_3}\Gt(t,t^\prime,x)\nonumber {\rm e}^{-\frac{i}{2}\check{\fee}(t^\prime,x)\s_3},
\label{gauge_transform}
\end{equation}
where $\Gt$ obeys Eq. (\ref{usadel_eq}) combined with Eqs. (\ref{10})-(\ref{12}). In the next subsection we will show that one can safely put  gauge invariant combinations $\A \to 0$ and $\Phi \to 0$ and choose $\Gt$ equal to the Green function $\L$ of a uniform superconductor in thermodynamic equilibrium, i.e.
\begin{equation}
	\Gt=\L=\begin{pmatrix}
		\Lambda^R & \Lambda^K \\
		0 & \Lambda^A
	\end{pmatrix},
\end{equation}
where
\begin{equation}
	\Lambda^R_\epsilon =\frac{1}{\sqrt{(\e+i0)^2-\Delta^2}}\begin{pmatrix}
		\e & \Delta\\
		-\Delta & -\e
	\end{pmatrix},
\label{lambda_r}
\end{equation}
$\Lambda^A=-\s_3 (\Lambda^R)^\dag \s_3$ and
\begin{equation}
	\Lambda^K_\epsilon=\Lambda^R_\epsilon F_\epsilon  - F_\epsilon \Lambda^A_\epsilon,\quad F_\epsilon =\tanh\left({\frac{\e}{2T}}\right).
\label{fermionic_fdt}
\end{equation}
Then we obtain
\begin{gather}
	\G(t,t^\prime,x)\simeq {\rm e}^{ \frac{i}{2}\check{\fee}(t,x)\s_3}\L(t-t^\prime ){\rm e}^{-\frac{i}{2}\check{\fee}(t^\prime,x)\s_3},
\label{Gre}
\end{gather}
where $\L(t-t^\prime)$ is the inverse Fourier transform of $\L_\epsilon$.

\subsection{Effective action for phase fluctuations}
In order to evaluate any physical observable one needs to average the corresponding variable over all possible phase configurations. This can be done by means of the path integral technique employing an effective action $S_{\rm eff}$ which controls phase fluctuations in our system at sufficiently low energies. This action was microscopically derived and analyzed elsewhere  \cite{book,AGZ,GZQPS,OGZB}. Here we recover this action from simple symmetry arguments. 

It is well known that  global $U(1)$ symmetry is broken in a superconductor whereas local gauge symmetry remains preserved. Within this picture the phase of the order parameter plays the role of a Goldstone boson implying that the action should be constructed from the gauge invariant quantities $\A$ and $\Phi$ rather than from the phase derivatives only. The simplest appropriate form of the action then reads
\begin{equation}
	S_{\rm le}[\varphi,V,A_x]=S_{\rm em}[V,A_x]+\int dt\int dx\left[ \zeta_1(2eV+\dot\varphi)^2- \zeta_2(\partial_x\varphi-2eA_x)^2\right],
\end{equation}
where $S_{\rm em}[V,A_x]$ is action for the electromagnetic field and $\zeta_{1,2}$ are some constants. In the case of a quasi-one-dimensional wire we have
\begin{equation}
	S_{\rm em}[V,A_x]=\frac12\int dt\int dx\left[ CV^2-\frac{A_x^2}{\mathcal L}\right]
\end{equation}
where  $C$ is the capacitance per unit wire length and $\mathcal L$ is the geometric inductance times unit length. The constants $\zeta_{1,2}$ can be identified evaluating the response to static fields $V$ and $A_x$ with the result
\begin{equation}
	\zeta_1=\frac{\nu_Fs}{4},\qquad \zeta_2=\frac{1}{8e^2\mathcal L_{\rm kin}}.
\end{equation}
Integrating out the electromagnetic potentials we arrive at the effective action for the phase variable
\begin{equation}
	S_{\rm eff}[\varphi]=\int dt\int dx\left[ \frac{\nu_FsC}{4(C+2e^2\nu_F s)}\dot\varphi^2- \frac{1}{8e^2(\mathcal L+\mathcal L_{\rm kin})}(\partial_x\varphi)^2\right].
\end{equation}
In the case of interesting for us here diffusive metallic wires with diameter of order superconducting coherence length $\xi=\sqrt{D/\Delta}$ one finds $C\ll 2e^2\nu_Fs$ and $\mathcal L\ll\mathcal L_{\rm kin}$. Then we obtain
\begin{equation}
	S_{\rm eff}[\varphi]= \frac{C}{8e^2} \int dt\int dx\left[\dot\varphi^2- \frac{1}{C\mathcal L_{\rm kin}}(\partial_x\varphi)^2\right].
\end{equation}
Following the same route one can link fluctuations of the variables $\Phi$ and $\dot\varphi$ to each other. From the equations of motion we get
\begin{equation}
	\Phi=\frac{C}{C+2e^2\nu_Fs}\dot\varphi\approx \frac{1}{4E_C\nu_Fs}\dot\varphi\ll\dot\varphi
\end{equation}
and, hence, the effects related to weak ($\propto \Phi$) penetration of the fluctuating electric field inside the wire can be safely neglected \cite{GZQPS,OGZB}. Here and below $E_C=e^2/(2C)$ stands for the charging energy of a unit wire length. As usually, magnetic effects related to fluctuations of $\A$ can also be neglected in the nonrelativistic limit considered here.

For the sake of completeness let us also comment on the differences between a superconducting metal considered here and a neutral superfluid. In the latter situation there exists no interaction with the electromagnetic field which can be formally achieved by taking the limit $\mathcal L \to 0$ and $C\to\infty$, i.e. the opposite condition $C\gg 2e^2\nu_Fs$ is realized. Accordingly, the gauge transformation trick would not work anymore in this case. This fact constitutes a clear manifestation of fundamental difference between the situations of global and local gauge symmetry breaking.

\subsection{Density of states}

In order to evaluate the electron DOS we need to average the Green function (\ref{Gre}) over all possible phase configurations. This averaging is conveniently accomplished by means of the path integral technique which yields
\begin{equation}
	\left\langle \check{G}\right\rangle_\fee (t-t^\prime)=\int D\fee \, \exp\left({iS^K_{\rm eff}[\fee]}\right)\G(t,t^\prime,x),
\end{equation}
where $S^K_{\rm eff}[\fee]$ is the effective Keldysh action which accounts for phase fluctuations in a superconducting wire. As we derived above at low enough energies it can be written in the form 
\begin{eqnarray}
	S^K_{\rm eff}[\fee]=\frac{1}{16E_C}\tr\left[\begin{pmatrix}
		\fee_{+} & \fee_{-}
	\end{pmatrix}\V^{-1}\begin{pmatrix}
		\fee_{+}\\
		\fee_{-}
	\end{pmatrix}\right],
\label{Seff}
\end{eqnarray}
where
\begin{equation}
	\V=\begin{pmatrix}
		\V^K & \V^R\\
		\V^A & 0
	\end{pmatrix}
\end{equation}
is the equilibrium Keldysh matrix propagator describing plasma modes and
\begin{gather}
	\V^{R,A}(\omega,k)={\displaystyle\frac{1}{(\omega\pm i0)^2-(vk)^2}},\\
	\V^K(\omega,k)=\frac12\left(\V^R(\omega,k)-\V^A(\omega,k)\right)\coth\left({\frac{\omega}{2T}}\right).
\label{bosonic_fdt}
\end{gather}

Making use of the structure of $\L$ in the Nambu space and performing Gaussian integration, we get
\begin{multline}
	\nu(E)=\nu_F\int d(t-t^\prime){\rm e}^{iE(t-t^\prime)}\tr\left\langle \frac{\t_3\s_3}{4}{\rm e}^{ \frac{i}{2}\check{\fee}(t,x)\s_3}\L(t-t^\prime){\rm e}^{ -\frac{i}{2}\check{\fee}(t^\prime,x)\s_3}\right\rangle_\fee\\
	=\nu_F\int dt\,{\rm e}^{iEt}\tr\left(\frac{\t_3\s_3}{4}\,\t_a \L(t)\t_b \bB^{ab}(t)\right),\hspace*{0.5cm}
\label{averaging_result}
\end{multline}
where $a,b=\{0,1\}$, $\tau_0\equiv \hat 1$,
\begin{multline}
	\bB(t)=\begin{pmatrix}
		\bB^K(t) & \bB^R(t)\\
		\bB^A(t) & 0
	\end{pmatrix}={\rm e}^{ 2iE_C(\V^K(t)-\V^K(0))}\\
	\times\begin{pmatrix}
		\cos\left(E_C(\V^R(t)-\V^A(t))\right) & i\sin\left(E_C\V^R(t)\right)\\
		i\sin\left(E_C\V^A(t)\right) & 0
	\end{pmatrix}\hspace*{0.25cm}
\label{24}
\end{multline}
and
\begin{equation}
	\V(t)=\V(t,0)=\int \frac{d\omega dk}{(2\pi)^2}\,{\rm e}^{-i\omega t}\V(\omega,k).
\end{equation}
Note that Eq. (\ref{averaging_result}) accounts for all emission and absorption processes of multiple plasmons in our system via an auxiliary propagator $\bB$. This propagator obeys the standard causality requirements and satisfies bosonic fluctuation-dissipation theorem (FDT) because plasmons remain in thermodynamic equilibrium, cf. Eq. (\ref{bosonic_fdt}).

Employing this theorem and taking traces both in Keldysh and in Nambu spaces, from Eq. (\ref{averaging_result}) we obtain
\begin{multline}
	\left\langle\nu\right\rangle_{\fee}(E)=\frac{\nu_F}{4}\int dt{\rm e}^{-iEt}\tr\left(\s_3\left(\Lambda^R(t)-\Lambda^A(t)\right)\bB^K(t)+\s_3\Lambda^K(t)\left(
\bB^R(t)-\bB^A(t)\right)
\right)\\
	=\int\frac{d\e}{2\pi}\nu_{BCS}(\e)\bB^K(E-\e)\left(1+F_\epsilon F_{E-\epsilon}\right),\hspace{0.55cm}
\label{average_dos_formula}
\end{multline}
where $\nu_{BCS}(\e)$ is the BCS density of states in a bulk superconductor.

Since for $\epsilon \gtrsim E+2T$ the combination $1+F_\epsilon F_{E-\epsilon}$ decays as $\propto \exp ((E-\epsilon)/T)$, the electron DOS at subgap energies  is suppressed by the factor $\sim \exp ((E-\Delta)/T)$ and at $T \to 0$ the superconducting gap $\Delta$ is not affected by Mooij-Sch\"on plasmons at all.

Evaluating $\bB^K$ in Eq. (\ref{24}), one finds
\begin{equation}
	\bB^K(t)=\exp \left(-\frac{1}{g} \int\limits_0^{\omega_c} d\omega\,\frac{1-\cos(\omega t)}{\omega}\coth\left(\frac{\omega}{2T}\right)\right)\cos\left(\frac{1}{g} \int\limits_0^{\omega_c} d\omega\,\frac{\sin(\omega t)}{\omega}\right),
\label{Bkexpression}
\end{equation}
where an exponential high frequency cutoff at $\omega_c \sim \Delta$ is implied. This cutoff procedure is consistent with the fact that the effective action (\ref{Seff})-(\ref{bosonic_fdt}) remains applicable only at energies well below the superconducting gap.

Equation (\ref{Bkexpression}) provides a lot of useful information about the effect of phase fluctuations on DOS. For instance, the identity 
\begin{equation}
	\int dE\, \left(\nu(E)-\nu_{BCS}(E)\right)=0
\end{equation}
(which follows directly from the condition $\bB^K(t=0)=1$) implies that phase fluctuations can only redistribute the electron states among different energies without affecting the energy integrated DOS.

In the low temperature limit $T \to 0$ Eq. (\ref{Bkexpression}) reduces to
\begin{equation}
	\bB^K(t)=\left(\frac{\sinh(\pi T t)}{\pi T t}\sqrt{1+(\omega_c t)^2}\right)^{-1/g}\cos\left(\frac{\arctan(\omega_c t)}{g}\right).
\end{equation}

It is also instructive to evaluate the Fourier transform of Eq. (\ref{Bkexpression}) $\bB^K_\omega$. 
After some algebra we obtain \cite{RSZ}
\begin{equation}
	\bB^K_\omega\simeq \cosh\left(\frac{\omega}{2T}\right)\left(\frac{2\pi T}{\omega_c}\right)^{1/g}\frac{\left|\GammaF\left(\d\frac{1}{2g}+\frac{i\omega}{2\pi T}\right)\right|^2}{2\pi T \GammaF(1/g)},
\label{34}
\end{equation}
where $\omega$ remains well below the superconducting gap $\Delta$ and $\GammaF(x)$ is Euler Gamma function. At low frequencies  $\omega\ll T$ Eq. (\ref{34}) reduces further to
\begin{equation}
	\bB^K_\omega\simeq \frac{1}{g\omega_c}\left(\frac{2\pi T}{\omega_c}\right)^{1/g}\frac{2\pi T}{\omega^2+(\pi T/g)^2},
\end{equation}
whereas for the frequency interval $T \ll \omega\ll \Delta$ we find
\begin{equation}
	\bB^K_\omega\simeq \frac{\pi}{\omega_c\GammaF(1/g)}\left(\frac{\omega}{\omega_c}\right)^{1/g-1}.
\end{equation}

Making use of the above expressions, at energies in the vicinity of the superconducting gap $\Delta$ we arrive at the following result for the electron DOS
\begin{equation}
	\nu(\Delta+\omega)=\frac{\nu_F\sqrt{\Delta}}{\sqrt{2}}\left(\frac{2\pi T}{\Delta}\right)^{1/g}\sum\limits_{k=0}^\infty \frac{\GammaF(k+1/g)}{k!\GammaF(1/g)}{\rm Re}\left(\frac{{\rm e}^{-\frac{i\pi}{2g}}}{\sqrt{\omega+2i\pi T(\frac{1}{2g}+k)}}\right).
\end{equation}

\begin{figure}
	\includegraphics[width=0.32\linewidth]{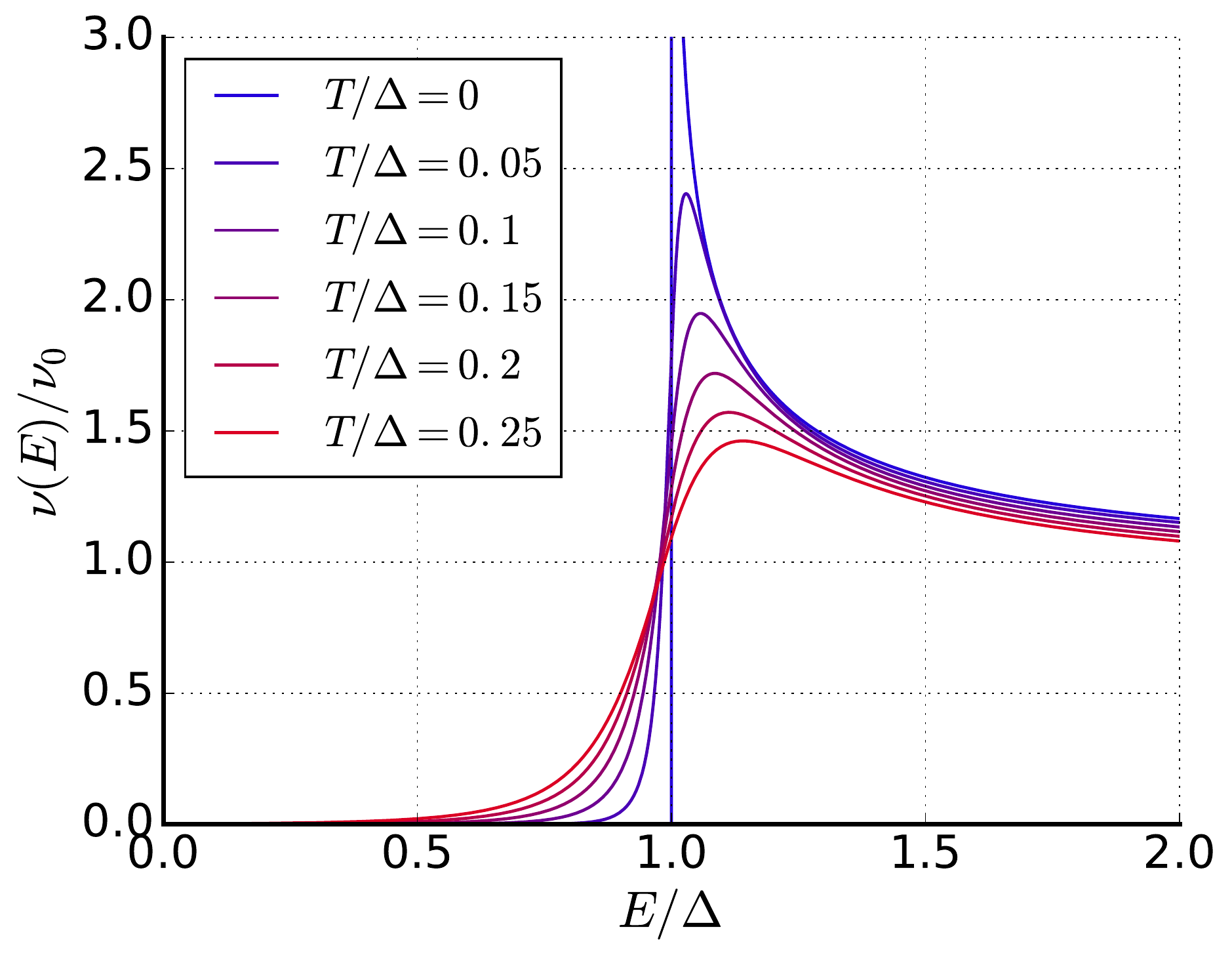}
	\includegraphics[width=0.32\linewidth]{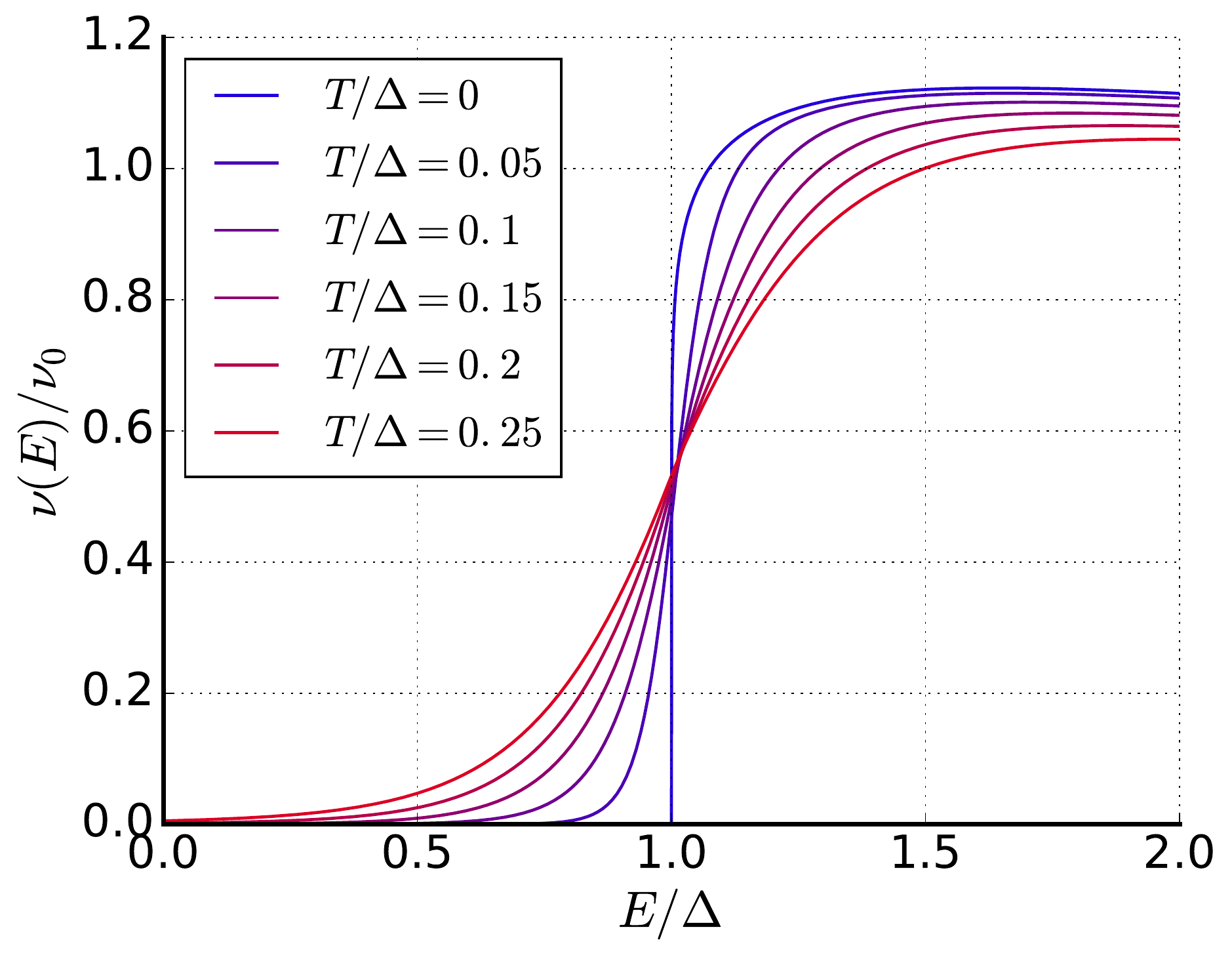}
	\includegraphics[width=0.32\linewidth]{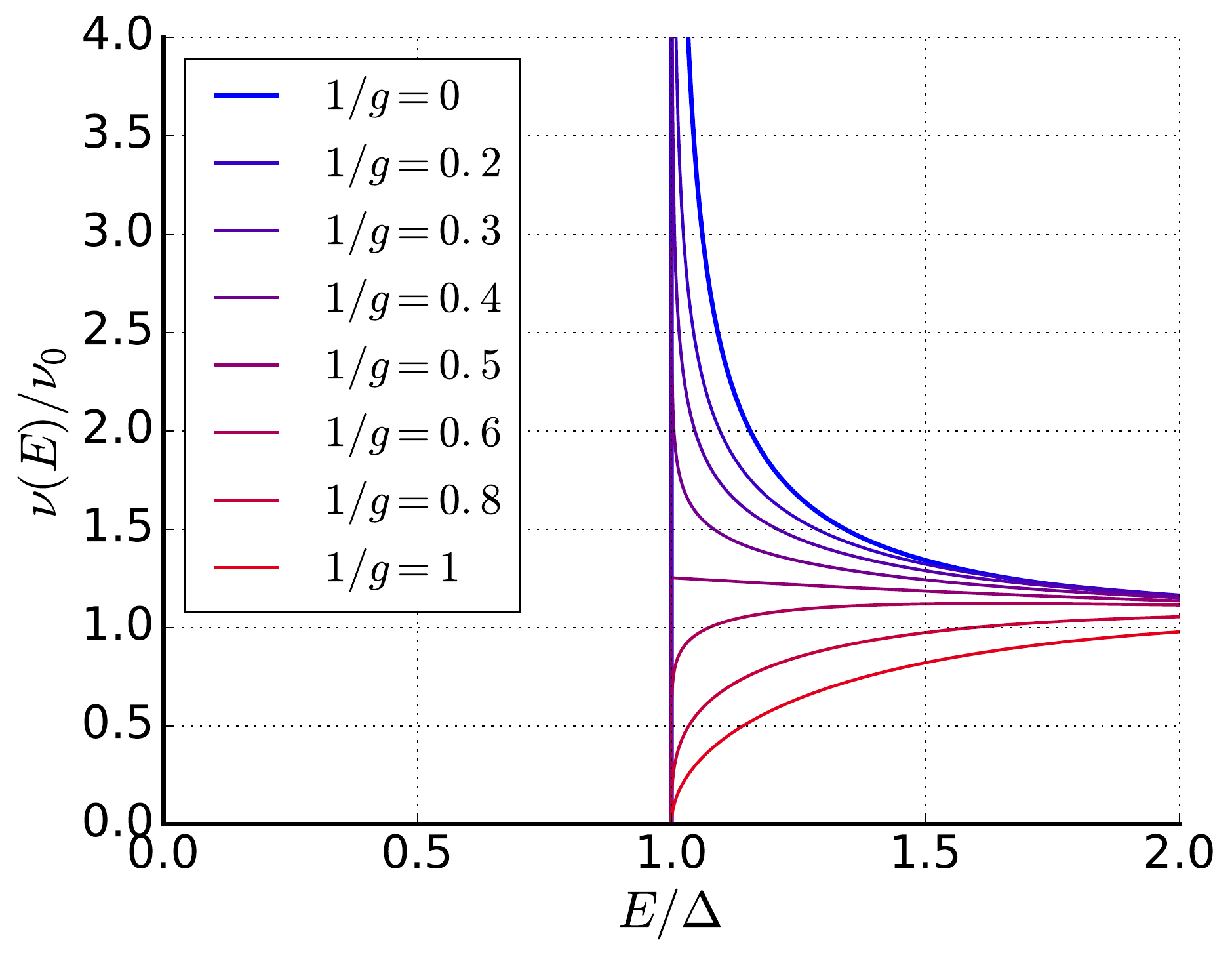}
\caption{The normalized energy dependent electron DOS $\nu(E)/\nu_F$ for superconducting nanowires at different temperatures and two values of $g=5$ (left panel) and $g=1.67$ (middle panel) as well as at $T=0$ and different values of $g$ (right panel). The energy $E$ and temperature $T$ are expressed in units of $\Delta$.}
\label{Fig:3}
\end{figure}

The energy dependent DOS $\nu (E)$ for superconducting nanowires in the presence of phase fluctuations is also displayed in Fig. \ref{Fig:3} at different temperatures and two different values of the parameter $g$. We observe that at $T\neq $ the BCS singularity at $E \to \Delta$ is smeared due to interactions between Mooij-Sch\"on plasmons and electrons propagating inside the wire. For the same reason, the electron DOS at subgap energies $0<E<\Delta$ remains non-zero at any non-zero $T$, i.e.
\begin{equation}
	\nu (E) \propto \exp ((E-\Delta)/T).
\label{38}
\end{equation}

We also note that at bigger values $g$  the function $\nu (E)$ demonstrates a non-monotonous behavior at energies slightly above the gap (top panel), whereas at smaller $g$ DOS decreases monotonously with decreasing energy at all $E$ not far from the gap (bottom panel). In the zero temperature limit $T \to 0$ and for $E-\Delta \ll \Delta$ we obtain
\begin{equation}
	\nu(E)\simeq \frac{\nu_F \sqrt{\pi} \theta(E-\Delta)}{\sqrt{2}\GammaF(\frac{1}{2}+\frac{1}{g})}\left(\frac{E-\Delta}{\Delta}\right)^{\frac{1}{g}-\frac{1}{2}}.
\label{DOST0}
\end{equation}
This result demonstrates that, while at $E<\Delta$ the electron DOS at $T=0$ vanishes at all values of $g$, the behavior of $\nu (E)$ (\ref{DOST0}) at overgap energies differs depending on the dimensionless conductance $g$. For relatively thicker wires with $g > 2$ the DOS singularity at $E \to \Delta$ survives becoming progressively weaker with decreasing $g$. In contrast,  for thinner wires with $g\leq 2$ the DOS singularity is washed out completely due to intensive phase fluctuations and $\nu (E)$ tends to zero at $E \to \Delta$ as a power law (\ref{DOST0}). This behavior is also illustrated in Fig. \ref{Fig:3} (right panel).

\subsection{Summary and comparison with experiments}
To summarize, we demonstrated that {\it local} properties of superconducting nanowires, such as the electron density of states, can be sensitive to phase fluctuations in such nanowires. At this stage we intentionally restricted our analysis to the effect of small phase fluctuations associated with low energy sound-like plasma modes propagating along the wire and forming an effective quantum dissipative environment for electrons inside the wire. 

The coupling strength between electrons inside the wire and such effective plasmon environment is controlled by the dimensionless parameter $g$. For relatively thick wires with $g \gg 1$ or, equivalently, provided the wire impedance $Z_{\rm w}$ remains much smaller than the quantum resistance unit $R_q$, phase fluctuations weakly affect the electron DOS except in the immediate vicinity of the superconducting gap $\Delta$. For larger values $Z_{\rm w} \sim R_q$ the effect of phase fluctuations becomes strong and has to be treated non-perturbatively in $1/g$ at all energies. 

At any nonzero $T$ the electron DOS depends on temperature and substantially deviates from that derived from the standard BCS theory. In particular, at $T >0$ the BCS square-root singularity in DOS at $E=\Delta$ gets totally smeared and $\nu (E)$ differs from zero also at subgap energies, cf. Eq. (\ref{38}).  This behavior can be interpreted in terms of a depairing effect due to the interaction between electrons and Mooij-Sch\"on plasmons. We also note that our results are consistent with the phenomenological Dynes formula \cite{dynes78}
\begin{equation}
	\nu (E)\simeq \nu_F{\rm Re}\left(\frac{E+i\Gamma }{\sqrt{(E+i\Gamma )^2-\Delta^2}}\right)
\label{Dynes}
\end{equation}
describing smearing of the BCS singularity in DOS in the immediate vicinity of the superconducting gap.

At $T=0$ and subgap energies the electron DOS vanishes as in the BCS theory, while the BCS singularity in DOS at $E \to \Delta$ becomes weaker for any finite $g >2$ and eventually disappears for $g \leq 2$. 

The local electron DOS in superconducting nanowires can be probed in a standard manner by performing a tunneling experiment, as it is also illustrated in Fig. \ref{Fig:2}. Attaching a normal or superconducting electrode to our wire and measuring the differential conductance of the corresponding tunnel junction one gets a direct access to the energy dependent electron DOS of a superconducting nanowire. For instance, in the case of a normal electrode at $T \to 0$ and $eV > \Delta$ one finds
\begin{equation}
	dI/dV \propto \nu (eV) \propto (V -\Delta /e)^{\frac{1}{g}-\frac{1}{2}}.
\label{CA}
\end{equation}
This power law dependence of the differential conductance resembles one encountered in small normal tunnel junctions at low voltages $dI/dV \propto V^{2/g_N}$ \cite{PZ88}, where $g_N$ is the dimensionless conductance of normal leads. In fact, both the dependence (\ref{CA}) and the zero bias anomaly in normal metallic junctions \cite{PZ88} are caused by Coulomb interaction and are controlled by the impedance of the corresponding effective electromagnetic environment. 

The experiments similar to that described above were performed with both thicker  ($\sqrt{s} >40$ nm) and ultrathin ($\sqrt{s} < 35$ nm) titanium nanowires \cite{Kostya}. While the shape of $I-V$ curves measured for thicker wires agrees well with the standard BCS-like DOS, qualitatively different behavior was found in thinner nanowires. Namely, upon decreasing of the wire diameter  (i) smaller and smaller values of the superconducting gap $\Delta$ in titanium were observed and (ii) progressively stronger smearing of the gap singularity was detected. The observation (i) is consistent with theoretical results \cite{GZTAPS} predicting suppression of the order parameter by quantum fluctuations. Indeed, as the inverse dimensionless conductance $1/g_\xi$ increases with decreasing wire diameter, the order parameter suppression becomes more pronounced (cf. Eq. (\ref{estim22})), as it was indeed observed in experiments \cite{Kostya}. 

The observed effect (ii) can be interpreted in terms of theoretical predictions \cite{RSZ} outlined in this section. For instance, the experimentally detected temperature dependence of the non-vanishing DOS tail at subgap energies \cite{Kostya} agrees well with our Eq. (\ref{38}). We conclude that experimental observations \cite{Kostya} clearly support our theory, thus, also serving as an independent confirmation of the existence of Mooij-Sch\"on plasmons in superconducting nanowires. Earlier such plasma modes were also detected within a different experimental scheme in Ref. \cite{Buisson}.

\section{Gaussian phase fluctuations in superconducting rings}

\begin{figure}
\includegraphics[width=0.7\linewidth]{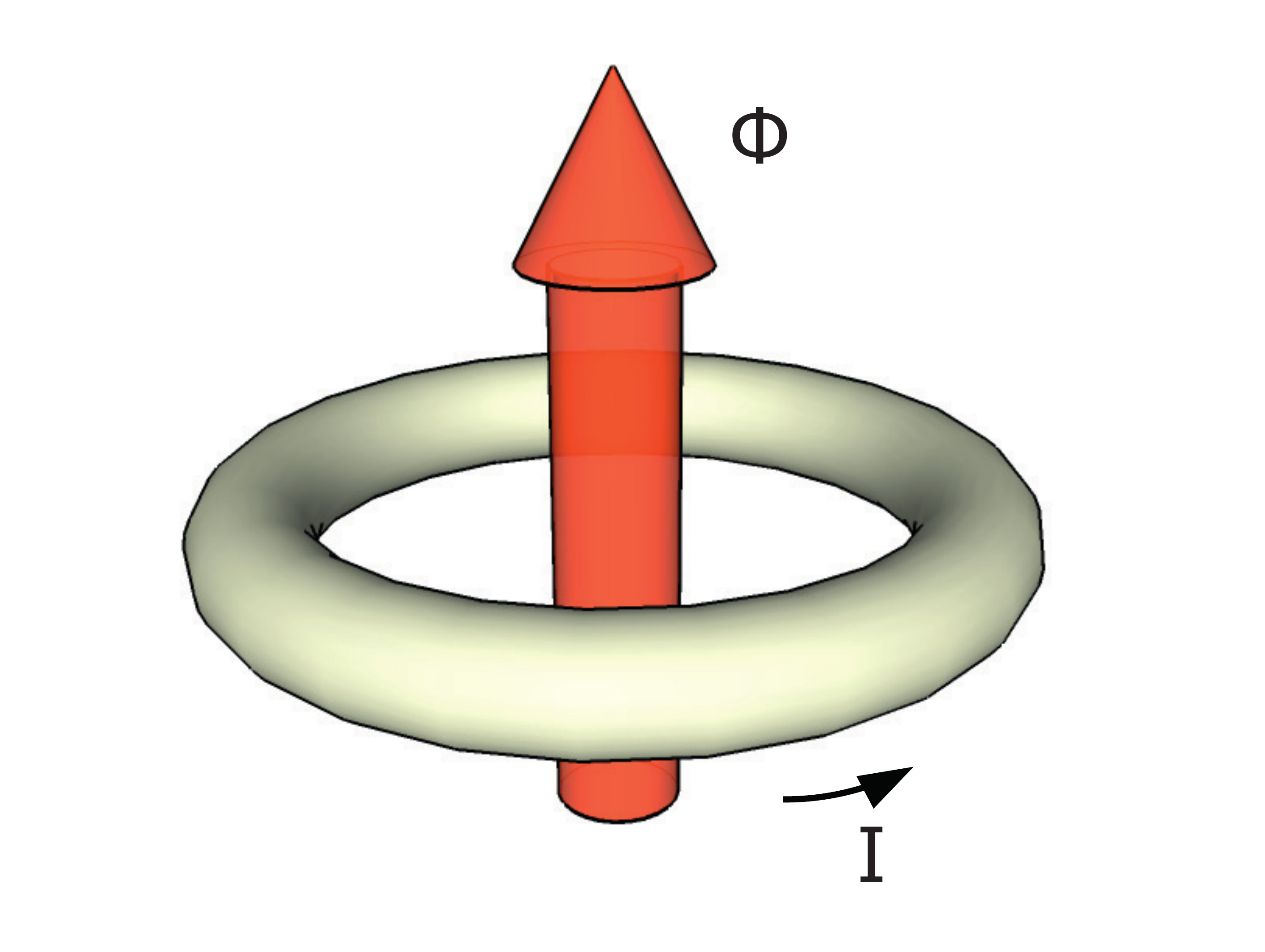}%
\caption{A superconducting ring threaded by the magnetic flux $\Phi$.}%
\label{Fig:4}%
\end{figure}

Let us now consider a somewhat different configuration for which small (Gaussian) fluctuations of the superconducting phase variable $\varphi (x, t)$ also yield interesting physical effects. We will consider a superconducting wire with cross section $s$ closed in the form of a ring of radius $R$ pierced by an external magnetic flux $\Phi_x$, see Fig. \ref{Fig:4}. As before, the wire is assumed to be thick enough (implying that $g_\xi \gg 1$) to be able to fully ignore fluctuations of the absolute value of the order parameter field $|\Delta (x,\tau )|$, where $x$ is now the coordinate along the ring and $\tau$ is the imaginary time, $0\geq \tau \geq \beta \equiv 1/T$. In this case one can construct a complete description of fluctuation effects.

\subsection{Grand partition function}
It will be convenient for us to define the grand partition function of our system $\mathcal  Z$ which can be expressed via the following path integral over the superconducting phase variable $\varphi (x,\tau )$:
\begin{equation}
\mathcal  Z=\sum\limits_{m,n}\int \mathcal D\varphi e^{-\frac{\lambda}{2\pi}\int dxd\tau\left(v(\partial_x\varphi)^2+v^{-1}(\partial_\tau\varphi)^2\right)},
\label{partf}
\end{equation}
where we introduced an effective coupling constant $\lambda \equiv g/8$ \cite{ZGOZ}. According to our assumptions, this partition function
includes only small fluctuations of the superconducting phase $\varphi$ described by the imaginary time version of the effective action (\ref{Seff}) in the exponent of Eq. (\ref{partf}).

The path integral (\ref{partf}) should be supplemented by proper boundary conditions which should keep track of (a)
periodicity of the phase variable $\varphi$ in space-time, (b) the fact that the phase is defined up to $2\pi m$, where $m$ 
is an arbitrary integer number (the so-called winding number) and (c) the magnetic flux $\Phi_x$
piercing the ring. Putting all these requirements together, we obtain
\begin{eqnarray}
\varphi(x,0)=\varphi(x,\beta)+2\pi m,\nonumber\quad \\
\varphi(L,\tau)=\varphi(0,\tau)+2\pi(\phi_x+n).
\label{boucond}
\end{eqnarray}
Here $L=2\pi R$ is the ring perimeter, $\phi_{x}=\Phi/\Phi_{0}$ and $\Phi_0=\pi c/e$ is the superconducting
flux quantum. Combining Eqs. (\ref{partf}) and (\ref{boucond}), after a simple calculation one finds
\begin{equation}
\mathcal Z=\sum\limits_{n=-\infty}^\infty e^{-\beta E_n(\phi_x)}=\sqrt{\frac{2\pi T}{E_R} }\vartheta_3(\pi\phi_x,e^{-\frac{2\pi^2T}{E_R}})
\end{equation}
where  are defined as
\begin{equation}
E_n(\phi_x)=\frac{E_R}{2}(n+\phi_x)^2
\end{equation}
are the flux-dependent energy levels of the ring, $\vartheta_k(u,q)$ is the third Jacobi Theta function and 
\begin{equation}
E_R=\frac{4\pi\lambda v}{L}=\frac{\pi^2\nu_FD\Delta s}{R}.
\end{equation}
In the limit $T \to 0$, the supercurrent $I(\phi_x)$ flowing around the ring in its ground state is obtained by means of a well known simple formula
\begin{equation}
I(\phi_x)=\partial E(\Phi_x)/\partial \Phi_x,
\label{pf}
\end{equation}
where $E(\Phi_x)=\mathrm{min}_{n}E_n(\phi_x)$ is the ground-state energy of the ring, which is periodic in $\Phi_x$ with the period $\Phi_0$.
Hence, the supercurrent $I$ is also periodic in $\Phi_x$, being defined as
\begin{equation}
I(\phi_{x})=\frac{eE_R}{\pi}\frac{\partial}{\partial \phi_x}\mathrm{min}_{n}\left(  n+\phi_{x}\right)^2.
\label{evcur}
\end{equation}

The supercurrent flowing across the ring can also be expressed in terms of the phase variable by means of the relation
\begin{equation}
I(\tau)=\frac{eE_R}{2\pi^2 }\left(\varphi(L,\tau)-\varphi(0,\tau)\right).
\label{current0}
\end{equation}
Combining Eq. (\ref{current0}) with the second Eq. (\ref{boucond}), in the zero temperature limit we again recover the expression for the expectation value of the current operator $I(\phi_{x})=\langle I(\tau )\rangle$.

\subsection{Coherent fluctuations of supercurrent}

Fluctuations of the phase should in general cause fluctuations of the supercurrent flowing inside the ring. It turns out that under the conditions formulated above one can derive formally exact expressions for all current correlators in our problem. Employing Eq. (\ref{current0}) it is straightforward to demonstrate that all these correlators do not depend on time and
can be expressed through the derivatives of the Theta function $\vartheta_3^{(k,0)}$ as \cite{SZ13}
\begin{multline}
\langle \hat I(\tau_1)...\hat I(\tau_k)\rangle=
\left(\frac{eE_R}{\pi}\right)^k\frac{\sum\limits_{n=-\infty}^\infty (n+\phi_x)^ke^{-\beta E_n(\phi_x)}}{\sum\limits_{n=-\infty}^\infty e^{-\beta E_n(\phi_x)}}\\
=\sum\limits_{n=0}^{2n\leq k}\frac{(-eT)^kk!}{n!(k-2n)!}\left(\frac{E_R}{2\pi^2T}\right)^n
\frac{\vartheta_3^{(k-2n,0)}(\pi\phi_x,e^{-\frac{2\pi^2T }{E_R}})}{\vartheta_3(\pi\phi_x,e^{-\frac{2\pi^2T }{E_R}})}.
\label{curcor000}
\end{multline}
At $T \to 0$ this expression obviously yields 
\begin{equation}
\langle I^k(\tau)\rangle =\langle I(\tau)\rangle^k
\label{Ik}
\end{equation}
for all integer $k \geq 0$.  Equation (\ref{Ik}) implies that {\it no supercurrent fluctuations can occur in the ground state of superconducting rings.}  This observation is fully consistent with a general theorem \cite{SZ10} stating that no persistent current fluctuations can occur at $T=0$ provided the current operator commutes with the total Hamiltonian of the ring. Sufficiently thick superconducting rings where QPS effects can be totally neglected represent an example of this physical situation.

Supercurrent fluctuations, however, may and do occur at non-zero temperatures. Below let us focus our attention on the supercurrent noise which power spectrum reads
\begin{equation}
S_\omega=\int dt e^{i\omega t} S(t),
\end{equation}
where
\begin{equation}
S(t)=\frac12\left\langle \hat I(t)\hat I(0)+\hat I(0)\hat I(t)\right\rangle-\left\langle\hat I\right\rangle^2
\label{soft}
\end{equation}
and $\hat I(t)=e^{it\hat H}\hat I e^{-it\hat H}$  is the current operator in the Heisenberg representation and $\hat H$ is the system Hamiltonian. In order to evaluate the above current-current correlation function it will be convenient for us to also define the irreducible Matsubara correlator
\begin{equation}
\Pi(\tau)=T\sum\limits_k e^{-i\omega_k\tau}\Pi_{i\omega_k}=\langle \hat I_M(\tau)\hat I_M(0)\rangle-\langle\hat I\rangle^2,
\label{pi}
\end{equation}
where $\hat I_M=e^{\tau\hat H}\hat Ie^{-\tau\hat H}$ is the current operator in the Matsubara representation and $\omega_k=2\pi k T$ is the Matsubara frequency. It is also important that from the expression for the imaginary time correlator
(\ref{pi}) one can directly recover the real time PC noise power spectrum

The quantities $S(t)$ and $\Pi(\tau)$ defined respectively in Eqs.  (\ref{soft}) and (\ref{pi}) can be related to each other through the appropriate analytic continuation procedure combined with the fluctuation-dissipation theorem. Expressing both correlators in terms of the exact eigenstates $E_m$ of the system Hamiltonian $\hat H|m\rangle=E_m|m\rangle$, we find
\begin{equation}
S_\omega=2\pi P\delta(\omega)+\frac{\pi}{\mathcal Z}\sum\limits_{m\neq n}|\langle m|\hat I|n\rangle|^2\left(e^{-\beta E_n}+e^{-\beta E_m}\right)\delta(\omega+E_n-E_m)
\label{soft2}
\end{equation}
and
\begin{equation}
 \Pi_{i\omega_k}=\beta P\delta_{k,0}+\frac{1}{\mathcal Z}\sum\limits_{m\neq n}|\langle m|\hat I|n\rangle|^2\frac{e^{-\beta E_m}-e^{-\beta E_n}}{i\omega_k+E_n-E_m},
 \label{pi0}
\end{equation}
where
\begin{equation}
P=\frac{1}{\mathcal Z}\sum\limits_n|\langle n|\hat I|n\rangle|^2e^{-\beta E_n}-I^2
\label{PPP}
\end{equation}
defines the zero-frequency contribution and  $I=\mathcal Z^{-1}\sum_n\langle n|\hat I|n\rangle e^{-\beta E_n}$ is the expectation value for the current. With the aid of the above general expressions one easily arrives at the relation
\begin{equation}
{\rm Im}\left[\left.\left(\Pi_{i\omega_k}-\beta P\delta_{k,0}\right)\right|_{i\omega_k\to\omega+i0}\right]=\tanh\left(\frac{\omega}{2T}\right)S_\omega. 
\label{cont}
\end{equation}
Equation (\ref{cont}) enables one to recover the current noise power spectrum $S_\omega$ directly from the imaginary time analysis.

It follows from Eqs. (\ref{soft2}), (\ref{PPP}) that in the zero temperature limit (i) $P\equiv0$, i.e. zero frequency the supercurrent noise vanishes identically and (ii) at non-zero frequencies this noise also vanishes  provided the current operator commutes with the system Hamiltonian $\hat H$. 

At non-zero temperatures the supercurrent noise power does not vanish being peaked at zero frequency,
\begin{equation}
S_\omega=2\pi P \delta(\omega),
\end{equation}
where from Eq. (\ref{curcor000}) one finds \cite{SZ13}
\begin{equation}
P=e^2T^2\left(\frac{E_R}{\pi^2 T}+\frac{\vartheta_3^{(2,0)}(\pi\phi_x,e^{-\frac{2\pi^2 T}{E_R}})}{\vartheta_3(\pi\phi_x,e^{-\frac{2\pi^2T}{E_R}})}\right)-e^2T^2\left(\frac{\vartheta_3^{(1,0)}(\pi\phi_x,e^{-\frac{2\pi^2T}{E_R}})}{\vartheta_3
(\pi\phi_x,e^{-\frac{2\pi^2T}{E_R}})}\right)^2.
\label{nzT}
\end{equation}
In the low and high temperature limits this expression reduces to
\begin{equation}
P\approx\begin{cases}
\frac{2e^2E_R^2}{\pi^2}e^{-\frac{E_R}{2T}}\cosh\left(\frac{\pi I(\phi_x)}{eT}\right), & T\ll E_R,\\
\frac{e^2E_R T}{\pi^2}-8e^2T^2e^{-\frac{2\pi^2T}{E_R}}\cos(2\pi\phi_x), & T\gg E_R,
\end{cases}
\end{equation}
where $I(\phi_x)$ is defined in Eq. (\ref{evcur}). Also, one can note that in the absence of phase slips $P$ can be related to the second derivative of free energy $\mathcal F$ over the flux
\begin{equation}
  P=\frac{e^2TE_R}{\pi^2}-T\frac{\partial^2\mathcal F}{\partial\Phi_x^2}
\end{equation}
and thus to the difference between Drude and Meissner weights of the system.\cite{SQZ1993,resta2018}

\begin{figure}
\includegraphics[width=0.8\linewidth]{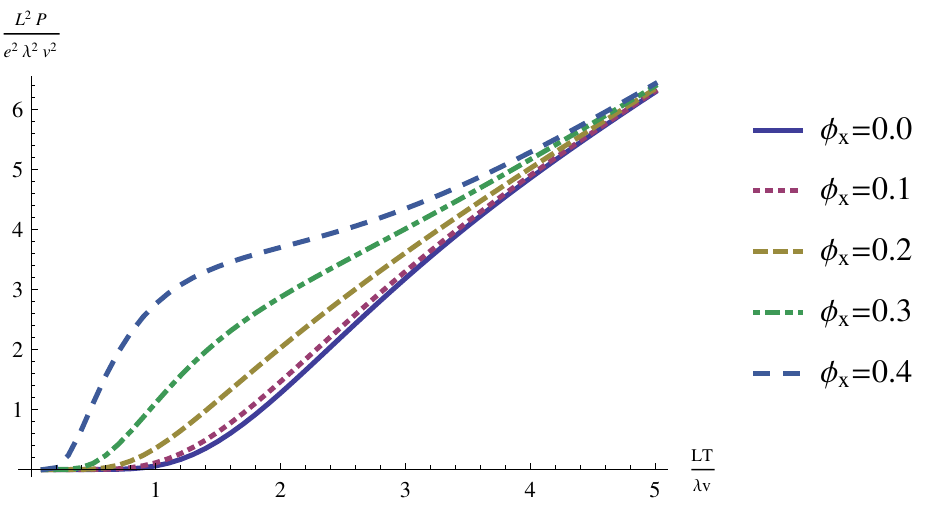}
\caption{Temperature dependent zero frequency supercurrent fluctuations in superconducting rings at different values of the magnetic flux $\phi_x$.}
\label{Fig:5}
\end{figure}
The dependence (\ref{nzT}) is also depicted in Fig. \ref{Fig:5} for different values of the magnetic flux. We observe that at sufficiently low temperatures the magnitude of PC fluctuations can be tuned by the external flux $\phi_x$, hence, indicating coherent nature of such fluctuations. At higher temperatures quantum coherence is destroyed and $P(T) \propto T$ becomes practically independent of $\phi_x$.

The above picture remains applicable as long as
the ring is sufficiently thick and one can essentially ignore quantum fluctuations of the absolute value of the order parameter field.
However, upon decreasing the wire diameter $\sim \sqrt{s}$, typically down to values in the $10$ nm
range, one eventually reaches the regime in which quantum fluctuations of $|\Delta (x)|$ gain importance and
may strongly modify the low temperature behavior of the system. This regime of strong quantum fluctuations in superconducting nanowires and nanorings will be considered in the forthcoming sections.

\section{Quantum phase slips and phase-charge duality}
 As we already discussed, at low temperatures the most significant non-Gaussian quantum fluctuations in superconducting nanowires are quantum phase slips. Provided such a wire is sufficiently thin quantum fluctuations may yield temporal local suppression of the absolute value of the superconducting order parameter field $\Delta (x)=|\Delta(x)|e^{i\varphi(x)}$ in different points along the wire. As soon as the modulus of the order parameter $|\Delta(x)|$ in the point $x$  vanishes, the
phase $\varphi(x)$ becomes unrestricted and can jump by the value $\pm 2\pi$. After this process the modulus $|\Delta(x)|$ gets restored, the phase becomes single valued again and the system returns to its initial state accumulating the net phase shift $\pm 2\pi$. 

Loosely speaking, each QPS event involves suppression of the order parameter {\it inside} the
phase slip core and a winding of the superconducting phase {\it around}
this core. This process can also be viewed as quantum tunneling of the
order parameter field through an effective potential barrier. As the phase $\varphi$ changes in time, according to the Josephson relation $V=\dot{\varphi}/2e$ each QPS event causes a voltage pulse inside the wire, thus essentially influencing the system electrodynamics.

An important property of superconducting nanowires is the so-called phase-charge duality. This property will be essentially explored below in this section. Note, that earlier duality between the phase and the charge variables was extensively discussed for 
ultrasmall Josephson junctions \cite{ZP87,PZ88,AverinOd,Z90,SZ90}. In particular, it was demonstrated that under a certain duality transformation the effective actions for Josephson tunnel junctions in the phase and in the charge representations are {\it exactly} transformed onto each other.
Furthermore, in the absence of a shunt resistor one can describe the Josephson junction in terms of an effective Hamiltonian for a "quantum particle" in the periodic potential in the (quasi)-charge space. Within this picture, the charge $q$ and the flux $\Phi$ are canonically conjugate variables exactly analogous to the momentum and coordinate variables in quantum mechanics. For more details on this issue we refer the reader to the book \cite{book} and the review paper \cite{SZ90}.

Later on it was pointed out \cite{MN} that all the same arguments remain applicable for short superconducting nanowires in the presence of quantum phase slips which properties are exactly dual to those of Josephson junctions. For this reason a short superconducting nanowire was named a {\it QPS junction} \cite{MN}. Note, that the "Josephson-junction-like" duality arguments are strictly applicable only to sufficiently short nanowires, i.e. as long as the coordinate dependence of both the phase and the charge variables can be neglected and the QPS junction can be effectively treated as a zero-dimensional object. It turns out that one can also modify and extend these arguments further to the case of long superconducting nanowires. This task can be accomplished either by means a rigorous path integral analysis \cite{SZ13} or in terms of simple quantum mechanical operator manipulations \cite{SZ17b}. Both methods will be outlined below.

\subsection{Phase-charge duality in the operator formalism}
We first consider a more intuitive operator approach. As before, we are going to deal with a uniform superconducting wire of length $L$ and cross section $s$. The effective Hamiltonian of the wire can be expressed in a simple form 
\begin{equation}
\hat H_{\rm eff} = \int_0^L dx \left[ \frac{\hat Q^2(x)}{2C} + \frac{1}{2{\mathcal L}_{\rm kin}}\left(\frac{\partial_x \hat \varphi (x)}{2e}\right)^2\right],
\label{Heff19}
\end{equation}
where $\hat Q(x)$ and $\hat \varphi(x)$ are canonically conjugate local charge and phase operators obeying the commutation relations
\begin{equation}
[\hat Q(x),\hat \varphi (x')]=-2ie\delta (x-x').
\end{equation}
Employing the Hamiltonian (\ref{Heff19}) one should arrive at the results exactly equivalent to those derived, e.g., within the effective action approach based on Eq. (\ref{Seff}) or Eq. (\ref{partf}). 

At this stage we will assume that our superconducting wire is isolated from any external circuit, in which case the current at its end points $x=0$ and $x=L$ vanishes and, hence, we can define the boundary conditions for the phase in the form
\begin{equation}
\partial_x \hat \varphi (0)=\partial_x\hat \varphi (L)=0.
\label{bouvarp19}
\end{equation}
Employing the Fourier series expansion, we get
\begin{equation}
\hat \varphi (x) = \hat \varphi_0 + \sqrt{\frac{2}{L}}\sum_{n=1}^{\infty}\hat \varphi_n\cos (\pi nx/L), \qquad
\hat Q (x) = \frac{\hat Q_0}{X} + \sqrt{\frac{2}{L}}\sum_{n=1}^{\infty}\hat Q_n\cos (\pi nx/L),
\end{equation}
where
\begin{equation}
[\hat Q_0,\hat \varphi_0]=-2{\rm i}e, \qquad [\hat Q_m,\hat \varphi_n]=-2{\rm i}e\delta_{mn}.
\end{equation}

Let us now introduce the following (dual) operators
\begin{equation}
\hat \Phi (x)=\frac{\partial_x\hat \varphi (x)}{2e}, \qquad \hat \chi (x)=- \frac{\pi}{e} \int_x^Ldx'\hat Q (x')+
\frac{\pi (L-x)}{eL} \int_0^Ldx'\hat Q (x'),
\end{equation}
which can also be expressed as
\begin{equation}
\hat \Phi (x) = - \sqrt{\frac{\pi^2}{2e^2L^3}}\sum_{n=1}^{\infty}n\hat \varphi_n\sin (\pi nx/L),\qquad
\hat \chi (x) = \sqrt{\frac{2L}{e^2}}\sum_{n=1}^{\infty}\frac{\hat Q_n}{n}\sin (\pi nx/L).
\end{equation}
These new canonically conjugate operators obey the commutation relations
\begin{equation}
[\hat\Phi (x),\hat\chi (x')]=- i \Phi_0\delta (x-x')
\end{equation}
and obvious boundary conditions
\begin{equation}
\hat\Phi (0)= \hat\Phi (L)=0, \qquad \hat\chi (0)=\hat\chi (L)=0.
\end{equation}
Substituting the relations
\begin{equation}
\partial_x\hat \varphi (x)= 2e\hat \Phi (x), \qquad \hat Q(x)=\frac{\hat Q_0}{L}+ \frac{e}{\pi}\partial_x \hat \chi (x)
\label{rel19}
\end{equation}
into Eq. (\ref{Heff19}), we obtain
\begin{equation}
\hat H_{\rm eff} = \frac{\hat Q_0^2}{2LC}+\hat H_{TL},
\label{Heff192}
\end{equation}
where
\begin{equation}
\hat H_{TL}=\int_0^Ldx \left(\frac{\hat\Phi^2}{2{\mathcal L}_{\rm kin}}+\frac{1}{2C}\left(\frac{\partial_x \hat\chi }{\Phi_0}\right)^2\right)
\label{HTL}
\end{equation}
is the Hamiltonian for a transmission line formed by a superconducting wire.

The above analysis does not yet include the effect of QPS.  In order to account for the QPS contribution to the wire Hamiltonian let us first define the phase field configurations as
\begin{equation}
\hat \varphi (x)|\varphi (x)\rangle =\varphi (x)|\varphi (x)\rangle
\end{equation}
and bear in mind that the phase of the superconducting order parameter is a compact variable implying that, e.g, the field configurations $\varphi (x)$ and $\varphi (x) +2\pi$
correspond to the same quantum state of our system. Furthermore, in the absence of QPS, i.e. provided the
absolute value of the order parameter $|\Delta (x,t)|$ does not fluctuate, also the states $\varphi$ and $\tilde \varphi (x)=\varphi (x)+2\pi \theta (x-x_1)$ (where $0 < x_1 < L$ and $\theta (x)$ is the Heaviside step function equal to 0 for $x\leq 0$ and to 1 for $x>0$) are physically indistinguishable. For instance, the supercurrent operator $\hat I$ proportional to the combination $|\Delta|^2 \exp (-{\rm i}\hat \varphi (x)) \partial_x \exp ({\rm i}\hat \varphi (x))$ remains the same in both cases.

Let us now slightly modify the step function making it continuous by effectively smearing it at the scale of the superconducting coherence length $\xi$.  We substitute $\theta (x) \to \theta_\xi (x)$, where field configuration $\tilde \varphi_\xi (x)=\varphi (x)+2\pi \theta_\xi (x-x_1)$, on one hand, remains very close
to $\tilde \varphi (x)$ and, on the other hand, is now physically distinguishable from the latter. The QPS process can be interpreted as quantum tunneling between these two different (though very close to each other) phase configurations.

Making use of the fact that any shift by a constant phase does not change the state of our system, without loss of generality we may set $\hat \varphi_0|\psi\rangle =0$ for any system state $\psi\rangle$. This condition applies for quantum dynamics controlled by the Hamiltonian (\ref{Heff19}) and it is also maintained in the presence of quantum phase slips. Hence, we conclude that the QPS process corresponds to quantum tunneling of the phase between the states $\varphi (x)$ and
\begin{equation}
\varphi'(x)=\varphi (x)+2\pi \theta_\xi (x-x_1)-2\pi \int_0^Ldx\theta_\xi (x-x_1).
\end{equation}
In the operator language this tunneling process can be denoted as $\hat U_\xi (x_1)|\varphi (x)\rangle =|\varphi'(x)\rangle$, where the operator $\hat U_\xi (x_1)$ can be established with the aid of the commutation relations. It reads
\begin{equation}
\hat U_\xi (x_1) = \exp \left(\frac{{\rm i}\pi}{e}\int_0^Ldx (\hat Q(x)-\hat Q_0/L)\theta_\xi (x-x_1)\right).
\end{equation}
As a result, the part of the Hamiltonian which explicitly accounts for the QPS contribution takes the form
\begin{equation}
\label{HQPS}
\hat H_{QPS}=-\gamma_{QPS}\int_0^Ldx_1 \cos \left(\frac{\pi}{e}\int_0^Ldx (\hat Q(x)-\hat Q_0/L)\theta_\xi (x-x_1)\right).
\nonumber
\end{equation}
Setting now $\xi \to 0$ and making use of the second Eq. (\ref{rel19}), we obtain
\begin{equation}
\hat H_{QPS}=-\gamma_{QPS}\int_0^Ldx \cos (\hat \chi(x)) .
\label{HQPS0}
\end{equation}

\begin{figure}
\includegraphics[width=0.9\linewidth]{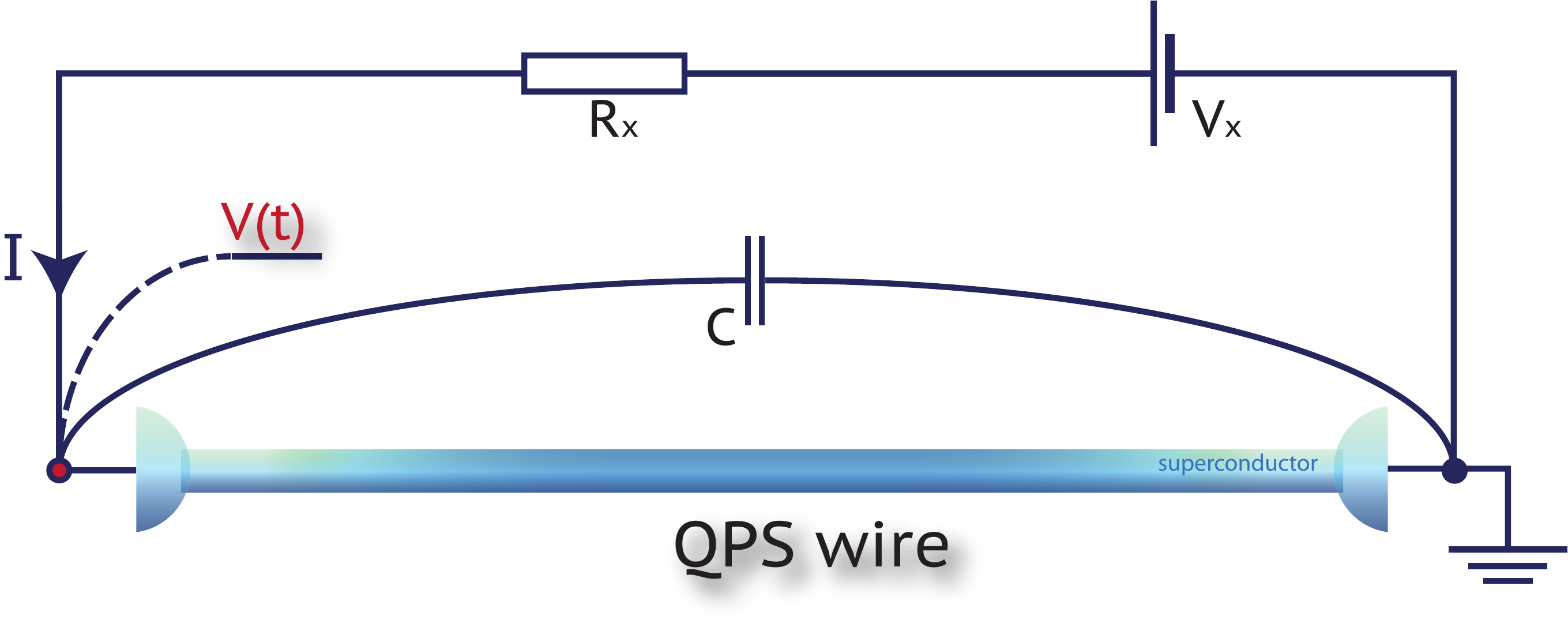}
\caption{A superconducting circuit embedded in an external circuit.}
\label{Fig:6}
\end{figure}

The above analysis can be easily generalized in order to include the effect of an external circuit. 
Let us consider one example of an external circuit displayed in Fig. \ref{Fig:6}.
The system consists of a superconducting nanowire and a capacitance $C_0$ (which also includes the wire capacitance $C$) switched in parallel to this wire. The right end of the wire ($x=L$) is grounded as shown in the figure. The voltage $V(t)$ at its left end $x=0$ can be measured by a detector. The whole system is biased by an external current $I=V_x/R_x$.

The system depicted in Fig. \ref{Fig:6} can be described by means of the effective Hamiltonian in the mixed phase-charge representation:
\begin{equation}
\hat H = \hat H_{\rm SW}+\frac{\hat Q_0^2}{2C_0}-\frac{I\hat\varphi}{2e},
\label{HAm}
\end{equation}
where the term
\begin{equation}
\hat H_{\rm SW} = \hat H_{TL}+ \hat H_{QPS}
\label{HAmw}
\end{equation}
defined by Eqs. (\ref{HTL}), (\ref{HQPS0}) accounts for the superconducting nanowire.  The last two terms in Eq. (\ref{HAm}) describe respectively the charging energy (which also includes the first term in the right-hand side of Eq. (\ref{Heff192})) and the potential energy tilt produced by an external current $I$. The operator $\hat \varphi \equiv \hat \varphi (0)$ corresponds to the phase of the superconducting order parameter field $\Delta (x,t)$ at $x=0$. Here we also set $\hat \varphi (L)\equiv 0$.

\begin{figure}
\includegraphics[width=0.7\linewidth]{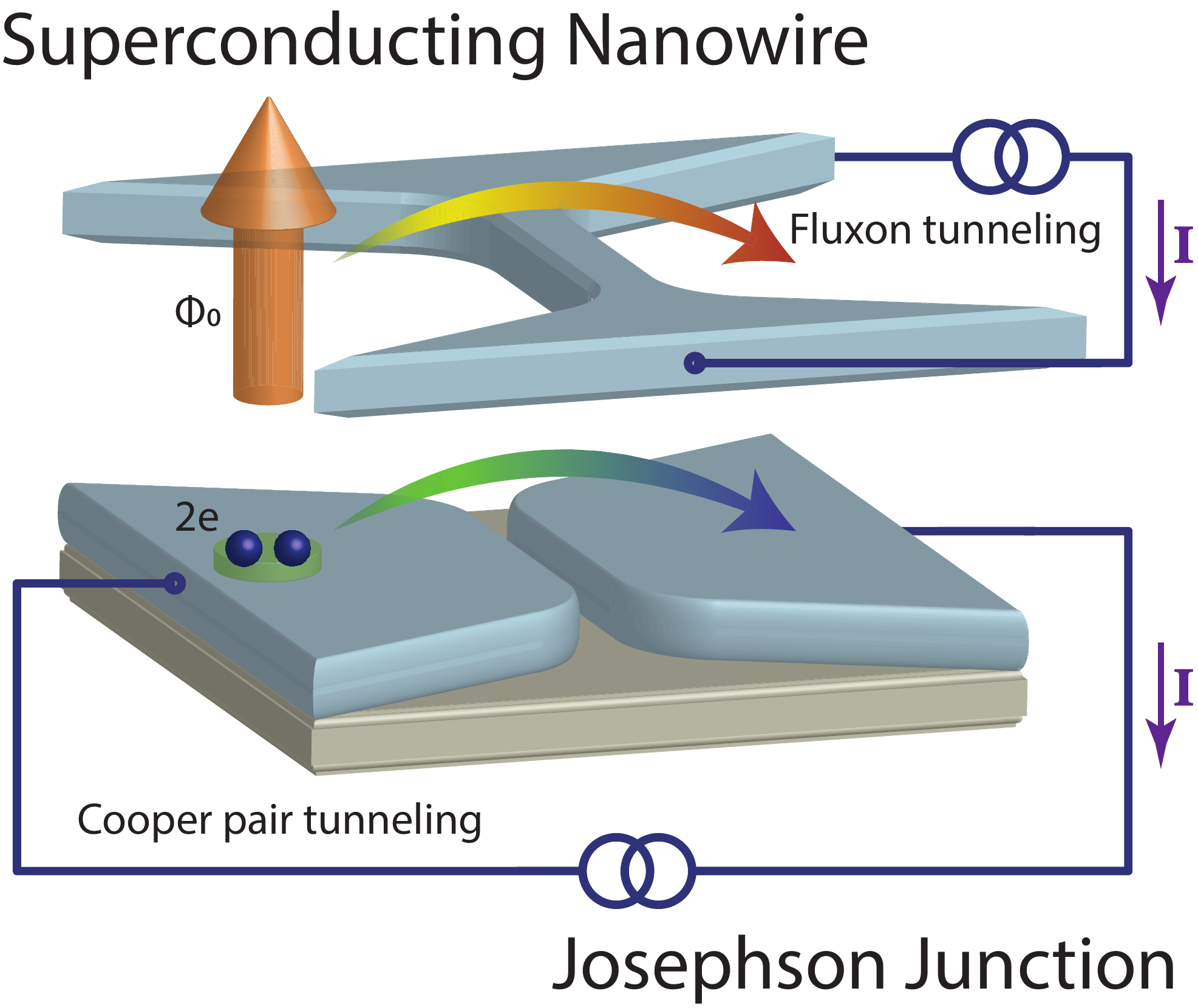}
\caption{Dual tunneling processes for a quantum fluxon (that tunnels through a superconducting nanowire) and for a Cooper pair (which tunnels across a Josephson junction).}%
\label{Fig:7}%
\end{figure}
  
Let us now take a quick look at Fig. \ref{Fig:7} where we display two complementary superconducting devices. One of them is an arbitrarily long  superconducting nanowire surrounded by the vacuum or an insulator (upper part of the figure). In the device depicted in the lower part of the figure the superconductor is interchanged with the vacuum/insulator, thus forming a spatially extended Josephson junction between two superconductors. As we just demonstrated, a superconducting nanowire in Fig. \ref{Fig:7} is described by the Hamiltonian  (\ref{HAmw}), whereas the Hamiltonian corresponding to a Josephson junction (of length $L$) in Fig. \ref{Fig:7} is well known to have the form  (see, e.g., \cite{BP})
\begin{equation}
\nonumber
\hat H_{\rm JJ} = \int_0^L dx \left[ \frac{\hat {\mathcal Q}^2(x)}{2C_J} + \frac{1}{2{\mathcal L}_J}\left(\frac{\partial_x \hat \phi (x)}{2e}\right)^2\right]-\frac{j_C}{2e} \int_0^L dx\cos (\hat \phi (x)),
\label{HJJ}
\end{equation}
where $\hat {\mathcal Q}(x)$ and  $\hat \phi (x)$ are the local charge and phase difference operators, $C_J$ and ${\mathcal L}_J$ represent respectively the Josephson junction capacitance and inductance per unit junction length and, finally, $j_C$ is the Josephson critical current density. We observe that under the transformation of the operators 
\begin{equation}
\hat\Phi (x)  \leftrightarrow \hat {\mathcal Q}(x), \qquad \hat \chi (x) \leftrightarrow \hat \phi (x)
\label{optrans}
\end{equation}
the Hamiltonians $\hat H_{\rm SW}$ (defined by Eqs. (\ref{HTL}) and (\ref{HQPS0}))  and $\hat H_{\rm JJ}$ (\ref{HJJ}) are {\it exactly dual} to each other provided we interchange
\begin{equation}
\Phi_0 \leftrightarrow 2e,\qquad \gamma_{QPS}  \leftrightarrow \frac{j_C}{2e},\qquad
{\mathcal L}_{\rm kin} \leftrightarrow C_J, \qquad C  \leftrightarrow {\mathcal L}_J.
\label{dualtrans}
\end{equation}
The above duality transformations -- on one hand -- interchange magnetic and charging energies in these two Hamiltonians
(cf. Eq.  (\ref{HTL}) and the first line in Eq. (\ref{HJJ})) and -- on the other hand -- establish the correspondence between the term 
(\ref{HQPS0}) describing the effect of QPS and the Josephson coupling energy in the second line in Eq. (\ref{HJJ})) that accounts 
for Cooper pair tunneling across the junction. 

We conclude that tunneling of a Cooper pair with charge $2e$ between two superconductors is a dual process to a QPS event that can be viewed as tunneling of a quantum fluxon (i.e. the flux quantum $\Phi_0$) across a superconducting wire, as it is illustrated in Fig. \ref{Fig:7}. Indeed, we note that the Hamiltonian (\ref{HQPS0}) contains a linear combination of creation ($e^{i\hat{\chi}}$) and annihilation ($e^{-i\hat{\chi}}$) operators for the flux quantum $\Phi_{0}$.  Each QPS event corresponds to the net phase jump by $2\pi$ associated with voltage pulse  $\delta V=\dot{\varphi}/2e$ and magnetic flux $\int |\delta V(t)|dt=\pi/e \equiv \Phi_0$ passing through the wire in the direction normal to its axis. 

Also the physical meaning of the quantum field $\chi (x,t)$ is transparent: It is proportional to the total electric charge $q(x,t)$ that has passed through the point $x$ up to the time moment $t$, i.e. $q(x,t)=\chi (x,t)/\Phi_0$. Accordingly, the local current $I(x,t)$ and the local charge density $\rho (x,t)$ are defined as
\begin{equation}
	I(x,t)=\partial_t\chi (x,t)/\Phi_0, \qquad \rho (x,t)=-\partial_x\chi (x,t)/\Phi_0,
\end{equation}
thereby satisfying the continuity equation.

The property of the phase-charge duality in superconducting nanowires was confirmed and illustrated in a number of experiments. For instance,  tunneling of magnetic flux quanta through such nanowires was detected in experiments \cite{AstNature,AstPRB}. Insulating behavior of these nanowires as well as Bloch steps (dual to Shapiro ones) on their $I-V$ curves were also reported experimentally \cite{Kostya2}.
Phase-charge duality-based single-charge transistor and charge quantum interference device were demonstrated respectively in Refs. \cite{HZ}  and \cite{Zhenya}. The duality property also enables one to investigate the possibility to employ superconducting nanowires  for creating a QPS-based standard of electric current \cite{Wang}.

\subsection{Path integral analysis}
We now turn to a more formal path integral analysis. For pedagogical purposes and also for the reasons which will be clear below in the next section we will now consider a closed ring (with perimeter $L=2\pi R$ and cross section $s$) made of a thin superconducting wire. The configuration remains essentially the same as that already treated in Section III (e.g., we again assume that the magnetic flux $\Phi_x$ pierces the ring) with the only important difference: Now we allow for quantum phase slips. We will perform the whole calculation for the ring geometry and in the very end of it we will explain how to apply our results to superconducting wires with open ends and/or attached to an external circuit.

In order to proceed we will again make use of the expression for the grand partition function ${\mathcal Z}$ (\ref{partf}) that accounts for the ring geometry as well as for an external magnetic flux inside the ring. It is also important to bear in mind that Eq. (\ref{partf}) remains valid only at length and time scales exceeding respectively the superconducting coherence length $\xi \sim \sqrt{D/\Delta}$ and the inverse gap $\Delta^{-1}$, i.e. outside the QPS core where only superconducting phase fluctuations  may occur. Within the semiclassical approximation
it suffices to take into account all relevant saddle point configurations of the phase variable $\varphi$ which satisfy the equation
\begin{equation}
(\partial_\tau^2+v^2\partial_x^2)\varphi(x,\tau)=0.
\label{speq}
\end{equation}
Apart from trivial solutions of this equation (linear in $\tau$ and $x$)
there exist nontrivial ones which correspond to virtual phase jumps by $\pm 2\pi$ at various points of a superconducting ring where the magnitude of the order parameter gets locally (at spatial scales $x_0 \sim \xi$) and temporarily (within the time interval $\tau_0 \sim 1/\Delta$) suppressed by quantum fluctuations. These quantum topological objects can be viewed as vortices in space-time and just represent quantum phase slips.
For sufficiently long wires or large rings and outside the QPS core $|x| > x_0$, $|\tau | > \tau_0$ (which position in space-time can be chosen, e.g., at $x=0$ and $\tau=0$) the saddle point solution $\tilde{\varphi}(x,\tau )$ corresponding to a single QPS event should satisfy the identity
\begin{equation}
\partial_{x}\partial_{\tau}\tilde{\varphi}-\partial_{\tau}\partial_{x}
\tilde{\varphi}=2\pi\delta(\tau,x)
\label{wind}
\end{equation}
implying that after a wind around the QPS center the phase
should change by $2\pi$. This saddle point solution has the form \cite{GZQPS}
\begin{equation}
\tilde{\varphi}(x,\tau)=-\arctan(x/v\tau).\label{QPSphi}
\end{equation}

Configurations $\varphi^{qps}(x,\tau)$ consisting of an arbitrary number of quantum phase slips can be treated analogously. Our goal here is to effectively sum up the contributions to the partition function ${\mathcal Z}$ (\ref{partf}) from all possible QPS configurations. This goal can be conveniently accomplished with the aid of the approach involving the so-called duality transformation.

Let us express the general solution of Eq. (\ref{speq}) in the form
\begin{equation}
  \varphi^{sp}(x,\tau)=a_{m}\tau+b_{n}x+\varphi^{qps}(x,\tau),
\end{equation}
where $a_m$ and $b_n$ are some constants fixed by the boundary conditions. We also introduce the vorticity field $\varpi(x,\tau)$ by means of the relations
\begin{equation}
 v\partial_x\varpi=\partial_\tau\varphi^{qps}\quad \partial_\tau\varpi=-v\partial_x\varphi^{qps}.
 \label{dualr}
\end{equation}
This field is single-valued obeying the equation
\begin{equation}
\partial^2_\tau\varpi+v^2\partial^2_x\varpi=-2\pi v\sum\limits_j\nu_j\delta(x-x_j)\delta(\tau-\tau_j),
\label{vorteq}
\end{equation}
where $x_j$ and $\tau_j$ denote respectively the space and time coordinates of the $j$-th phase slip, while  $\nu_j=\pm 1$  is its topological charge corresponding to the phase jump by $\pm 2\pi$. It follows from the boundary conditions (\ref{boucond})  that the vorticity field derivatives are periodic functions in both space and time implying, in turn, that $\sum_j\nu_j=0$.

Let us define the function
\begin{equation}
\varpi^{qps}(x,\tau)=\frac{\beta Lv}{2\pi }\sum\limits_{|m|+|n|>0}\frac{e^{\frac{2\pi im\tau}{\beta}+\frac{2\pi inx}{L}}}{m^2L^2+n^2v^2\beta^2}.
\end{equation}
One can verify that the function
\begin{equation}
\varpi(x,\tau)=\sum\limits_j\nu_j\varpi^{qps}(x-x_j,\tau-\tau_j)
\end{equation}
satisfies Eq. (\ref{vorteq}) and by virtue of the duality relations (\ref{dualr}) it yields the saddle point configuration $\varphi^{qps}$. Combining the boundary conditions (\ref{boucond}) with the above equations, one finds
\begin{equation}
\varphi^{qps}(L,\tau)-\varphi^{qps}(0,\tau)=-\frac{1}{v}\int\limits_0^Ldx\partial_\tau\varpi(x,\tau)
=2\pi\sum\limits_j\nu_j\left(\theta(\tau-\tau_j)+\frac{\tau_j}{\beta}\right),
\end{equation}
\begin{equation}
\varphi^{qps}(x,\beta)-\varphi^{qps}(x,0)=v\int\limits_0^\beta d\tau\partial_x\varpi(x,\tau)
=-2\pi\sum\limits_j\nu_j\left(\theta(x-x_j)+\frac{x_j}{L}\right)
\end{equation}
and, hence,
\begin{equation}
a_m=\frac{2\pi}{\beta}\left(m+\sum\limits_j\nu_j\frac{x_j}{L}\right),
\end{equation}
\begin{equation}
 b_n=\frac{2\pi}{L}\left(n+\phi_x
-\sum\limits_j\nu_j\frac{\tau_j}{\beta}\right).
\end{equation}
We also note that for each of the above saddle point configurations the expression for the current (\ref{current0}) takes the form
\begin{equation}
I(\tau)=\frac{4 ev\lambda}{L}\left(n+\phi_x+\sum\limits_j\nu_j\theta(\tau-\tau_j)\right).
\end{equation}

Let us now carry out the summation over all possible saddle point configurations. Expanding the
partition function in powers of $\gamma_{QPS}$ we get
\begin{widetext}
\begin{multline}
\mathcal Z[J(\tau)]=\sum\limits_{N=0}^\infty\frac{1}{N!}\sum\limits_{\nu_1,..,\nu_N=\pm1}\delta_{\sum_j\nu_j,0}\int dx_1d\tau_1...dx_Nd\tau_N\sum\limits_{m,n=-\infty}^\infty e^{-\frac{\lambda}{2\pi}\left(\frac{\beta L}{v}a_m^2+\beta L v b_n^2\right)}\\
\times\left(\frac{\gamma_{QPS}}{2}\right)^N e^{-\frac{\lambda}{2\pi}\int dxd\tau\left(v(\partial_x\varpi)^2+v^{-1}(\partial_\tau\varpi)^2\right)+i\int d\tau J(\tau)I(\tau)}.
\label{pfqps1}
\end{multline}
Here $J(\tau)$ is the source variable which we introduced introduced for our future purposes.
Rewriting the sum over $m,n$ with the aid of the relations
\begin{equation}
\sum\limits_{m=-\infty}^\infty e^{-\frac{\lambda\beta L}{2\pi v}a_m^2}\sim\sum\limits_{m=-\infty}^\infty e^{-\frac{\pi v\beta m^2}{2\lambda L}+2\pi i m\sum\limits_j\nu_j\frac{x_j}{L}},
\end{equation}
\begin{equation}
\sum\limits_{n=-\infty}^\infty e^{-\frac{\lambda\beta vL}{2\pi }b_n^2+\frac{4 i ev\lambda(n+\phi_x)}{L}\int d\tau J(\tau)}\sim\sum\limits_{n=-\infty}^\infty e^{-\frac{\pi L }{2\lambda \beta v}\left(n-\frac{2ev\lambda}{\pi L}\int d\tau J(\tau)\right)^2+2\pi in\phi_x -2\pi i \left(n-\frac{2ev\lambda}{\pi L}\int d\tau J(\tau)\right)\sum\limits_j\nu_j\frac{\tau_j}{\beta}},
\end{equation}
employing the Kronecker delta-function representation $\delta_{m,n}=\int_0^{2\pi} dz e^{iz(m-n)}/(2\pi)$
and formally inserting the path integral over the $\varpi$-field,
we find
\begin{multline}
\mathcal Z[J(\tau)]\sim\sum\limits_{N=0}^\infty\frac{1}{N!}\sum\limits_{\nu_1,..,\nu_N=\pm1}\int\limits_0^{2\pi}\frac{dz}{2\pi}\int dx_1d\tau_1...dx_Nd\tau_N\sum\limits_{m,n=-\infty}^\infty e^{2\pi in\phi_x-\frac{\pi v\beta m^2}{2\lambda L}-\frac{\pi L }{2\lambda \beta v}\left(n-\frac{2ev\lambda}{\pi L}\int d\tau J(\tau)\right)^2}\\
\times\left(\frac{\gamma_{QPS}}{2}\right)^N e^{2\pi i m\sum\limits_j\nu_j\frac{x_j}{L} -2\pi i \left(n-\frac{2ev\lambda}{\pi L}\int d\tau J(\tau)\right)\sum\limits_j\nu_j\frac{\tau_j}{\beta}}
\int\mathcal D\varpi
e^{-\frac{\lambda}{2\pi}\int dxd\tau\left(v(\partial_x\varpi)^2+v^{-1}(\partial_\tau\varpi)^2\right)}\\
\times e^{iz\sum\limits_j\nu_j+\frac{4iev\lambda}{L}\sum\limits_j\nu_j\int d\tau J(\tau)\theta(\tau-\tau_j)}\delta\left(\partial^2_\tau\varpi+v^2\partial^2_x\varpi+2\pi v\sum\limits_j\nu_j\delta(x-x_j)\delta(\tau-\tau_j)\right),
\label{funcdf}
\end{multline}
where the functional delta-function follows from Eq. (\ref{vorteq}). Expressing this delta-function via the integral over the dual field $\eta(x,\tau)$ with the periodic boundary conditions and evaluating the gaussian integral over $\varpi$, from Eq. (\ref{funcdf}) we get
\begin{multline}
\mathcal Z[J(\tau)]\sim\sum\limits_{N=0}^\infty\frac{1}{N!}\sum\limits_{\nu_1,..,\nu_N=\pm1}\int\limits_0^{2\pi}\frac{dz}{2\pi}\int dx_1d\tau_1...dx_Nd\tau_N\sum\limits_{m,n=-\infty}^\infty e^{2\pi in\phi_x-\frac{\pi v\beta m^2}{2\lambda L}-\frac{\pi L }{2\lambda \beta v}\left(n-\frac{2ev\lambda}{\pi L}\int d\tau J(\tau)\right)^2}\\
\times\left(\frac{\gamma_{QPS}}{2}\right)^N e^{2\pi i m\sum\limits_j\nu_j\frac{x_j}{L} -2\pi i \left(n-\frac{2ev\lambda}{\pi L}\int d\tau J(\tau)\right)\sum\limits_j\nu_j\frac{\tau_j}{\beta}}
\int\mathcal D\eta
e^{-\frac{\pi v}{2\lambda}\int dxd\tau\left((\partial_\tau\eta)^2+v^2(\partial_x\eta)^2\right)}\\
\times e^{2\pi i v\sum\limits_j\nu_j\eta(x_j,\tau_j)+iz\sum\limits_j\nu_j+\frac{4iev\lambda}{L}\sum\limits_j\nu_j\int d\tau J(\tau)\theta(\tau-\tau_j)}.
\end{multline}
Introducing now the field
\begin{equation}
\chi(x,\tau)=-\frac{2\pi m x}{L}+\frac{2\pi\tau}{\beta}\left(n-\frac{2ev\lambda}{\pi L}\int d\tau J(\tau)\right)-2\pi v\eta(x,\tau)-z-\frac{4ev\lambda}{L}\int d\tau' J(\tau')\theta(\tau'-\tau),
\label{chi}
\end{equation}
\end{widetext}
which obeys the boundary conditions 
\begin{eqnarray}
\chi(x,\beta)-\chi(x,0)=2\pi n, \qquad
\chi(0,\tau)-\chi(L,\tau)=2\pi m, 
\end{eqnarray}
we obtain
\begin{equation}
\mathcal Z[J(\tau)]\sim\sum\limits_{m,n=-\infty}^\infty e^{2\pi i n\phi_x}\int^{mn} \mathcal D\chi e^{-S_{\rm eff}[\chi(x,\tau),J(\tau)]}
\label{ZJ}
\end{equation}
with the effective action
\begin{multline}
S_{\rm eff}=
\frac{1}{8\pi\lambda v}\int\limits_0^\beta d\tau\int\limits_0^L dx[(\partial_\tau\chi(x,\tau)-4ev\lambda J(\tau)/L)^2\\+v^2(\partial_x\chi(x,\tau))^2]
-\gamma_{QPS}\int\limits_0^\beta d\tau\int\limits_0^Ldx\cos(\chi(x,\tau)).
\label{efacchi}
\end{multline}
These expressions define the generating functional
and the effective action for superconducting nanorings in the dual representation. In other words, our original problem of a nanoring with quantum phase slips was exactly mapped onto a sine-Gordon model on torus. Equations (\ref{ZJ}), (\ref{efacchi}) keep track of interactions between different QPS and serve as a convenient starting point for the analysis of the ground state properties of superconducting nanorings that will be carried out in the next section.

In the absence of the source $J(\tau) \to 0$ the effective action (\ref{efacchi}) turns out to be exactly dual to that for spatially extended quasi-one-dimensional Josephson barriers described by the Hamiltonian (\ref{HJJ}), i.e. we again arrive at the duality transformations (\ref{optrans}) and (\ref{dualtrans}) already derived within the operator formalism for superconducting nanowires. In particular, the Josephson phase $\phi (x,\tau)$ in the latter model is dual to the field $\chi (x,\tau)$ (\ref{chi}).

Finally, we point out that all the above arguments developed for superconducting nanorings equally apply for nanowires with open ends. In that 
case one should set $\phi_x=0$, $J =0$ and abandon the second boundary condition (\ref{boucond}). Removing the summation over $n$ in Eq. (\ref{partf}) and repeating all the same steps one again arrives at the sine-Gordon action (\ref{efacchi}) (with $J(\tau)=0$) describing a superconducting nanowire of length $L$ in the presence of quantum phase slips.

\section{Superconducting nanorings with quantum phase slips}
In Sec. III we already considered some quantum coherent effects associated with fluctuations of the phase variable in superconducting rings threaded by the magnetic flux. That analysis was performed for sufficiently thick rings, thus enabling one to fully neglect QPS effects. Our main goal here is to include quantum phase slips into our consideration.

\begin{figure}
\includegraphics[width=0.45\columnwidth]{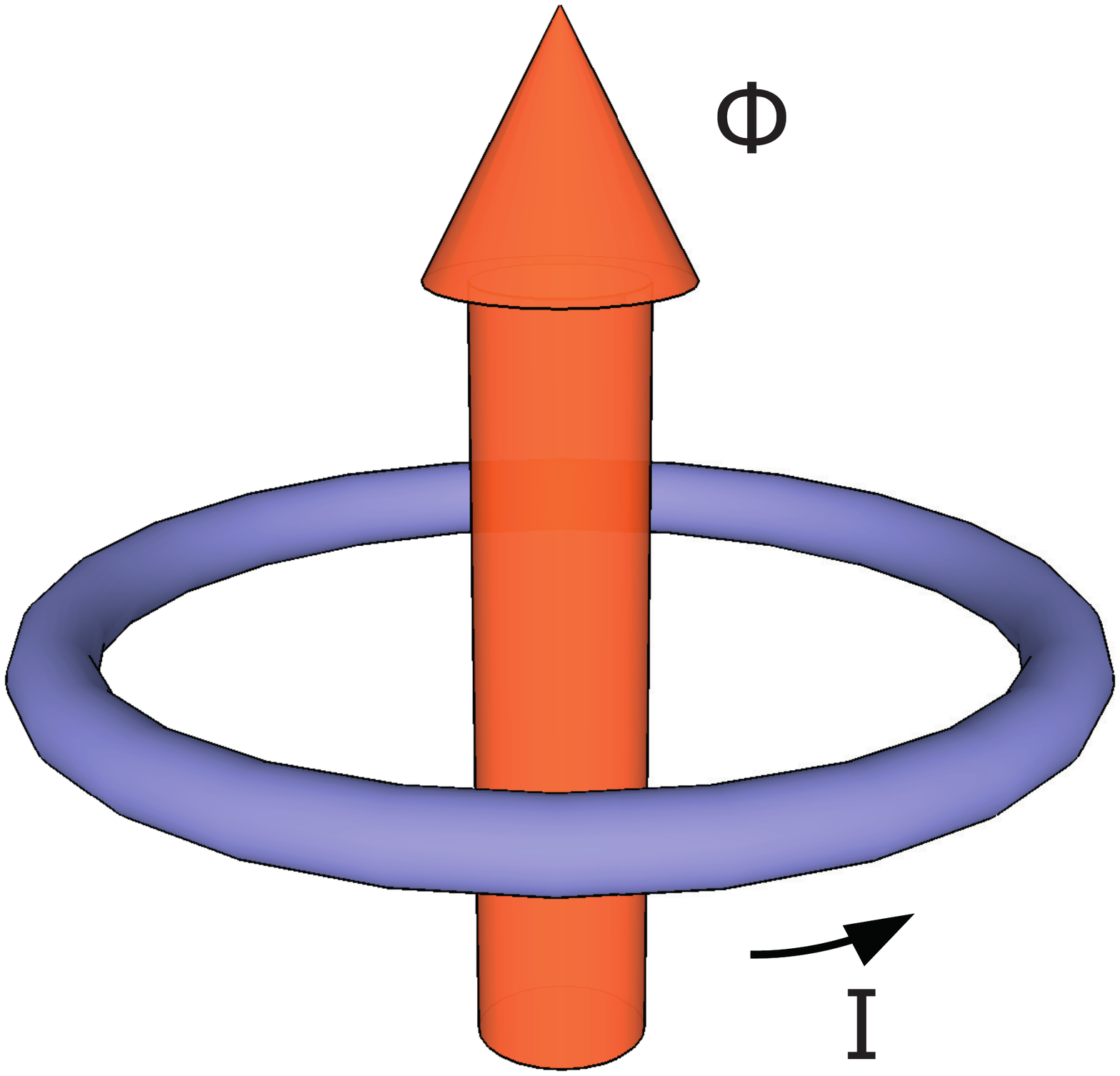}
\includegraphics[width=0.45\columnwidth]{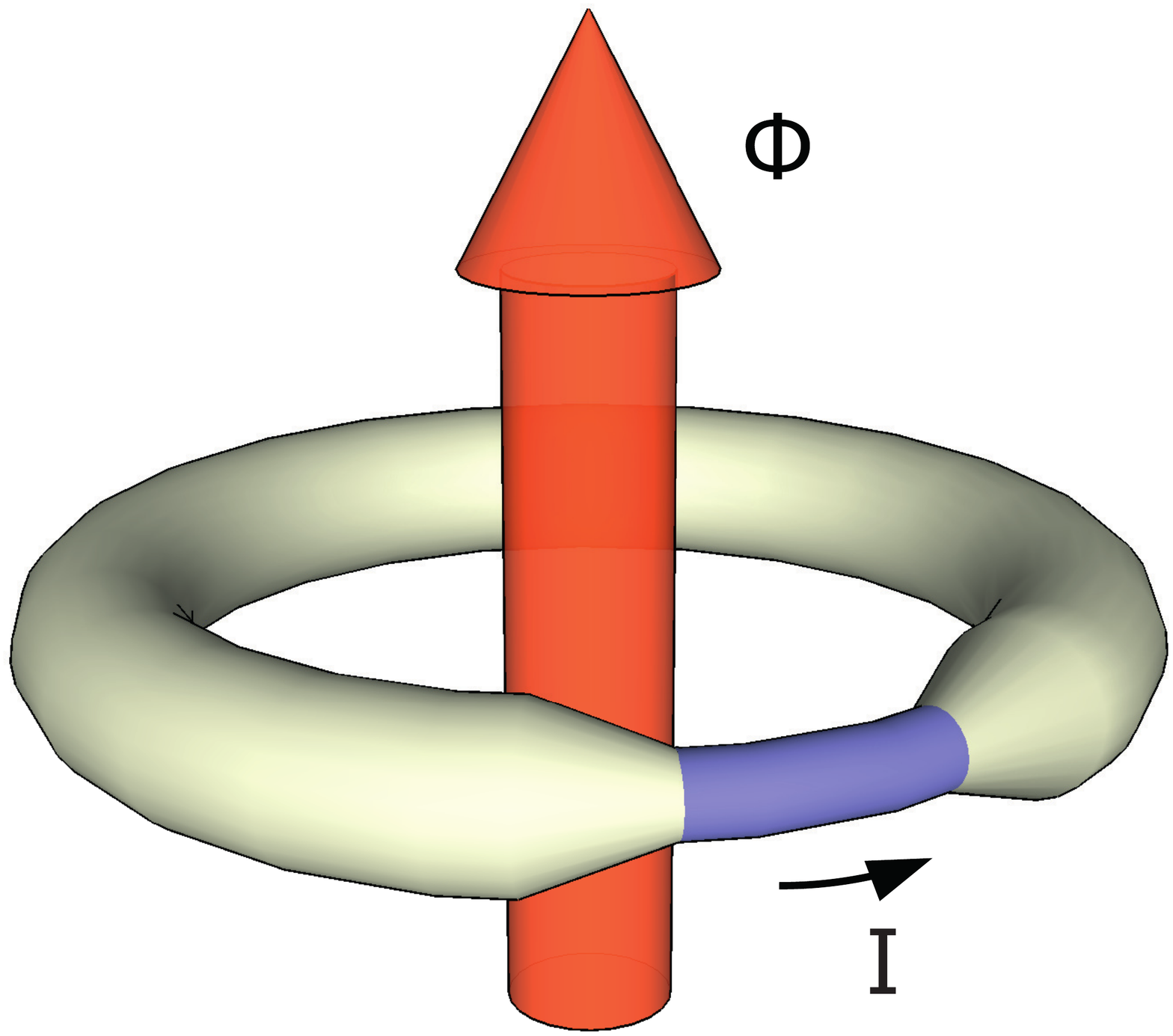}
\caption{Left: Quantum phase slip ring, i.e. an ultrathin superconducting ring threaded by an external magnetic flux.
Right: Quantum phase slip junction embedded in a thick superconducting ring.}
\label{Fig:8}
\end{figure}

According to the existing microscopic theory \cite{book,AGZ,Z10,ZGOZ,GZQPS} QPS represent
quantum coherent objects \cite{FN} which may significantly affect not only transport but also equilibrium ground state properties of superconducting nanowires and nanorings. Coherent nature of quantum phase slips was also demonstrated in a number of experiments \cite{AstNature,AstPRB,Zhenya}. A fundamental manifestation of a quantum coherent ground state is the possibility for a non-vanishing supercurrent to flow around a superconducting ring pierced by an external magnetic flux. Provided such a ring is sufficiently thin, as displayed in the left side of Fig. \ref{Fig:8}, quantum phase slips proliferate and may drastically modify both the supercurrent magnitude and its dependence on the magnetic flux \cite{AGZ,Z10,SZ13}. Below we will also demonstrate that quantum phase slips
cause non-vanishing supercurrent noise \cite{SZ12} which otherwise would be totally absent in the limit $T \to 0$, see Sec. III. 

All the same phenomena can also occur in rings consisting of thicker and thinner
parts, as it is shown in the right part of Fig. \ref{Fig:8}. In this case QPS effects are negligible in a thicker part of the
ring and may only occur in its thinner part which was named
a {\it quantum phase slip junction} \cite{MN}. It was proposed to employ such QPS junctions as central elements of the so-called quantum phase slip flux qubits \cite{MH}.

Bearing in mind that QPS effects are basically the same in uniform superconducting nanorings  and QPS
junctions (Fig. \ref{Fig:8}), in what follows we will merely address only the former systems.
In order to distinguish them from thicker superconducting rings (where QPS effects are negligible) and, on the other hand, to stress their similarity to QPS junctions we will denote such systems {\it quantum phase slip rings}.

\subsection{Supercurrent in quantum phase slip rings}
In order to proceed we will make use of the effective action  (\ref{efacchi}) derived in the previous section.
At low enough temperatures $T \ll v/L$  and provided the ring perimeter $L=2\pi R$ remains sufficiently small 
one can ignore the spatial dependence of the field $\chi$ and, hence, neglect the term $v^2(\partial_x\chi)^2$ in the effective action (\ref{efacchi}). Then our problem reduces to a zero-dimensional one with an effective Hamiltonian \cite{SZ10,SZ11}
\begin{equation}
   \hat H=\frac{E_R}{2}(\hat \phi -\phi_x)^2+U_0 (1-\cos (\hat \chi) )
\label{H1}
\end{equation}
describing a fictitious quantum particle on a ring in the presence of the cosine external potential.  Here we identify \cite{SZ12}
\begin{equation}
E_R=\frac{\pi^2 N_0 D\Delta s}{R}\sim \frac{g_{\xi}\Delta\xi}{R}
\label{ER}
\end{equation}
and
\begin{equation}
U_0=2\pi R\gamma_{QPS} \sim \frac{g_{\xi}\Delta R}{\xi} e^{-ag_\xi}.
\label{U0}
\end{equation}

The average value of the supercurrent $I$ flowing across the ring can be obtained by means of the standard formula 
\begin{equation}
I=-\frac{e}{\pi\beta}\frac{\partial\ln\mathcal Z}{\partial\phi_x}
\label{PCav0}
\end{equation}
In the zero dimensional limit described by the effective Hamiltonian (\ref{H1})
and at low temperatures $T \to 0$ Eq. (\ref{PCav0}) reduces to
\begin{equation}
I=\frac{e}{\pi}\frac{\partial E_0(\phi_{x})}{\partial\phi_{x}},
\end{equation}
where $E_0(\phi_{x})$ is the flux-dependent ground state energy. For smaller rings with $U_0\ll E_{R}$ one has
\begin{equation}
E_{0}(\phi_x)=\frac{E_{R}}{2\pi^{2}}\arcsin^{2}\left[\left( 1-\frac{\pi^{2}
}{2}\left(\frac{U_0}{E_{R}}\right)^{2}\right)\sin(\pi\phi_{x})\right],
\end{equation}
i.e. the ground state energy is almost parabolic except in the vicinity of the crossing points $\phi_x=1/2+n$ where the gap to the first excited energy band $\delta E_{01}=U_0$  opens up due to level repulsion. Accordingly, not too close to the points $\phi_x=1/2+n$
the supercurrent is not affected by QPS and is again defined by Eq.
(\ref{evcur}).

For larger $U_0$ the bandwidth shrinks while the gaps become bigger. In the limit $U_0 \gg E_R$, 
i.e. for 
\begin{equation}
R\gg R_c \sim \xi \exp (ag_\xi /2), 
\label{Rc}
\end{equation}
we obtain \cite{AGZ,Z10}
\begin{equation}
I=I_{0}\sin(2\pi\phi_{x}), \qquad I_0\sim eE_R^{1/4} U_0^{3/4} e^{-R/R_{c}}.
\label{tok}
\end{equation}
This result demonstrates that quantum phase slips yield exponential suppression of the supercurrent even at $T=0$ provided the ring radius $R$ exceeds the critical value $R_c$ (\ref{Rc}). 

Let us now generalize our analysis by including the spatial derivative term $v^2(\partial_x\chi)^2$ in Eq. (\ref{efacchi}) or, in other words,
by taking into account logarithmic interactions between quantum phase slips described by the sine-Gordon effective action. This task can be accomplished by means of the standard Berezinskii-Kosterlitz-Thouless (BKT) renormalization group (RG) approach \cite{BKT}. Adapting the corresponding RG equations to our problem we get
\begin{equation}
\frac{d\zeta}{d\ln \Lambda}=(2-\lambda)\zeta\qquad \frac{d\lambda}{d\ln\Lambda}=-32\pi^2\zeta^2 \lambda^2 K(\lambda),
\label{BKT}
\end{equation}
where $\zeta=\gamma_{QPS}\Lambda^2$ is the dimensionless coupling parameter, $\Lambda$ is the renormalization scale and $K(\lambda)$ is some non-universal function (which depends on the renormalization scheme) equal to one at the quantum BKT phase transition point $\lambda=2$ which separates and superconducting (ordered) phase $\lambda > 2$ with bound QPS-antiQPS pairs and a disordered phase $\lambda < 2$ with unbound QPS \cite{ZGOZ}.

Starting renormalization at the shortest scale $\Lambda \sim \xi_c
=\sqrt{\xi^2+v^2/\Delta^2}$ we, as usually, proceed to bigger scales. Within the first order  perturbation theory in $\zeta$ it suffices to ignore weak renormalization of the parameter $\lambda$. Then the solution of Eqs. (\ref{BKT}) takes the simple form $\gamma_{QPS}(\Lambda)=\gamma_{QPS}(\xi_c/\Lambda)^\lambda$. Our RG procedure should be stopped at the scale corresponding to the ring perimeter $\Lambda \sim L=2\pi R$. As a result, we arrive at the renormalized QPS amplitude
\begin{equation}
\tilde \gamma_{QPS}=\gamma_{QPS}(\xi_c/L)^\lambda .
\label{ren}
\end{equation}
This result allows to conclude that inter-QPS interaction effects remain weak and can be disregarded only for very small values of \begin{equation}
\lambda \ll 1/\ln (L/\xi_c ).
\end{equation}
This inequality, in turn, may severely restrict both the wire length and cross section values at which the system can still be treated as effectively
zero-dimensional and analyzed by means of a simplified Hamiltonian (\ref{H1}).

Substituting the renormalized QPS amplitude (\ref{ren}) instead of the bare one into Eq.
(\ref{U0}), we again reproduce the same expressions for the supercurrent, now with $\gamma_{QPS} \to \tilde \gamma_{QPS}$. As before, for smaller rings
with $R \ll \tilde R_c$ and at $T \to 0$ the current is defined
by Eq. (\ref{evcur}) at all values of the flux except for $\phi_x \approx 1/2+n$ where QPS effects with the effective rate (\ref{ren})
become significant. The critical radius $\tilde R_c$ is now
determined by the condition $E_R \sim 2\pi \tilde\gamma_{QPS}$ which yields
\begin{equation}
\tilde R_c \sim \xi \exp \left(\frac{ag_\xi}{2-\lambda}\right)
\left(\frac{\xi}{\xi_c}\right)^{\frac{\lambda}{2-\lambda}},
\label{rc}
\end{equation}
where $\lambda$ is supposed not to exceed 2.
In the opposite limit of bigger rings $R \gg \tilde R_c$ we again
reproduce Eq. (\ref{tok}), where now
\begin{equation}
I_0 \sim e\left(\frac{\xi_c}{\xi}\right)^{3\lambda/4} g_{\xi}\Delta\left(\frac{R}{\xi}\right)^{1/2-3\lambda /4} e^{-3ag_\xi /4-(R/\tilde R_{c})^{1-\lambda/2}}.
\label{lsc2}
\end{equation}

We observe that the critical radius $\tilde R_{c}$ (\ref{rc}) increases with increasing $\lambda$ and eventually diverges at the
quantum BKT phase transition point $\lambda = 2$. In the ordered phase $\lambda > 2$ QPS are bound in "neutral" pairs and, hence, become practically irrelevant. In this case the supercurrent is determined by Eq. (\ref{evcur}) for any value of $R$.

\subsection{Supercurrent noise in QPS rings}
Now let us analyze fluctuations of supercurrent in QPS rings. Taking the derivatives of the generating functional $\mathcal Z[J(\tau)]$ over the source variable $J(\tau)$ and setting this variable equal to zero afterwards, we obtain
\begin{equation}
I\equiv\langle I(\tau)\rangle =-\frac{i e}{\pi}\langle\partial_\tau\chi(x,\tau)\rangle,
\end{equation}
whereas the irreducible Matsubara current-current correlator (\ref{pi}) now reads
\begin{equation}
\Pi(\tau_1-\tau_2)=\frac{4e^2\lambda v}{\pi L}\delta(\tau_1-\tau_2)-\frac{e^2}{\pi^2L^2}\int dx_1 dx_2\langle
\partial_{\tau_1}\chi(x_1,\tau_1)\partial_{\tau_2}\chi(x_2,\tau_2)\rangle-I^2
\label{pi1}
\end{equation}
One can also decompose the source variable as $J(\tau)=J_0+\partial_\tau J_1(\tau)$ and perform a shift under the functional integral
\begin{equation}
\mathcal Z[J_0+\partial_\tau J_1(\tau)]=\sum_{mn}e^{2\pi i n\phi_x+2eJ_0n-\frac{2e^2v\lambda J_0^2\beta}{\pi L}}\int^{mn} \mathcal D\chi e^{-S_{\rm eff}[\chi(x,\tau)+4ev\lambda J_1(\tau)/L,0]}
\end{equation}
Expanding both sides in powers of $J_0$ and $J_1(\tau)$ we again recover Eq. (\ref{PCav0}) for the current
and arrive at the following exact relations
\begin{equation}
\int d\tau \Pi(\tau)=\frac{4e^2\lambda v }{\pi L}\left(1+\frac{L}{4\pi \lambda\beta v}\frac{\partial^2\ln\mathcal Z}{\partial\phi_x^2}\right),
\end{equation}
\begin{multline}
\Pi''(\tau)=-\frac{16 \gamma_{QPS}e^2\lambda^2v^2}{L^2}\int dx\langle\cos(\chi(x,\tau_1))\rangle\delta(\tau)\\+\frac{16 \gamma_{QPS}^2e^2\lambda^2v^2}{L^2}\int dx_1dx_2\langle\sin(\chi(x_1,\tau))\sin(\chi(x_2,0)) \rangle .
\label{pi2}
\end{multline}

Let us restrict our attention to the low temperature limit $T \to 0$. We are going to evaluate the imaginary time current-current correlator  (\ref{pi}), (\ref{pi1}) and then perform its analytic continuation to real times. Setting $R \ll \tilde R_c$ and proceeding perturbatively in $\gamma_{QPS}$, in the leading approximation one can reduce Eq. (\ref{pi2}) to the form
\begin{multline}
\Pi''(\tau)=\frac{8\gamma_{QPS}^2e^2\lambda^2v^2}{L}\int dx\langle\cos(\chi(x,\tau)-\chi(0,0))\rangle_0\\
-\frac{8\gamma_{QPS}^2e^2\lambda^2v^2}{L}\delta(\tau)\int d\tau_1 dx\langle\cos(\chi(x,\tau_1)-\chi(0,0))\rangle_0,
\label{ccni}
\end{multline}
where averaging $\langle ...\rangle_0$ is now performed with the
non-interacting effective action 
\begin{equation}
S_0= \frac{1}{8\pi\lambda v}\int\limits_0^\beta d\tau\int\limits_0^L dx\left((\partial_\tau\chi)^2+v^2(\partial_x\chi)^2\right).
\label{S000}
\end{equation}
The task at hand is to evaluate the correlation function
\begin{equation}
\langle e^{i(\chi(x,\tau)-\chi(0,0))} \rangle_0=\sum\limits_{m,n=-\infty}^\infty e^{2\pi in\phi_x} \int^{mn}\mathcal D\chi
e^{i(\chi(x,\tau)-\chi(0,0))-S_0},
\label{niav}
\end{equation}
which can be rewritten through the zero topological sector $m=n=0$ as
\begin{equation}
\langle e^{i(\chi(x,\tau)-\chi(0,0))} \rangle_0=\frac{1}{\mathcal Z_0}\frac{\int^{00}\mathcal D\chi e^{-S_0+i(\chi(x,\tau)-\chi(0,0))}}{\int^{00}\mathcal D\chi e^{-S_0}}\sum\limits_{mn} e^{-\frac{2\pi gv\beta(\phi_x+m+\tau/\beta)^2}{L}-\frac{\pi\beta v n^2}{2\lambda L}+\frac{2\pi i n x}{L}},
\label{B2}
\end{equation}
where
\begin{equation}
\mathcal Z_0=\sum_{mn}e^{-\frac{2\pi\lambda v\beta(m+\phi_x)^2}{L}-\frac{\pi\beta v n^2}{2\lambda L}}.
\end{equation}
Performing gaussian integration in Eq. (\ref{B2}), we obtain
\begin{equation}
 \langle e^{i(\chi(x,\tau)-\chi(0,0))} \rangle_0=\frac{e^{G(x,\tau)-G(0,0)}}{\mathcal Z_0}\sum\limits_{mn} e^{-\frac{2\pi gv\beta(\phi_x+m+\tau/\beta)^2}{L}-\frac{\pi\beta v n^2}{2\lambda L}+\frac{2\pi i n x}{L}},
\end{equation}
where $G(x,\tau)$ is the non-interacting Green function (with subtracted zero mode) obeying the equation
\begin{equation}
\left(-\partial_{\tau}^2-v^2\partial_{x}^2\right)G(x,\tau)=4\pi \lambda v(\delta(\tau)\delta(x)-1)
\end{equation}
with periodic boundary conditions. The solution of this equation reads
\begin{equation}
G(x,\tau)-G(0,0)=\frac{2\pi \lambda v \tau^2}{\beta L}-\lambda\sum\limits_{m=-\infty}^\infty\ln\left(\frac{\cosh(2\pi v(\tau+\beta m)/L  )-\cos(2\pi x/L)}{\cosh(2\pi v(\beta m)/L  )-1}\right)
\end{equation}
The divergent term with $m=0$ in this sum is regularized by means of the replacement $G(0,0)\to G(x_0,\tau_0)$  which is appropriate since   the above expressions do apply only at the space and time scales exceeding respectively $x_0\sim \xi$ and $\tau_0 \sim 1/\Delta$. With this in mind we obtain the $m=0$ term in the form $$\sim\ln\left(\frac{\cosh(2\pi v\tau/L  )-\cos(2\pi x/L)}{4\pi^2\xi_c^2/L^2}\right).$$

In the zero temperature limit the above equations yield
\begin{equation}
\langle\cos(\chi(x,\tau)-\chi(0,0))\rangle_0= \left(\frac{4\pi^2\xi_c^2}{L^2}\right)^\lambda\frac{\cosh(4\pi \lambda v\phi_x\tau/L)}{(\cosh(2\pi v\tau/L)-\cos(2\pi x/L))^\lambda}.
\end{equation}
Integrating this expression over $x$, we get
\begin{multline}
\int\limits_0^L dx \langle\cos(\chi(\tau,x)-\chi(0,0))\rangle_0
= L\left(\frac{2\pi^2\xi_c^2}{L^2}\right)^\lambda\\\times\sum\limits_{n=0}^\infty\frac{\GammaF(n+1/2)\GammaF(\lambda+n)}{\sqrt{\pi}\GammaF(\lambda)\GammaF^2(n+1)}\frac{\cosh(4\pi \lambda v\phi_x\tau/L)}{\cosh^{2n+2\lambda}(\pi v\tau/L)},
\label{cosint}
\end{multline}
where again $\GammaF(x)$ is the Gamma function. Performing now the Fourier transformation in Eq. (\ref{cosint}) with the aid of the relation
\begin{multline}
 \int\limits_{-\infty}^\infty d\tau e^{i\omega\tau}\frac{\cosh(4\pi \lambda v\phi_x\tau/L)}{\cosh^{2n+2\lambda}(\pi v\tau/L)}=\frac{2^{2n+2\lambda}L}{4\pi v}\sum\limits_{m=0}^\infty\frac{\GammaF(2\lambda+2n+m)(-1)^m}{\GammaF(2\lambda+2n)\GammaF(m+1)}\\\times\left(\frac{1}{\lambda(1-2\phi_x)+n+m-\frac{iL\omega}{2\pi v}}+\frac{1}{\lambda(1+2\phi_x)+n+m-\frac{iL\omega}{2\pi v}}\right.
\\\left.+\frac{1}{\lambda(1-2\phi_x)+n+m+\frac{iL\omega}{2\pi v}}+\frac{1}{\lambda(1+2\phi_x)+n+m+\frac{iL\omega}{2\pi v}}\right)
\end{multline}
and combining the result with Eq. (\ref{ccni}) after a simple algebra we arrive at the following expression
\begin{multline}
\Pi_{i\omega}=\frac{8\gamma_{QPS}^2 e^2 \lambda^2 v^2  }{\GammaF^2(\lambda)}\left(\frac{8\pi^2\xi_c^2}{L^2}\right)^\lambda
\sum\limits_{k=0}^\infty\frac{\GammaF^2(\lambda+k)}{\GammaF^2(1+k)}
\left(\frac{1}{\Omega_k(\phi_x)(\omega^2+\Omega_{k}^2(\phi_x))}\right.
\\\left.
+\frac{1}{\Omega_k(-\phi_x)(\omega^2+\Omega_k^2(-\phi_x))}\right),
\label{piom}
\end{multline}
where
\begin{equation}
\Omega_k(\phi_x)=\frac{2\pi \lambda v(1+2\phi_x)}{L}+\frac{4\pi v k}{L}
\label{diff}
\end{equation}
denote the energy differences between the exited states and the ground state of our ring. Eq. (\ref{diff}) applies for $-1/2<\phi_x \leq 1/2$. Outside this interval $\Omega_k(\phi_x)$ should be continued periodically with the period equal to unity.

What remains is to perform an analytic continuation of Eq. (\ref{piom}) with the aid of Eq. (\ref{cont}). As a result, we obtain the expression for the supercurrent noise power spectrum at $T=0$:
\begin{multline}
{\mathcal S}_\omega=\frac{4\pi\gamma_{QPS}^2 e^2 \lambda^2 v^2  }{\Gamma^2(\lambda)}\left(\frac{8\pi^2\xi_c^2}{L^2}\right)^\lambda
\sum\limits_{k=0}^\infty\frac{\GammaF^2(\lambda+k)}{\GammaF^2(1+k)}
\left(\frac{1}{\Omega_{k}^2(\phi_x)}
\delta(\omega-\Omega_k(\phi_x))\right.\\+\frac{1}{\Omega_{k}^2(\phi_x)}
\delta(\omega+\Omega_k(\phi_x))
+\frac{1}{\Omega_{k}^2(-\phi_x)}\delta(\omega-\Omega_k(-\phi_x))
\left.+\frac{1}{\Omega_{k}^2(-\phi_x)}
\delta(\omega+\Omega_k(-\phi_x))\right).
\label{final}
\end{multline}
In accordance with general considerations \cite{SZ10,SZ11} the spectrum ${\mathcal S}_\omega$ depends periodically on the external magnetic flux $\phi_x$ and consists of sharp peaks at the
frequencies $\Omega_k$ corresponding to the system eigenmodes.
These features clearly illustrate coherent nature of supercurrent noise.
\begin{figure}
\includegraphics[width=0.45\columnwidth]{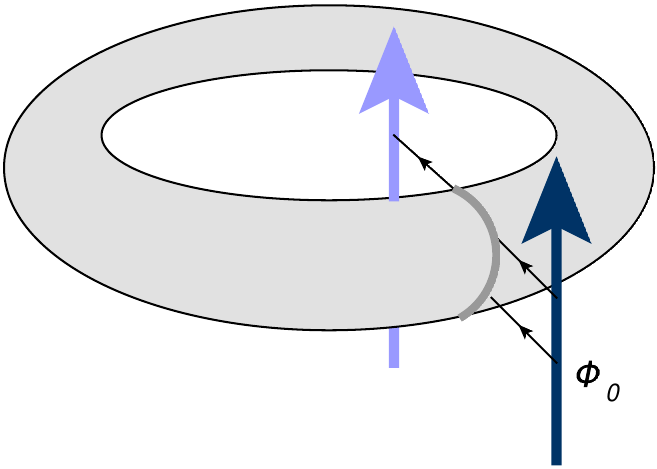}
\includegraphics[width=0.45\columnwidth]{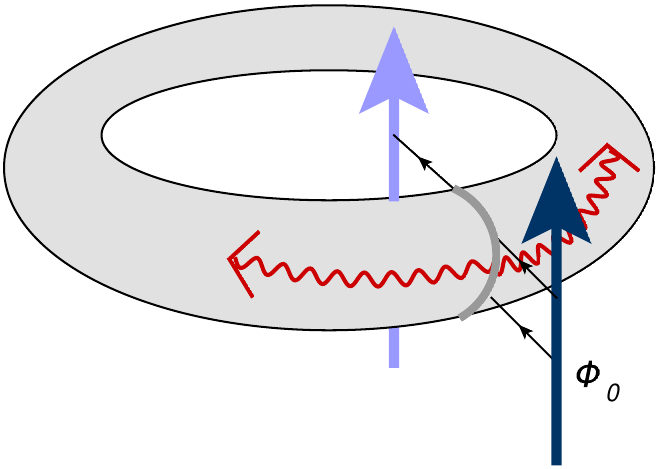}
\caption{The process of coherent flux tunneling without and with excitation of a pair of Mooji-Sch\"on plasma modes schematically displayed respectively in the left and right parts of the figure.}
\label{Fig:9}
\end{figure}
The effect of inter-QPS interactions on supercurrent noise turns out to be
richer than that for the average supercurrent value analyzed above. 
In contrast to the latter, fluctuations of supercurrent cannot in general be correctly described by the Hamiltonian (\ref{H1}) even if
the renormalization of $\gamma_{QPS}$ (\ref{ren}) is taken into  account. This is because virtual tunneling of flux quanta across the
superconducting wire in general leads to creation of plasmon modes, thereby causing extra peaks with $k\neq 0$ in the supercurrent noise power spectrum (\ref{final}). The simplest process of this kind is associated with simultaneous creation of two plasmons with opposite momenta values propagating clockwise and counterclockwise around the ring. Such a process is illustrated in Fig. \ref{Fig:9}. Thus, by experimentally detecting these peaks one can directly demonstrate the existence of Mooij-Sch\"on plasma modes in superconducting nanorings.

Let us emphasize again that -- owing to its coherent nature -- supercurrent noise can be tuned by the external magnetic flux piercing the ring. Both the positions of the peaks and the magnitude of this noise essentially depend on $\phi_x$. Here we evaluated the dependence ${\mathcal S}_\omega (\phi_x)$ in the experimentally relevant limit $R < \tilde R_c$ in which case one can
proceed perturbatively in the QPS rate $\gamma_{QPS}$. In the opposite
limit $R>\tilde R_c$ supercurrent noise also has the form of sharp peaks although its dependence on the magnetic flux becomes much weaker \cite{SZ12}. As follows from Eq. (\ref{final}), in the immediate vicinity of level degeneracy points $\phi_x =\pm 1/2$ supercurrent fluctuations become strong and our perturbative in $\gamma_{QPS}$ analysis fails even for $R \ll \tilde R_c$. In this case it is necessary to account for level splitting and regularize the corresponding terms in Eq. (\ref{final}) by substituting the value $\sim U_0$ instead of $\Omega_0(\phi_x)$ whenever the former exceeds the latter.

At non-zero temperatures supercurrent noise is modified in two ways: (i) a zero frequency peak (\ref{nzT}) appears which is not related to QPS and (ii) numerous extra QPS-related peaks at non-zero
frequencies emerge, cf. Eq. (\ref{soft2}). At low enough $T$ quantum coherence is still maintained, however with increasing temperature
the dependence on $\phi_x$ gets less pronounced and supercurrent noise eventually becomes incoherent.

Finally, we point out a certain physical similarity between
supercurrent noise studied here and that in superconducting weak links analyzed elsewhere \cite{Averin,Madrid,GZ10}. Also in the latter case the noise power spectrum depends on the phase difference across the weak link and has the form of peaks both at zero and non-zero frequencies. Similarly to our problem, at $T\to 0$ the zero frequency peak
disappears, while all other peaks persist except in the limit
of fully transparent barriers \cite{Averin,Madrid,GZ10}. Unlike here, however, in the case of superconducting weak links non-zero frequency peaks originate from subgap Andreev levels and are not related to quantum phase slips.

\section{Shot noise from quantum phase slips}

In the previous section we demonstrated that quantum phase slips may strongly affect equilibrium properties of superconducting nanorings and even cause supercurrent noise in the ground state of such systemss. Here we will continue studying QPS-generated noise in a different physical situation. To this end we will get back to a thin superconducting superconducting wire  embedded in an external circuit as shown in Fig. \ref{Fig:6}, see Sec. IVA.

Can such a superconducting wire generate voltage fluctuations? Furthermore, can this wire produce shot noise provided it is biased by an external current $I=V_x/R_x$? Posing these questions we imply that temperature $T$, typical values of voltage, frequency and all other
relevant energy parameters remain well below the superconducting gap $\Delta$, i.e. the superconductor is either in or
sufficiently close to its quantum ground state.

At the first sight, positive answers to both these questions can be rejected on fundamental grounds because a superconducting state is characterized by zero resistance. Hence, the system can sustain a non-dissipative current below some critical value and neither non-zero average voltage nor voltage fluctuations can be expected.

These simple considerations -- although applicable to bulk superconductors -- become obviously insufficient in the case of ultrathin superconducting wires because of the presence of quantum phase slips. As we already discussed in Sec. IV, each QPS event corresponds to the net phase jump by $\delta \varphi =\pm 2\pi$ implying positive or negative voltage pulse  $\delta V=\dot{\varphi}/2e$ and tunneling of one magnetic flux quantum $\Phi_0\equiv \pi/e =\int |\delta V(t)|dt$ across the wire perpendicular
to its axis. This process is illustrated in the upper part of Fig. \ref{Fig:7}.  Biasing the wire by an external current $I$ one breaks the symmetry between positive and negative voltage pulses making the former more likely than the latter. As a result, the net voltage drop $V$ occurs across the wire also implying non-zero resistance $R=V/I$ which may not vanish down to lowest temperatures \cite{ZGOZ,GZQPS}. Thus, according, e.g., to the fluctuation-dissipation theorem (FDT), in the presence of QPS one should also expect voltage fluctuations to occur in the system.

While these general arguments suggest a positive answer to the first of the above questions they do not specifically address the issue of shot noise. Let us recall that two key pre-requisits of shot noise are (see, e.g., Ref. \cite{BB}):  ($i$) the presence of discrete charge carriers (e.g., electrons) in the system and ($ii$) scattering of such carriers at disorder. As for superconducting nanowires, although discrete charge carriers -- Cooper pairs -- are certainly present there, they form a superconducting condensate flowing along the wire {\it without any scattering}. For this reason the possibility for shot noise to occur in superconducting nanowires needs special analysis that we are going to outline further below.

\subsection{Keldysh technique and perturbation theory}

An effective Hamiltonian $\hat H_{\rm eff}$ for the structure depicted in Fig. \ref{Fig:6} was already derived in Sec. IVA, see Eq. (\ref{Heff192}).  In order to proceed we will follow Ref. \cite{SZ16} and employ the Keldysh path integral technique. As usually, we define our variables of interest on the forward and backward time branches of the Keldysh contour, i.e. we introduce the variables $\varphi_{F,B} (t)$ and $\chi_{F,B}(x,t)$. We also routinely define the ``classical'' and ``quantum'' variables, respectively $\varphi_+ (t)= (\varphi_F (t)+\varphi_B (t))/2$ and $\varphi_- (t)= \varphi_F (t)-\varphi_B (t)$ (and similarly for the $\chi$-fields). Making use of the Josephson relation between the voltage and the phase one can formally express the expectation value of the voltage operator across the the superconducting wire in the form
\begin{equation}
\langle V(t_1)\rangle=\frac{1}{2e}\left\langle\dot \varphi_{+}(t_1) e^{iS_{QPS}}\right\rangle_0
\label{V}
\end{equation}
where
\begin{equation}
 S_{QPS}=-2 \gamma_{QPS}\int dt\int\limits_0^L dx\sin(\chi_{+})\sin (\chi_-/2)
\label{sqps}
\end{equation}
and
\begin{equation}
\langle ...\rangle_0 =\int\mathcal D^2\varphi (t) \mathcal D^2\chi (x,t) (...) e^{iS_0[\varphi , \chi ]}
\end{equation}
implies averaging with the Keldysh effective action $S_0$ corresponding to the non-interacting Hamiltonian $\hat H_{0} = \hat H_{\rm eff}- \hat H_{QPS}$, where  $H_{QPS}$ is defined in Eq. (\ref{HQPS0}). Similarly, for
the symmetrized voltage-voltage correlator 
\begin{equation}
\langle V(t_1)V(t_2)\rangle =\frac12\langle (\hat V(t_1)\hat V(t_2) + \hat V(t_2)\hat V(t_1))\rangle
\label{V222}
\end{equation}
we obtain
\begin{equation}
\langle V(t_1)V(t_2)\rangle =\frac{1}{4e^2}\left\langle\dot \varphi_{+}(t_1)\dot \varphi_{+}(t_2) e^{iS_{QPS}}\right\rangle_0.
\label{VV}
\end{equation}

Eqs. (\ref{V}) and (\ref{VV}) are formally exact expressions which we are now going to evaluate perturbatively in $\gamma_{QPS}$. In the zero order in $\gamma_{QPS}$ all averages can be handled exactly with the aid of the Green functions
\begin{eqnarray}
G^R_{ab}(X,X')=-i\langle a_{+}(X)b_-(X')\rangle , \nonumber\\
G^K_{ab}(X,X')=-i\langle a_{+}(X)b_{+}(X')\rangle ,
\label{GRK}
\end{eqnarray}
where $a(X)$ and $b(X)$ denote one of the fields $\varphi (t)$ and $\chi (x,t)$. As both these fields are
real, the advanced  and retarded Green functions obey the condition $G^A_{ab}(\omega)=G^R_{ba}(-\omega)$, and 
the Keldysh function $G^K$ can be expressed in the form
\begin{eqnarray}
 G^K_{ab}(\omega)=\frac12\coth\left(\frac{\omega}{2T}\right)\left(G^R_{ab}(\omega)-G^R_{ba}(-\omega)\right).
\label{GK}\end{eqnarray}

\begin{figure}
\includegraphics[width=0.75\columnwidth]{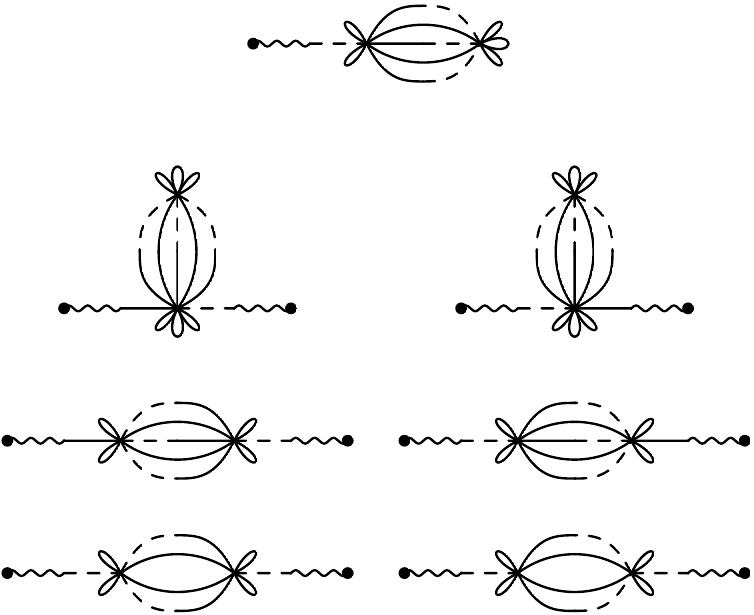}
\caption{Candy-like diagrams which determine both average voltage (\ref{V}) (upper diagram) and voltage noise (\ref{VV}) (six remaining diagrams) in the second order in $\gamma_{QPS}$. The fields $\varphi_+$, $\chi_+$ and $\chi_-$ in the propagators (\ref{GRK}) are denoted respectively by wavy, solid and dashed lines.}
\label{Fig:10}
\end{figure}

Expanding Eqs. (\ref{V}) and (\ref{VV}) up to the second order in $\gamma_{QPS}$ and performing all necessary averages
we evaluate the results in terms of the Green functions (\ref{GRK}) expressing them in the form of ``candy'' diagrams displayed in Fig. \ref{Fig:10}. They involve four different propagators ($G_{\chi\chi}^{R,K}$ and $G_{\varphi\chi}^{R,K}$) and plenty of vertices originating from  Taylor expansion of the cosine terms. Summing up all the diagrams in the same order in $\gamma_{QPS}$ one arrives at the final expression containing the exponents of the Green functions.

\subsection{I-V curve and voltage noise}

Let us first re-derive the results \cite{ZGOZ}
for the average voltage within the framework of our technique. We obtain
\begin{equation}
 \langle V\rangle=\frac{i\gamma_{QPS}^2}{8e}\int\limits_{0}^L dx\int\limits_0^L dx'\left(\lim_{\omega\to 0}\omega G_{\varphi\chi}^R(x;\omega)\right)\left({\mathcal P}_{x,x'}(-I\Phi_0)-{\mathcal P}_{x,x'}(I\Phi_0)\right),
\label{V1}
\end{equation}
where ${\mathcal P}_{x,x'}(\omega)=P_{x,x'}(\omega)+\bar P_{x,x'}(\omega)$ and
\begin{equation}
 \label{P00}
 P_{x,x'}(\omega)=\int\limits_{0}^\infty dt e^{i\omega t}e^{i{\mathcal G}(x,x';t,0)},
\end{equation}
\begin{equation}
{\mathcal G}(x,x';t,0)= G^K_{\chi\chi}(x,x';t,0)-\frac 12G^K_{\chi\chi}(x,x;t,t)
-\frac 12G^K_{\chi\chi}(x',x';0,0)+\frac 12 G^R_{\chi\chi}(x,x';t,0).
\end{equation}
Bearing in mind that $\lim_{\omega\to 0}\omega G_{\varphi\chi}^R(x;\omega)=2\pi i$, Eq. (\ref{V1})
can be cast to the form
\begin{equation}
\langle V \rangle =\Phi_0\left(\Gamma_{QPS}(I\Phi_0)-\Gamma_{QPS}(-I\Phi_0)\right),
\label{V2}
\end{equation}
where we identify $\Gamma_{QPS}$ as
\begin{equation}
 \Gamma_{QPS}(\omega)=\frac{\gamma_{QPS}^2}{4}\int\limits_{0}^L dx\int\limits_0^L dx'{\mathcal P}_{x,x'}(\omega).
\end{equation}
Comparing the result (\ref{V2}) with that found in Ref. \cite{ZGOZ} we immediately conclude that
$\Gamma_{QPS}(I\Phi_0)$ defines the quantum decay rate of the current state due to QPS. In \cite{ZGOZ} this rate was evaluated from the imaginary
part of the free energy $\Gamma_{QPS}(I\Phi_0)=2{\rm Im}F$. Here we derived the expression for $\Gamma_{QPS}$ by means of the real time technique without employing the ${\rm Im}F$-method \cite{Weiss}.

Making use of the above results, evaluating the Green functions (\ref{GRK}) and keeping in mind the detailed balance condition
\begin{equation}
{\mathcal P}_{x,x'}(\omega)=e^{\frac{\omega}{T}}{\mathcal P}_{x,x'}(-\omega)
\label{dbc}
\end{equation}
we obtain
\begin{equation}
\langle V\rangle =\frac{\Phi_0 L v \gamma_{QPS}^2}{4}\varsigma^2\left(\frac{I\Phi_0}{2}\right)\sinh\left(\frac{I\Phi_0}{2T}\right),
\label{V3}
\end{equation}
where 
\begin{equation}
\varsigma(\omega)=\tau_0^\lambda (2\pi T)^{\lambda -1}\frac{\GammaF\left(\frac{\lambda}{2}-\frac{i\omega}{2\pi T}\right)\GammaF\left(\frac{\lambda}{2}+\frac{i\omega}{2\pi T}\right)}{\GammaF(\lambda)}.
\label{vsi}
\end{equation}
 Here for the sake of simplicity we assumed that $v\tau_0 \sim x_0$. It is satisfactory to observe that the result (\ref{V3}), (\ref{vsi}) matches with that derived in Ref. \cite{ZGOZ} by means of the ${\rm Im}F$-technique. A detailed analysis of the relation between the ${\rm Im}F$-approach and the Keldysh technique employed here can be found in Ref. \cite{SZ17}.

Let us now turn to the voltage-voltage correlator. Our perturbative analysis allows to recover three different contributions
to the noise power spectrum, i.e.
\begin{equation}
 {\mathcal S}_{\Omega}=\int dt e^{i\Omega t}\langle V(t)V(0)\rangle ={\mathcal S}^{(0)}_{\Omega}+{\mathcal S}^{r}_{\Omega}+{\mathcal S}^{a}_{\Omega}.
\label{VV2}
\end{equation}
The first of these contributions ${\mathcal S}^{(0)}_{\Omega}$  defines equilibrium voltage noise for a transmission line and has nothing to do with QPS. This contribution reads
\begin{equation}
{\mathcal S}^{(0)}_{\Omega}= \frac{i\Omega^2\coth\left(\frac{\Omega}{2T}\right)}{16e^2}\left(G_{\varphi\varphi}^R(\Omega)-
G_{\varphi\varphi}^R(-\Omega)\right).
 \end{equation}
The remaining two terms are due to QPS effects. The term ${\mathcal S}^{r}_{\Omega}$ is also proportional to $\coth\left(\frac{\Omega}{2T}\right)$ and contains the products of two retarded (advanced) Green functions:
\begin{multline}
{\mathcal S}^{r}_{\Omega}=\frac{\gamma_{QPS}^2\Omega^2\coth\left(\frac{\Omega}{2T}\right)}{16e^2}\int\limits_{0}^L dx\int\limits_0^L dx' {\rm Re}\left[G_{\varphi\chi}^R(x;\Omega) \right.\\
 \left.\times ({\mathcal F}_{x,x'}(\Omega )G_{\varphi\chi}^R(x';\Omega)
 -{\mathcal F}_{x,x'}(0)G_{\varphi\chi}^R(x;\Omega))\right],
 \label{Sr}
 \end{multline}
where
\begin{equation}
{\mathcal F}_{x,x'}(\Omega )=
- P_{x,x'}(\Omega +I\Phi_0)-P_{x,x'}(\Omega -I\Phi_0)
+\bar P_{x,x'}(-\Omega +I\Phi_0)+\bar P_{x,x'}(- \Omega -I\Phi_0).
\end{equation}
The remaining term ${\mathcal S}^{a}_{\Omega}$ contains the product of one retarded and one advanced Green functions and scales with
the combinations ${\mathcal C}_{\pm} = \coth\left(\frac{\Omega\pm I\Phi_0}{2T}\right)-\coth\left(\frac{\Omega}{2T}\right)$ as
\begin{multline}
\label{Sa}
{\mathcal S}^{a}_{\Omega}=\frac{\gamma_{QPS}^2\Omega^2}{32e^2}\int\limits_{0}^L dx\int\limits_0^L dx'G_{\varphi\chi}^R(x;\Omega)G_{\varphi\chi}^R(x';-\Omega)\\
\times \left[\sum_{\pm}{\mathcal C}_{\pm}\left(
 {\mathcal P}_{x,x'}(\Omega\pm I\Phi_0)-{\mathcal P}_{x,x'}(-\Omega\mp I\Phi_0)\right)\right].
\end{multline}
Eqs. (\ref{VV2})-(\ref{Sa}) together with the expressions for the Green functions fully account for the voltage
noise power spectrum of a superconducting nanowire in the perturbative in $\gamma_{QPS}$ regime.

In the zero bias limit $I\to 0$
the term ${\mathcal S}^{a}_{\Omega}$ vanishes, and the equilibrium noise spectrum ${\mathcal S}_{\Omega}={\mathcal S}^{(0)}_{\Omega}+{\mathcal S}^{r}_{\Omega}$ is determined from FDT, see also \cite{SZ13}.
At non-zero bias values the QPS noise turns non-equilibrium. In the zero frequency limit $\Omega \to 0$
the terms ${\mathcal S}^{(0)}_{\Omega}$ and ${\mathcal S}^{r}_{\Omega}$ vanish, and the voltage noise ${\mathcal S}_{\Omega \to 0}\equiv {\mathcal S}_0$ is determined solely by ${\mathcal S}^{a}_{\Omega}$. Then from Eq. (\ref{Sa}) we get \cite{SZ16}
\begin{equation}
{\mathcal S}_{0}= \Phi_0^2\left(\Gamma_{QPS}(I\Phi_0)+\Gamma_{QPS}(-I\Phi_0)\right)
 =\Phi_0\coth\left(\frac{I\Phi_0}{2T}\right)\langle V \rangle,
\label{shot1}
\end{equation}
where $\langle V \rangle$ is specified in Eqs. (\ref{V2}), (\ref{V3}).  Combining
Eq. (\ref{shot1}) with Eqs. (\ref{V3}), (\ref{vsi}) we obtain
\begin{equation}
\label{S0lim}
{\mathcal S}_{0}\propto
\begin{cases}
T^{2\lambda -2}, & T\gg I\Phi_0,
\\
I^{2\lambda -2}, & T\ll I\Phi_0.
\end{cases}
\end{equation}
At higher temperatures $T\gg I\Phi_0$ (though still $T \ll \Delta_0$) Eq. (\ref{S0lim})
just describes equilibrium voltage noise $S_0=2TR$ of a linear Ohmic resistor $R=\langle V \rangle /I \propto T^{2\lambda -3}$ \cite{ZGOZ}.
In the opposite low temperature limit $T\ll I\Phi_0$ it accounts for QPS-induced {\it shot noise} ${\mathcal S}_0=\Phi_0\langle V \rangle$ obeying {\it Poisson statistics} with an effective ``charge'' equal to the flux quantum $\Phi_0$. This result makes the physical origin of shot noise in superconducting nanowires transparent: It is produced by coherent tunneling of magnetic flux quanta $\Phi_0$ across the wire, cf. also Fig. \ref{Fig:7}.

Another interesting limit is that of sufficiently high frequencies  and/or long wires $v/L\ll\Omega\ll \Delta_0$. In this case we obtain
\begin{equation}
{\mathcal S}^{(0)}_{\Omega}=\frac{\lambda}{8\pi e^2}\frac{\Omega\coth\left(\frac{\Omega}{2T}\right)}{(\Omega/2E_C)^2+(\lambda /\pi )^2}.
\label{S0L}
\end{equation}
Note that this contribution is independent of the wire length $L$. At low $T$ and $\Omega /\lambda \gtrsim  E_C=e^2/2C$ we have ${\mathcal S}^{(0)}_{\Omega} \propto 1/\Omega$, i.e. in a certain regime the wire may generate $1/f$ voltage noise.
Evaluating the QPS terms ${\mathcal S}^{r}_{\Omega}$ and ${\mathcal S}^{a}_{\Omega}$ we observe that the latter scales linearly with the wire length $L$ whereas the former does not. Hence,
the term ${\mathcal S}^{r}_{\Omega}$ can be disregarded in the long wire limit. For the remaining QPS term ${\mathcal S}^{a}_{\Omega}$ we get
\begin{equation}
{\mathcal S}^{a}_{\Omega}=\frac{L\lambda^2v\gamma_{QPS}^2}{8e^2}\left[\varsigma\left(\frac{I\Phi_0}{2}-\Omega\right)
-\varsigma\left(\frac{I\Phi_0}{2}+\Omega\right)\right]\frac{\sinh\left(\frac{I\Phi_0}{2T}\right)\varsigma\left(\frac{I\Phi_0}{2}\right)}{\left((\Omega/2E_C)^2+(\lambda /\pi)^2\right)\sinh\left(\frac{\Omega}{2T}\right)}.
\label{Sa3}
\end{equation}
At $T\to 0$ from Eq. (\ref{Sa3}) we find
\begin{equation}
{\mathcal S}^{a}_{\Omega}\propto
\begin{cases}
I^{\lambda -1}(I-2\Omega /\Phi_0)^{\lambda -1}, & \Omega < I\Phi_0/2,
\\
0, & \Omega > I\Phi_0/2.
\end{cases}
\label{Sa4}
\end{equation}
This result can be interpreted as follows.
At $T=0$ each QPS event excites (at least) two plasmons with total energy $E=I\Phi_0$ and zero total momentum propagating in the
opposite directions along the wire. One plasmon (which energy equals to $E/2$) gets dissipated at the grounded end of the wire
while another one (also with energy $E/2$) reaches its opposite end causing voltage fluctuations (emits a photon) with frequency $\Omega$ measured by a detector. Clearly, at $T=0$ this process is only possible at
$\Omega < E/2$ in the agreement with Eq. (\ref{Sa4}).
\begin{figure}
\includegraphics[width=0.8\columnwidth]{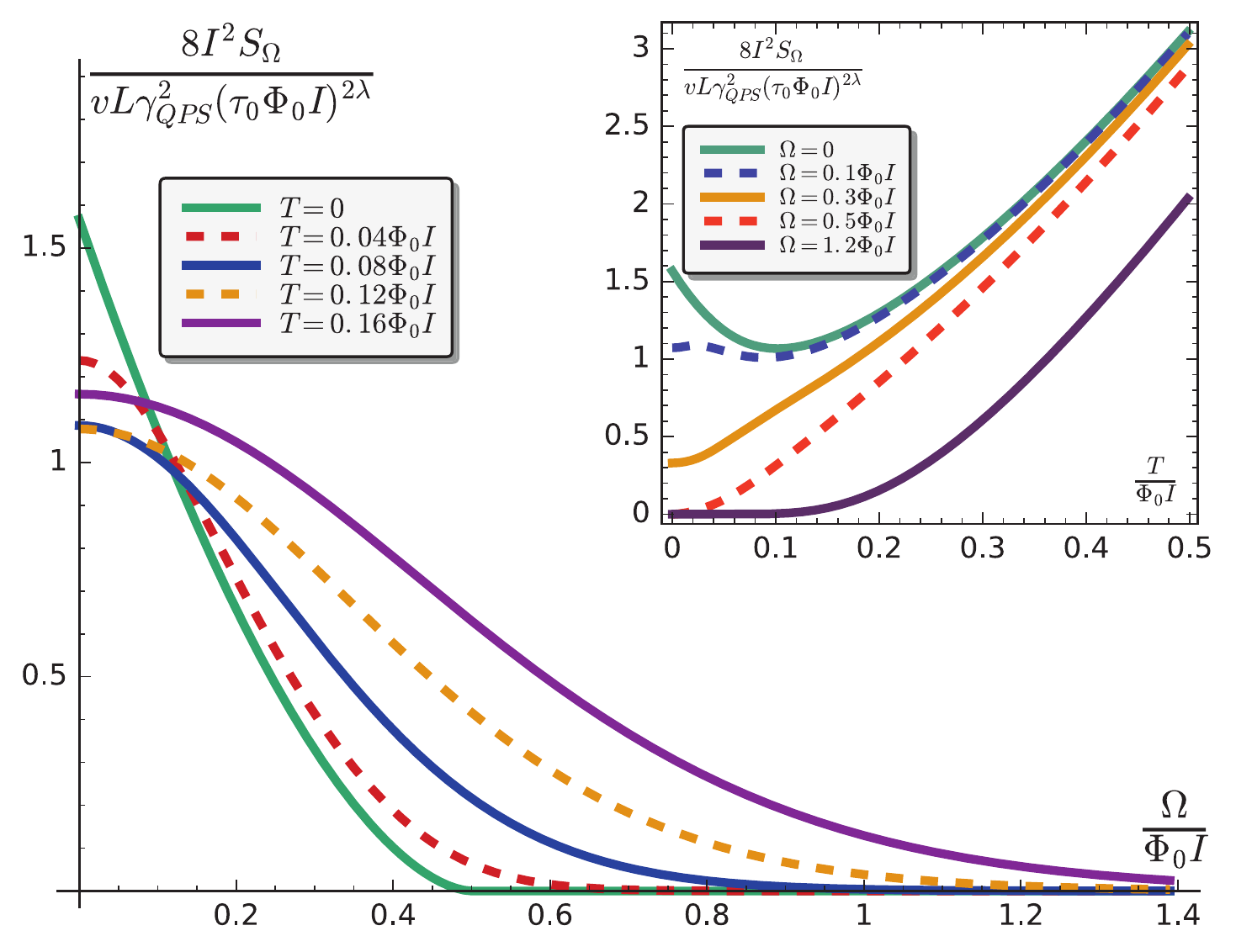}
\caption{The frequency dependence of the QPS noise spectrum ${\mathcal S}_{\Omega}$ (\ref{Sa3}) at $ \lambda =2.7$, large $E_C$ and different values of $T$ in the long wire limit. The inset shows ${\mathcal S}_{\Omega}$ as a function of $T$.}
\label{Fig:11}
\end{figure}

The result (\ref{Sa3}) is also illustrated in Fig. \ref{Fig:11}. At sufficiently small $\Omega$ (though still we keep $\Omega \gg v/L$) we observe a non-monotonous dependence of ${\mathcal S}_{\Omega}$ on temperature which serves as a clear manifestation of quantum coherent nature of QPS noise. 

To conclude our analysis of QPS-induced shot noise in superconducting nanowires, we point out that the perturbative in $\gamma_{QPS}$ approach employed here is fully justified in the "superconducting" regime, i.e. for
not too thin wires with $\lambda >  2$. In wires with  $\lambda < 2$
(characterized by unbound QPS-anti-QPS pairs) the perturbation theory becomes obsolete at low enough energies and large distances since $\gamma_{QPS}$ gets effectively renormalized to higher values, see, e.g., BKT-like RG equations (\ref{BKT}). 
Even in this case, however, our results may still remain applicable at relatively high temperature, frequency and/or current values. In the low energy limit long wires with $\lambda <2$ show an insulating behavior, as follows from the exact solution of the
corresponding sine-Gordon model \cite{CET}. This solution suggests that also voltage fluctuations become large in this limit.

Further interesting features of shot noise of the voltage in superconducting  nanowires under different measuring schemes were analyzed in Refs. \cite{SZ17,SZ17a,SZ17b,SZ18}.

\section{Full counting statistics of quantum phase slips}
In order to fully describe voltage fluctuations in the system under consideration it is in general necessary to
evaluate all cumulants of the voltage operator. Various aspects of voltage fluctuation statistics were already discussed in the case of quasi-one-dimensional wires \cite{GZTAPS} and resistively shunted Josephson junctions \cite{GMU2010, ZBN2015}. The authors  \cite{GZTAPS,GMU2010,ZBN2015} restricted their analysis to thermal fluctuations and, hence, their results remain applicable only at sufficiently high temperatures. Here, in contrast, we will set up a fully quantum mechanical treatment of the problem that essentially operates with interacting quantum phase slips and allows to fully describe full counting statistics of voltage fluctuations at any temperature down to $T \to 0$. 
In our further analysis we will to a large extent follow Ref. \cite{SZ19}.

As in Sec. VIA, we can again make use of the Josephson relation between the voltage and the phase variables. Then, analogously to the averages in Eqs.  (\ref{V}) and (\ref{VV}),  we can express the general correlator of voltages in the form
\begin{equation}
\langle V(t_1)V(t_2)...V(t_n)\rangle =\frac{1}{(2e)^n}   
\left\langle\dot \varphi_{+}(t_1)\dot \varphi_{+}(t_2)... \dot\varphi_{+}(t_n) e^{iS_{QPS}}\right\rangle_0.
\label{VVn}
\end{equation}
Here we stress that Eq. (\ref{VVn}) defines the symmetrized voltage correlators. E.g., for $n=2$ this equation is equivalent to 
Eq. (\ref{V2}), whereas for $n=3$ one can verify that \cite{3rdcum,SZ17}
\begin{multline}
\langle V(t_1)V(t_2)V(t_3)\rangle =\frac{1}{8}
 \big\{\langle\hat V(t_1)\big({\cal T}\hat V(t_2)\hat V(t_3)\big)\rangle\\
+\langle\big(\tilde{\cal T}\hat V(t_2)\hat V(t_3)\big)\hat V(t_1)\rangle
+\,\langle\hat V(t_2)\big({\cal T}\hat V(t_1)\hat V(t_3)\big)\rangle\\
+\langle\big(\tilde{\cal T}\hat V(t_1)\hat V(t_3)\big)\hat V(t_2)\rangle
+\,\langle\hat V(t_3)\big({\cal T}\hat V(t_1)\hat V(t_2)\big)\rangle\\
+\langle\big(\tilde{\cal T}\hat V(t_1)\hat V(t_2)\big)\hat V(t_3)\rangle
+\,\langle{\cal T}\hat V(t_1)\hat V(t_2)\hat V(t_3)\rangle\\
+\langle\tilde{\cal T}\hat V(t_1)\hat V(t_2)\hat V(t_3)\rangle\big\},
\label{tri}
\end{multline}
where ${\cal T}$ and $\tilde{\cal T}$ are, respectively, the forward and backward time ordering operators. 

\subsection{Cumulant generating function \label{sectFCS}}

In order to proceed with our calculation of higher voltage correlators it will be convenient for us to define the cumulant generating function 
\begin{equation}
  \mathcal W[J]=\ln(\mathcal Z[J])=\ln\left\langle e^{i\int dt J(t)V(t)} \right\rangle,
  \label{cumulantgenf}
\end{equation}
where
\begin{equation}
V=\frac{1}{\Phi_0C}\bigl(\partial_x \chi(-L/2)-\partial_x \chi(L/2)\bigr)
\label{vt}
\end{equation}
is a voltage drop across the wire and $\left\langle ... \right\rangle$ denotes the quantum average fulfilled with the total Hamiltonian of our system. Evaluating the $N$-th variational derivative of $\mathcal W[J]$ with respect to $J(t)$ one
recovers the $N$-th cumulant of the voltage operator.

The function $\mathcal Z[J]$ can be conveniently derived with the aid of Keldysh path integral technique already described in Sec. VI. Defining
the voltage values $V$ (\ref{vt})  on both forward and backward time branches of the Keldysh contour, respectively $V_F$ and $V_B$, and introducing, as before, ``classical'' and ``quantum'' variables $v_+=(V_F+V_B)/2$ and $V_-=V_F-V_B$, one can verify that the voltage cumulants of (\ref{VVn}) are exactly equivalent to the cumulants of the "classical" variable $v_+(t)$ in our path integral formalism. With this in mind the function $\mathcal Z[J]$ can be expressed as
\begin{equation}
  \mathcal Z[J]=\left\langle e^{iS_{QPS}[\chi_+,\chi_-]}e^{i\int dt J(t)v_+(t)}\right\rangle_0,
  \label{partf1}
\end{equation}
where $S_{QPS}$ is defined in Eq. (\ref{sqps}). The function (\ref{partf1}) generates voltage correlators
\begin{equation}
\langle v_+(t_1)v_+(t_2)...v_+(t_n)\rangle =
\left\langle  v_{+}(t_1)v_{+}(t_2)...
v_{+}(t_n) e^{iS_{QPS}}\right\rangle_0.
\label{VVnn}
\end{equation}

Let us eliminate the second exponent in Eq. (\ref{partf1}) by making a linear substitution $\chi_{i}=\lambda_{i}+\tilde\chi_i$ and imposing the condition
\begin{equation}
  \left\langle \tilde \chi_i e^{i\int dt J(t)v_+(t)}\right\rangle_0=0
\end{equation}
implying that
\begin{equation}
  \lambda_+(x,t)=\chi_0(x,t)-\int dt'G^K_{\chi v}(x;t,t')J(t'),
\end{equation}
\begin{equation}
  \lambda_-(x,t)=-\int dt' G^A_{\chi v}(x;t,t')J(t').
\end{equation}
Here we denoted $\chi_0\equiv \langle \chi_+ \rangle_0$ and introduced both Keldysh and advanced Green functions
\begin{equation}
  G^K_{\chi v}(x;t,t')=-i\langle \chi_+(x,t) v_+(t') \rangle_0
\label{KeldGF}
\end{equation}
and
\begin{equation}
  G^A_{\chi v}(x;t,t')=-i\langle \chi_-(x,t) v_+(t') \rangle_0.
\end{equation}
The latter function coincides with the transposed version of the retarded Green function
\begin{equation}
  G^R_{\chi v}(x;t,t')=-i\langle \chi_+(x,t) v_-(t') \rangle_0.
\end{equation}

As a result of the above manipulations, we get
\begin{equation}
  \mathcal Z[J]=e^{-\frac{i}{2}\int dtdt'J(t)G^K_{vv}(t,t')J(t')}\left\langle e^{iS_{QPS}[\lambda_++\tilde\chi_+,\lambda_-+\tilde\chi_-]}\right\rangle_0,
\end{equation}
where Keldysh Green function $G^K_{vv}(t,t')$ is defined analogously to that in Eq. (\ref{KeldGF}). The remaining average can be performed with the aid of the Wick's theorem and expressed via the two Green functions,
\begin{eqnarray}
  G^K_{\chi \chi}(x,x';t,t')=-i\langle \tilde \chi_+(x,t) \tilde \chi_+(x',t') \rangle_0,\\
  G^R_{\chi \chi}(x,x';t,t')=-i\langle \tilde \chi_+(x,t) \tilde \chi_-(x',t') \rangle_0,
\end{eqnarray}
while all averages of the type $\langle \tilde\chi_-\tilde\chi_- \rangle_0$ vanish identically due to causality.

Let us now evaluate the cumulant generating function by expanding $\mathcal Z[J]$ up to the second order in $\gamma_{QPS}$. We obtain
\begin{multline}
  \mathcal W[J]=-\frac{i}{2}\int dtdt'J(t)G^K_{vv}(t,t')J(t')+i\langle S_{QPS}[\lambda_++\tilde\chi_+,\lambda_-+\tilde\chi_-]\rangle_0\\-\frac{1}{2}\langle S^2_{QPS}[\lambda_++\tilde\chi_+,\lambda_-+\tilde\chi_-]\rangle_0+\frac{1}{2}\langle S_{QPS}[\lambda_++\tilde\chi_+,\lambda_-+\tilde\chi_-]\rangle_0^2.
\label{W2}
\end{multline}
Substituting now the QPS action $S_{QPS}$ (\ref{sqps}) into Eq. (\ref{W2}) one can verify that the first order contribution in $\gamma_{QPS}$ vanishes, while the second order one takes the form
\begin{widetext}
\begin{multline}
\mathcal W[J]\approx-\frac{i}{2}\int dtdt'J(t)G^K_{vv}(t,t')J(t')+\gamma_{QPS}^2\int\limits_{-L/2}^{L/2} dxdx'\int dt\int\limits^tdt'\bigl(P(x,x';t,t')-P(x',x;t',t)\bigr)\\\times\sin\biggl(\lambda_{+}(x,t)-\lambda_{+}(x',t')\biggr)\sin\biggl(\frac{\lambda_{-}(x,t)}{2}\biggr)\cos\biggl(\frac{\lambda_{-}(x',t')}{2}\biggr)
  \\-\frac{\gamma_{QPS}^2}{2}\int\limits_{-L/2}^{L/2} dxdx'\int dt\int dt'\bigl(P(x,x';t,t')+P(x',x;t',t)\bigr)\\\times\cos\biggl(\lambda_{+}(x,t)-\lambda_{+}(x',t')\biggr)\sin\biggl(\frac{\lambda_{-}(x,t)}{2}\biggr)\sin\biggl(\frac{\lambda_{-}(x',t')}{2}\biggr),
\label{cumgenf}
\end{multline}
where the function $P(x,x';t,t')$ is defined in 
\begin{multline}
  P(x,x';t,t')=\left\langle e^{i\left(\tilde\chi_{+}(x,t)-\tilde\chi_{+}(x',t')-\frac12\tilde\chi_{-}(x,t)-\frac12\tilde\chi_{-}(x',t')\right)}\right\rangle_0=\left\langle e^{i\left(\tilde\chi_{+}(x',t')-\tilde\chi_{+}(x,t)+\frac12\tilde\chi_{-}(x,t)+\frac12\tilde\chi_{-}(x',t')\right)}\right\rangle_0\\=e^{iG^K_{\chi\chi}(x,x';t,t')-\frac{i}2G^K_{\chi\chi}(x,x;t,t)-\frac{i}2G^K_{\chi\chi}(x',x';t',t')+\frac{i}2G^R_{\chi\chi}(x,x';t,t')-\frac{i}2G^A_{\chi\chi}(x,x';t,t')}.
\label{P}
\end{multline}
\end{widetext}

Equation (\ref{cumgenf}) enables one to directly evaluate all voltage correlators by taking variational derivatives
of $\mathcal W$ with respect to $J(t)$.  It follows immediately from Eq. (\ref{cumgenf}) that in the absence of QPS all voltage cumulants except for the second one (describing Gaussian noise) vanish identically.  We conclude, therefore, that at low enough temperatures only quantum phase slips give rise to both shot noise of the voltage and to all higher cumulants of the voltage operator in superconducting nanowires.

We also note that in the case of a constant in time current bias $I$ we have $\chi_0(x,t)=I\Phi_0t$ and the function
$P$ depends only on the time difference, i.e. $P(x,x';t,t')=P(x,x';t-t')$. This property will be employed in our subsequent
calculations.

\subsection{Voltage cumulants in the zero frequency limit\label{sectZF}}

As a first step, let us make use of the above general results and evaluate all cumulants of the voltage operator the zero frequency limit.
As we already know, an instantaneous voltage value $V(t)$ fluctuates in time due to a sequence of voltage pulses produced by QPS. It is instructive to define the time average
\begin{equation}
  \bar v =\frac{1}{\tau}\int\limits_{-\tau/2}^{\tau/2}d\tau V(\tau),
\end{equation}
where the time interval $\tau$ exceeds any relevant time scale for our problem.  It is easy to demonstrate that the cumulants of $\bar v$ are identical to the corresponding cumulants of the voltage operator evaluated in the zero frequency limit. For instance, for the first two cumulants one readily finds
\begin{gather}
  \langle\bar v\rangle=\langle V(t)\rangle=V(I),\\
  \langle(\bar v-\langle\bar v\rangle)^2\rangle=\frac{1}{\tau}\int dt\left(\langle V(t) V(0)\rangle-V^2\right)=\frac{1}{\tau}{\mathcal S}_0(I),
\end{gather}
where $S_\omega(I)$ is the frequency dependent voltage noise power for our wire already evaluated in Sec. VI.

In order to derive the cumulant generating function for $\bar v$
\begin{equation}
  w(j)=\ln\left\langle e^{ij\bar v}\right\rangle
\label{w0}
\end{equation}
it suffices to make use of Eq. (\ref{cumgenf}) and set $J(t)=j/\tau$ for $-\tau/2<t<\tau/2$ and $J(t)=0$ otherwise. For large enough time intervals $\tau$ the combination $\lambda_+(x,t)-\chi_0(x,t)$ becomes practically independent of both $x$ and $t$ implying that $\lambda_{+}(x,t)-\lambda_{+}(x',t')\approx I\Phi_0(t-t')$. Employing the equation of motion
\begin{equation}
  \biggl(\partial^2_t-v^2\partial_x^2\biggr)\hat\chi(x,t)=0,
\label{we}
\end{equation}
we obtain
\begin{equation}
\lim_{\omega\to 0}G^A_{\chi v}(x;\omega)=\lim_{\omega\to 0}G^R_{v\chi}(x;\omega)=\Phi_0
\end{equation}
and, hence, we have $\lambda_-(x,t)\approx -\Phi_0j/\tau$. As a result we get
\begin{multline}
  \frac{w(j)}{\tau}=-\frac{ij^2}{2\tau^2}G^K_{vv}(0)-\frac{\gamma^2_{QPS}}{2}\sin\bigg(\frac{\Phi_0j}{\tau}\bigg)\int\limits_{-L/2}^{L/2} dx dx'\int\limits_0^\infty dt \bigl(P(x,x';t)-P(x',x;-t)\bigr)\sin(I\Phi_0 t)\\
  -\gamma^2_{QPS}\sin^2\bigg(\frac{\Phi_0 j}{2\tau}\bigg) \int\limits_{-L/2}^{L/2} dx dx' \int\limits_{0}^\infty dt \bigl(P(x,x';t)+P(x',x;-t)\bigr)\cos(I\Phi_0 t).
\end{multline}
Performing the Fourier transformation
\begin{equation}
  P(x,x';\omega)=\int\limits_0^\infty dt e^{i\omega t}P(x,x';t)
\end{equation}
and introducing
\begin{equation}
  \Gamma(\omega)=\frac{\gamma_{QPS}^2}{4}\int\limits_{-L/2}^{L/2}dxdx'\bigl(P(x,x';\omega)+P^*(x',x;\omega)\bigr)
  \label{gamma}
\end{equation}
we find
\begin{equation}
  \frac{w(j)}{\tau}=-\frac{ij^2}{2\tau^2}G^K_{vv}(0)+\Gamma(I\Phi_0)\left(e^{\frac{i\Phi_0j}{\tau}}-1\right)+\Gamma(-I\Phi_0)\left(e^{-\frac{i\Phi_0j}{\tau}}-1\right).
\label{wf}
\end{equation}
This expression fully describes the statistics of QPS-induced voltage fluctuations in superconducting nanowires in the zero frequency limit.

It follows immediately from Eq. (\ref{wf}) that this statistics is Poissonian in the above limit. In particular, combining Eqs. (\ref{w0}) and (\ref{wf}) and evaluating the first and the second derivatives of $w$ with respect to $j$, we again recover the first two voltage cumulants in Eqs. (\ref{V2}) and (\ref{shot1}).

Higher voltage cumulants  in the zero frequency limit can be found in the same manner.  Let us define them as
\begin{equation}
\mathcal C_N(I)=(-i)^N\tau^{N-1}\left.\partial_j^Nw(j)\right|_{j\to0}.
\end{equation}
After a simple algebra all zero frequency cumulants can be expressed through the current-voltage characteristics for our system. In particular, for odd cumulants one has
\begin{equation}
\mathcal C_{2N+1}(I)=\Phi_0^{2N}V(I),
\label{oddc}
\end{equation}
whereas for even ones we obtain
\begin{equation}
\mathcal C_{2N}(I)=\Phi_0^{2N-1}V(I)\coth\left(\frac{I\Phi_0}{2T}\right).
\label{evenc}
\end{equation}
The above results demonstrate that in the long time limit the effect of interacting QPS reduces to that of independent sharp voltage pulses which occur with the effective rate $\Gamma(I\Phi_0)$ and are described by Poisson statistics. Note that this conclusion holds only for not too thin wires with $\lambda > 2$ described by quantum phase slips bound in close pairs.

\subsection{Noise power in the short wire limit \label{sectNP}}
At non-zero frequencies the system behavior becomes more involved and the statistics of voltage fluctuations deviates from Poissonian, as it will be demonstrated below. 

The general expression for the noise power is defined as
\begin{equation}
  {\mathcal S}_\omega(I)=-\int dt e^{i\omega t}\left.\frac{\delta^2\mathcal W[J]}{\delta J(t)\delta J(0)}\right|_{J\to0}.
\end{equation}
Making use of Eq. (\ref{cumgenf}) we find
\begin{widetext}
\begin{multline}
\label{np}  
{\mathcal S}_\omega(I)=iG^K_{vv}(\omega)\\+\frac{\gamma_{QPS}^2}{2}\Biggr[\int\limits_{-L/2}^{L/2}dxdx'G^K_{v\chi}(x;\omega)G^R_{v\chi}(x';\omega)\int\limits_0^\infty
  dt \bigl(P(x,x';t)-P(x',x;-t) \bigr)\cos(I\Phi_0 t)\left(e^{i\omega
      t}-1\right)\\+\frac{1}{4}\int\limits_{-L/2}^{L/2}dxdx'G^R_{v\chi}(x;\omega)G^R_{v\chi}(x';-\omega)\int\limits_{-\infty}^\infty
  dt \bigl(P(x,x';t)+P(x',x;-t) \bigr)\cos(I\Phi_0 t)e^{i\omega t}+\{\omega\to-\omega\}\Biggl].
\end{multline}
\end{widetext}
With the aid of FDT this result can be transformed to that already derived in Sec. VI where we merely addressed the long wire limit. Here, in contrast, we will specify the expression for the noise power for shorter wires. This limit also includes the case of Josephson junctions and other types of short superconducting weak links.

Let us note that each term in the square brackets in Eq. (\ref{np}) contains the combination of the form-factors $P(x,x';\omega)$ describing intrinsic dynamics of a superconducting nanowire during the QPS process, as well as two Green functions of the $v\chi$-type indicating how the detector "feels" voltage fluctuations inside the nanowire. Provided the wire is short enough one can retain only the dependence of these Green functions on frequency and ignore their spatial coordinates. Then one gets
\begin{equation}
G^R_{v\chi}(x;\omega)\approx\Phi_0(1-i\omega\tau_R+...),
\end{equation}
 where $\tau_R$ is the effective RC-time of the system. Accordingly we obtain
 \begin{equation}
 G^K_{v\chi}(x,\omega)\approx -i\omega\tau_R\coth(\omega/(2T))
 \end{equation}
Employing these approximations, from Eq. (\ref{np}) we obtain
\begin{multline}
{\mathcal S}_{\omega}(I)=iG^K_{vv}(\omega)-i\Phi_0^2\tau_R\omega\coth\left(\frac{\omega}{2T}\right)\bigl(\Gamma^R(\omega+I\Phi_0)\\+\Gamma^R(\omega-I\Phi_0)+\Gamma^R(-\omega+I\Phi_0)+\Gamma^R(-\omega-I\Phi_0)\\-2\Gamma^R(I\Phi_0)-2\Gamma^R(-I\Phi_0)\bigr)+\frac{1}{2}\Phi_0^2\bigl(\Gamma(\omega+I\Phi_0)\\+\Gamma(\omega-I\Phi_0)+\Gamma(-\omega+I\Phi_0)+\Gamma(-\omega-I\Phi_0)\bigr),
\end{multline}
where we introduced the function
\begin{equation}
  \Gamma^R(\omega)=\frac{\gamma_{QPS}^2}{4}\int\limits_{-L/2}^{L/2}dxdx'\bigl(P(x,x';\omega)-P^*(x',x;-\omega)\bigr)
\end{equation}
related to $\Gamma (\omega)$ (\ref{gamma}) as
\begin{equation}
  \Gamma^R(\omega)=\int \frac{dz}{2\pi i}\frac{\Gamma(z)-\Gamma(-z)}{z-\omega-i0}.
\end{equation}

For an illustration let us consider a short superconducting nanowire embedded in a linear dissipative external circuit modeled by an Ohmic shunt resistor $R_S$. This situation is equally relevant for resistively shunted Josephson junctions in the limit of large Josephson coupling energies $E_J$. In this limit one has
\begin{equation}
  G^R_{\chi\chi}(x,x';\omega)\approx -\frac{2\pi i\mu}{\omega+i0},
\end{equation}
where  $\mu=R_Q/R_S$ is the shunt dimensionless conductance and  $R_Q=\pi/(2e^2)$ is the "superconducting" resistance
quantum.  The QPS rate then equals to
\begin{equation}
  \Gamma(\omega)=\gamma_{QPS}^2(2\pi T\tau_R)^{2\mu}e^{\frac{\omega}{2T}}\frac{\GammaF\left(\mu+\frac{i\omega}{2\pi T}\right)\GammaF\left(\mu-\frac{i\omega}{2\pi T}\right)}{8\pi T \GammaF(2\mu)},
\label{G1}
\end{equation}
where $\GammaF(y)$ is the Euler gamma-function and $\tau_R^{-1}$ plays the role of effective high-energy cutoff frequency. Evaluating the corresponding integrals in the short wire limit $\omega,T,I\Phi_0\ll \tau_R^{-1}$ and also for $1<\mu<3/2$, we obtain
\begin{equation}
  \Gamma^R(\omega)={\rm const}-i\Gamma(\omega)e^{-\frac{\omega}{2T}}\frac{\sin\left(\pi\mu+\frac{i\omega}{2 T}\right)}{\cos(\pi\mu)}.
\end{equation}
These expressions can be simplified in certain limits. For example, by setting $0<\mu-1\ll 1$ we get 
\begin{equation}
  \Gamma(\omega)\approx\frac{\gamma_{QPS}^2(2\pi T\tau_R)^{2\mu}e^{\frac{\omega}{2T}-2{\bf C}(\mu-1)}\sqrt{\omega^2+4\pi^2 T^2(\mu-1)^2}}{16\pi T^2 \GammaF(2\mu)\sqrt{\sin\left(\pi\mu+\frac{i\omega}{2 T}\right)\sin\left(\pi\mu+\frac{i\omega}{2 T}\right)}},
\end{equation}
where $\bf C$ is Euler-Mascheroni constant. Also the expressions for both QPS rates get significantly simplified in the limit $|\omega|\gg T$. One has
\begin{equation}
\Gamma(\omega)\approx\pi\gamma_{QPS}^2\theta(\omega)\frac{(\omega\tau_R)^{2\mu}}{2\omega\GammaF(2\mu)},
\end{equation}
\begin{equation}
\Gamma^R(\omega)\approx{\rm const}+\pi\gamma_{QPS}^2\frac{|\omega\tau_R|^{2\mu}e^{-i\pi\mu{\rm sign}(\omega)}}{4\omega\GammaF(2\mu)\cos(\pi\mu)}.
\label{GR2}
\end{equation}
Accordingly, in the zero-temperature limit one finds
\begin{equation}
\mathcal C_N(I)=\pi\gamma_{QPS}^2{\rm sign}^N(I)\frac{\Phi_0^{N+2\mu-1}\tau_R^{2\mu}}{2\GammaF(2\mu)}|I|^{2\mu-1}.
\end{equation}
The above results are consistent with ones previously derived for ultrasmall Josephson junctions \cite{ANO}.

\subsection{Higher voltage cumulants \label{sectSW}}
We now turn to higher voltage cumulants at non-zero frequencies. We adopt the following definition for the $N$-th voltage cumulant:
\begin{multline}
{\mathcal S}_{\omega_1,...,\omega_{N-1}}(I)=\int dt_1...dt_{N-1} e^{i\omega_1t_1+...+i\omega_{N-1}t_{N-1}}(-i)^N\left.\frac{\delta^N \mathcal W[J]}{\delta J(t_{N-1})...\delta J(t_1)\delta J(0)}\right|_{J\to 0}.
\end{multline}
In the zero frequency limit this definition yields
\begin{equation}
{\mathcal S}_{\underbrace{00...0}_N}(I)=\mathcal C_N(I).
\end{equation}
Under the condition $T,\omega\ll \tau_R^{-1}$ or, in other words, provided the detector immediately "feels" QPS-generated voltage fluctuations, one can set $\tau_R\to 0$ and explicitly evaluate all voltage cumulants at non-zero frequencies. In this case the cumulant generating function takes the form
\begin{multline}
  \mathcal W[J]\approx-\frac{i}{2}\int dtdt'J(t)G^K_{vv}(t-t')J(t')\\-\gamma_{QPS}^2\int\limits_{-L/2}^{L/2} dxdx'\int dt\int\limits^tdt'\bigl(P(x,x';t-t')-P(x',x;t'-t)\bigr)\\\times\sin\bigl(I\Phi_0(t-t')\bigr)\sin\biggl(\frac{\Phi_0J(t)}{2}\biggr)\cos\biggl(\frac{\Phi_0 J(t')}{2}\biggr)
  \\-\frac{\gamma_{QPS}^2}{2}\int\limits_{-L/2}^{L/2} dxdx'\int dt\int dt'\bigl(P(x,x';t-t')+P(x',x;t'-t)\bigr)\\\times\cos\bigl(I\Phi_0 (t-t')\bigr)\sin\biggl(\frac{\Phi_0 J(t)}{2}\biggr)\sin\biggl(\frac{\Phi_0J(t')}{2}\biggr).
  \label{cumgenfs}
\end{multline}

One can observe that the second term in Eq. (\ref{cumgenfs}) can only contribute to odd cumulants, whereas the last term determines all even cumulants. After some algebra we arrive at the following expressions for both even and odd voltage cumulants, respectively
\begin{multline}
  {\mathcal S}_{\omega_{1},...,\omega_{2M}}(I)=\frac{\Phi_0^{2M+1}}{2^{2M}(2M)!}\sum_{p\in{\rm perm}} \sum_{m=0}^{2M} \binom{2M}{m} \\ \times\Bigl(\Gamma^R\bigl(I\Phi_0-(-1)^m(\omega_{p_1}+...+\omega_{p_m})\bigr)\\-\Gamma^R\bigl(-I\Phi_0-(-1)^m(\omega_{p_1}+...+\omega_{p_m})\bigr)\Bigr)
\end{multline}
and
\begin{multline}
  {\mathcal S}_{\omega_{1},...,\omega_{2M+1}}(I)=\frac{\Phi_0^{2M+2}}{2^{2M+1}(2M+1)!}\sum_{p\in{\rm perm}} \sum_{m=0}^{M} \binom{2M+1}{2m+1}\Bigl(\Gamma\bigl(I\Phi_0+(\omega_{p_1}+...+\omega_{p_{2m+1}})\bigr) \\+\Gamma\bigl(-I\Phi_0+(\omega_{p_1}+...+\omega_{p_{2m+1}})\bigr)
 +\Gamma\bigl(I\Phi_0-(\omega_{p_1}+...+\omega_{p_{2m+1}})\bigr)
 +\Gamma\bigl(-I\Phi_0-(\omega_{p_1}+...+\omega_{p_{2m+1}})\bigr),
  \Bigr).
\end{multline}
where the sum is taken over all permutations of frequencies.

The above results allows us to extend the relation between the voltage cumulants and the current-voltage characteristics of our device to non-zero frequencies. For the odd cumulants we obtain
\begin{widetext}
\begin{multline}
 {\mathcal S}_{\omega_{1},...,\omega_{2M}}(I)=\frac{\Phi_0^{2M+2}I}{2^{2M-1}(2M)!}\int \frac{dI'}{2\pi i} V(I') \\\times\sum_{p\in{\rm perm}} \sum_{m=0}^{2M} \binom{2M}{m}
  \frac{1}{(I'\Phi_0+(-1)^m(\omega_{p_1}+...+\omega_{p_m})-i0)^2-(I\Phi_0)^2},
\label{evencw}
\end{multline}
while the expression for the even cumulants reads
\begin{multline}
{\mathcal S}_{\omega_{1},...,\omega_{2M+1}}(I)=\frac{\Phi_0^{2M+1}}{2^{2M+1}(2M+1)!}\sum_{p\in{\rm perm}} \sum_{m=0}^{M} \binom{2M+1}{2m+1}\\\times\Biggl(\coth\biggl(\frac{I\Phi_0+(\omega_{p_1}+...+\omega_{p_{2m+1}})}{2T}\biggr)V\biggl(I+\frac{\omega_{p_1}+...+\omega_{p_{2m+1}}}{\Phi_0}\biggr)
  \\+\coth\biggl(\frac{I\Phi_0-(\omega_{p_1}+...+\omega_{p_{2m+1}})}{2T}\biggr)V\biggl(I-\frac{\omega_{p_1}+...+\omega_{p_{2m+1}}}{\Phi_0}\biggr)
  \Biggr).
  \label{oddcw}
  \end{multline}
\end{widetext}

\begin{figure}
\includegraphics[width=0.45\linewidth]{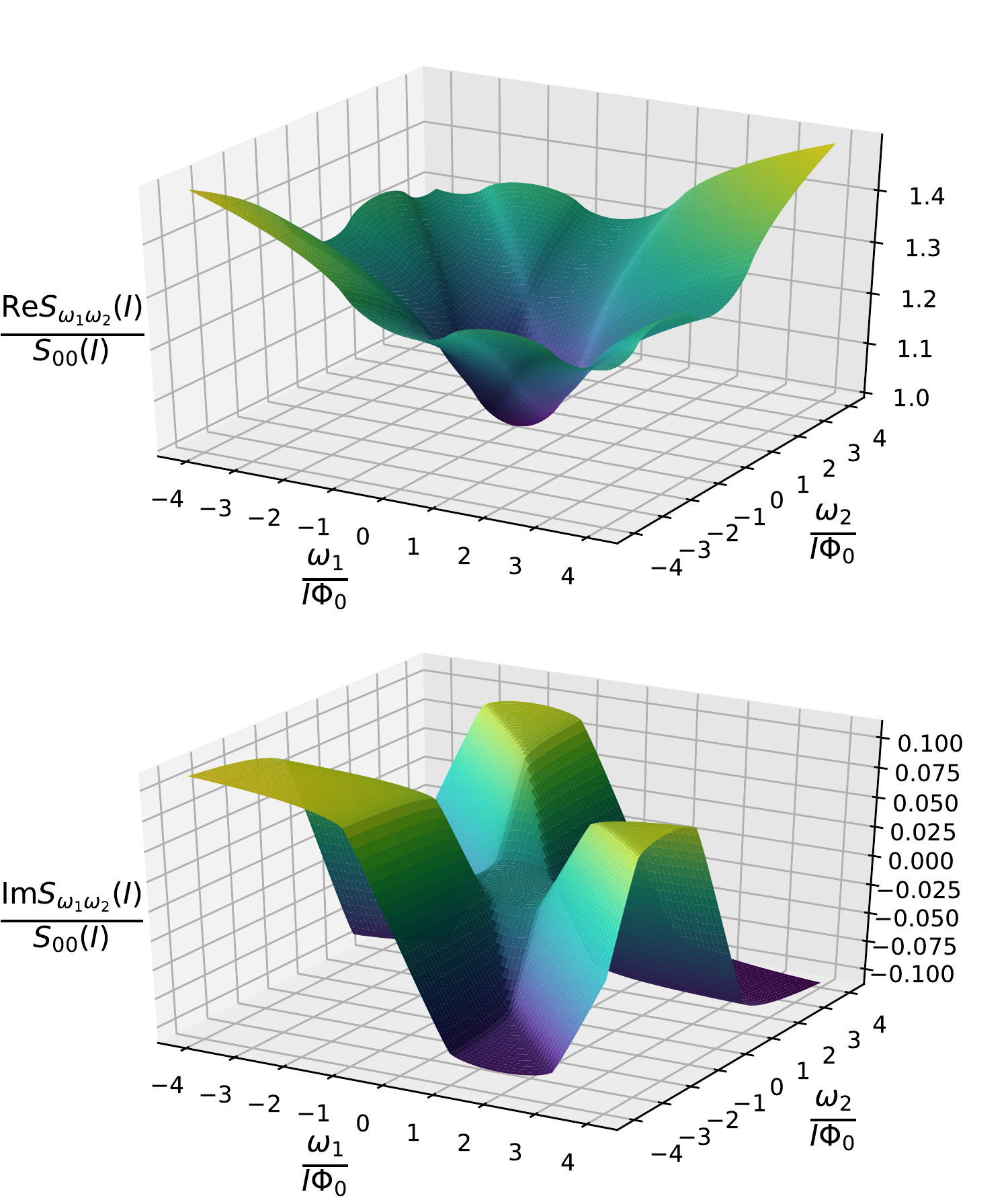}
\includegraphics[width=0.45\linewidth]{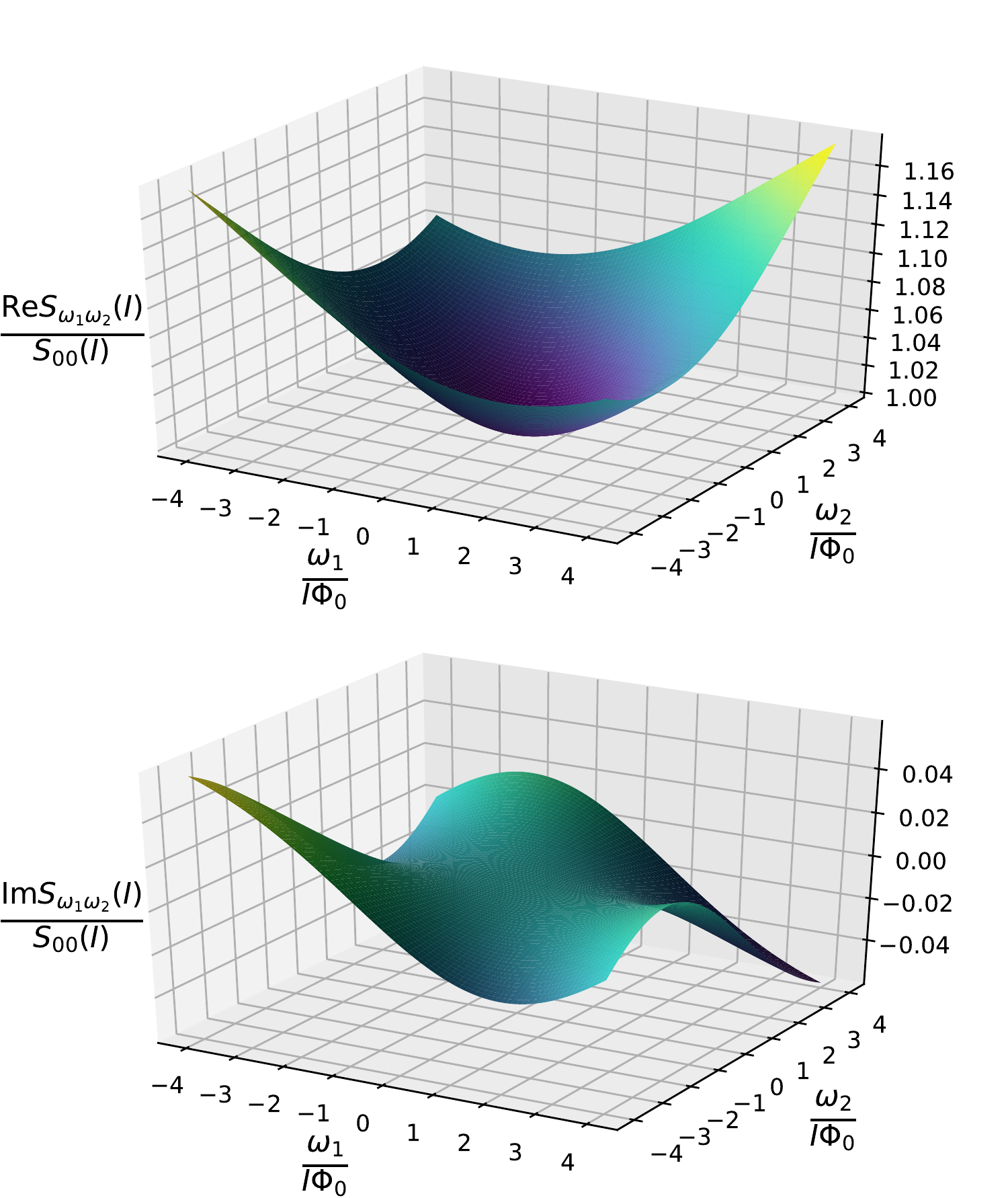}
\caption{Real and imaginary parts of the third voltage cumulant at $T \to 0$ (left panels) and $T=I\Phi_0$ (right panels) for $\mu=1.1$.}
\label{Fig:12}
\end{figure}

These expressions can be evaluated numerically with the aid of Eqs. (\ref{G1})-(\ref{GR2}) for $\Gamma(\omega)$ and $\Gamma^R(\omega)$ derived above. The corresponding results for the third voltage cumulant as a function of two frequencies are displayed in Fig. \ref{Fig:12} at low and high temperatures. The third voltage cumulant generally consists of real and imaginary parts 
\begin{equation}
{\mathcal S}_{\omega_{1},\omega_{2}}(I)={\rm Re}{\mathcal S}_{\omega_{1},\omega_{2}}(I)+i {\rm Im}{\mathcal S}_{\omega_{1},\omega_{2}}(I).
\end{equation}
As could be seen in the plots, both these functions become considerably smoother at higher $T$.

To conclude our analysis of voltage fluctuation statistics in superconducting nanowires we emphasize again that in the zero-frequency limit the this statistics reduces to Poissonian similarly to the situation encountered in a number of other tunneling-like problems. In this limit all (symmeterized) cumulants of the voltage operator can be expressed in a simple manner through the current-voltage characteristics of the system $V(I)$, cf. Eqs. (\ref{oddc}),  (\ref{evenc}). At non-zero frequencies, quantum voltage fluctuations in superconducting nanowires are not anymore described by Poisson statistics. This is because inter-QPS interaction produced by an effective environment (due to the wire itself and/or an external dissipative circuit) starts playing a more important role at shorter time scales making the whole problem much more involved.  Remarkably, also in this case it is possible to establish a relation between the voltage cumulants and the current-voltage characteristics of our device $V(I)$, though in a much more complicated form as compared to that in the zero frequency limit, cf. Eqs.  (\ref{evencw}),  (\ref{oddcw}).
The latter observation could be important for possible experimental verification of the above results.

\section{Topology-controlled phase coherence in superconducting nanowires}

It is by now well established that at $ T \to 0$ a long superconducting nanowire suffers a quantum phase transition (QPT) controlled by
the wire cross section $s$ or, equivalently by the parameter  $\lambda \equiv g/8$ which we already introduced above and which sets the magnitude of (logarithmic in space-time) interaction between different quantum phase slips \cite{ZGOZ}. For $\lambda >2$ this interaction  is strong enough and close QPS-anti-QPS pairs are formed in the wire which then demonstrates vanishing {\it linear} resistance $R \propto T^{2\lambda - 3}$, cf. Eqs. (\ref{V3}), (\ref{vsi}). Hence, as long as $\lambda >2$ (or, equivalently, $g >16$) the ground state of the system  can be considered  superconducting. In contrast, for  $\lambda < 2$ inter-QPS interaction is weak, quantum phase slips are unbound, and the wire acquires non-zero resistance which tends to {\it increase} with decreasing $T$. The latter feature allows one to call the wire behavior insulating provided $\lambda <2$. Thus, at $g=16$ and $T \to 0$ one expects a superconductor-to-insulator quantum phase transition (SIT) to occur in the systems under consideration.

Let us emphasize that possible insulating behavior of superconducting nanowires is essentially linked to a certain type of experiment performed with such nanowires and may not always be realized. For example, an ultrathin superconducting nanowire forming a closed ring does not loose the ability to carry supercurrent even for $\lambda <2$, as we have already demonstrated in Sec. V. In this case a characteristic length scale 
(\ref{rc}) emerges beyond which phase coherence (and, hence, supercurrent) gets exponentially suppressed by QPS. In what follows
\begin{equation}
R_c \sim L_c \propto \exp \left(\frac{ag_\xi}{2-\lambda}\right)
\label{Lc}
\end{equation}
emerges beyond which phase coherence (and, hence, supercurrent) gets exponentially suppressed by QPS. The correlation length  (\ref{Lc}) diverges at $\lambda \to 2$, thus signaling the transition to the ordered phase $\lambda > 2$ with bound QPS-anti-QPS pairs and more robust superconductivity.

\begin{figure}
\includegraphics[width=0.6\linewidth]{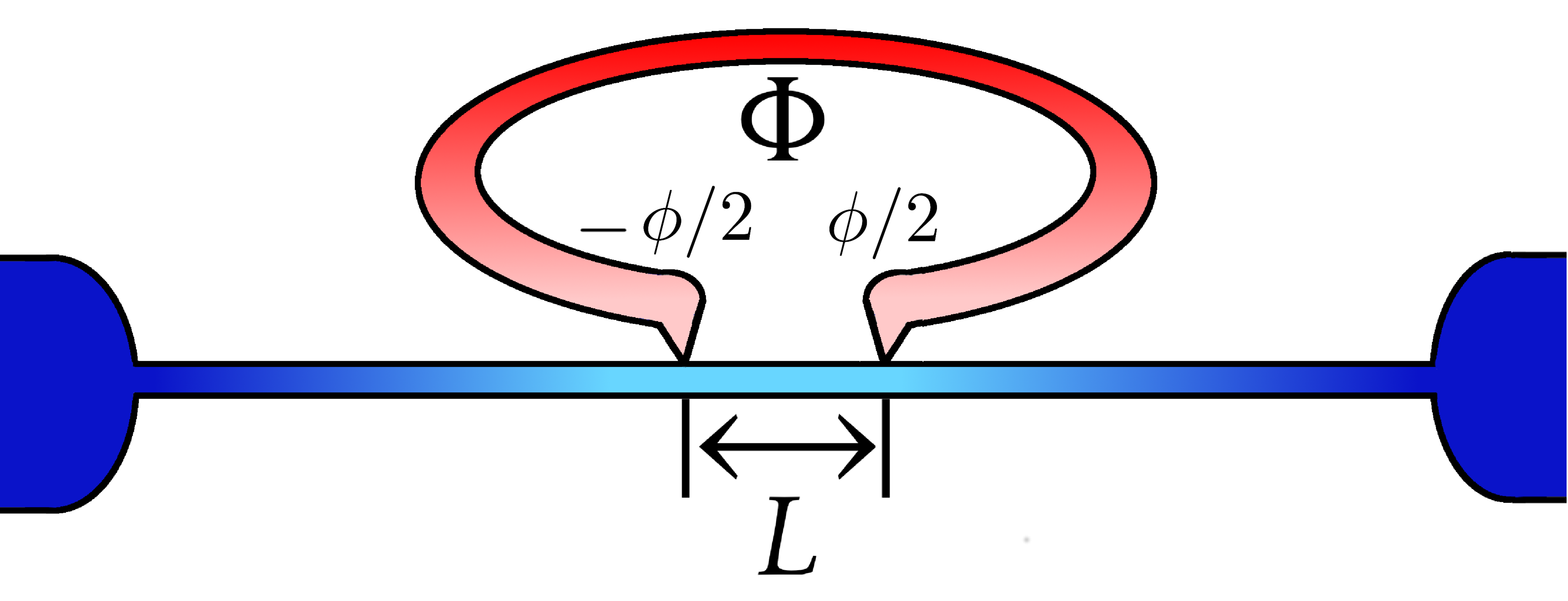}
\caption{Bulk open superconducting ring attached to a superconducting nanowire via two ultrasmall tunnel barriers located at a distance $L$ from each other. The ring is threaded by the magnetic flux $\Phi$.}
\label{Fig:13}
\end{figure}

On the other hand, a closed ring geometry seriously restricts the space for phase fluctuations, thereby enhancing the tendency towards superconductivity, see, e.g., Ref. \cite{BFSZ} for further discussion of this point. For this reason it is highly desirable to analyze ground state properties of superconducting nanowires  where no fluctuation configurations are suppressed by geometry constraints and/or boundary conditions. An example of the system of that kind is depicted in Fig. \ref{Fig:13}. A long superconducting nanowire with sufficiently small cross section $s$ geometric capacitance (per length) $C$ and kinetic inductance (times length) $\mathcal{L}_{\rm kin}$ is attached to two big superconducting reservoirs at its ends. In order to probe fluctuation effects inside the wire it is connected to a bulk superconductor forming an open ring by two identical small area tunnel junctions with Josephson energy $E_J$ located at a distance $L$ from each other at the points $x=0$ and $x=L$. External magnetic flux $\Phi$  piercing the ring controls the phase difference $\phi=2\pi \Phi/\Phi_0$ between the bulk sides of the point contacts. As desired, within this setup fluctuations of the superconducting phase $\varphi (x,\tau)$ remain unrestricted in any point $x$ of the wire.

Below we will analyze the effect of these fluctuations on the supercurrent $I(\phi)$ flowing through the wire segment of length $L$ between the two Josephson contacts \cite{RSZ19}. As $I(\phi)$ is a $2\pi$-periodic function of $\phi$ in what follows it suffices to restrict the phase interval to $|\phi|\leq \pi$. Below we will demonstrate that -- in the setup displayed Fig. \ref{Fig:13} -- a "disordered" phase $\lambda <2$ (or $g<16$) for a superconducting nanowire itself consists of two different phases: A non-superconducting one with $g<2$ as well as a "mixed" one with $2<g<16$ characterized by two different correlation lengths, $L_c$ (\ref{Lc}) and 
\begin{equation}
L^* \propto g^{\frac{1}{1-2/g}}, \quad g \geq 2.
\label{L***}
\end{equation}
The latter phase is characterized by a non-trivial interplay between supercurrent and quantum fluctuations resulting in superconducting behavior of the wire at shorter length scales combined with its vanishing superconducting response in the long scale limit. 

\subsection{Effective action}
Low energy processes in the system under consideration can be described by the effective action
\begin{equation}
S[\varphi]=S_{TL}[\varphi]+S_{J}[\varphi (0), \varphi (L)],
\label{system_action}
\end{equation}
where, as before,
\begin{equation}
S_{TL}[\varphi]=\frac{C}{8e^2}\int\limits_0^{1/T} d\tau \int dx \left[
(\partial_\tau\varphi)^2+v^2(\partial_x\varphi)^2
\right]
\label{wire_action}
\end{equation}
is the low energy effective action for a superconducting nanowire in the $\varphi$-representation \cite{ZGOZ,GZQPS,OGZB} and 
\begin{equation}
S_{J}[\varphi_1, \varphi_2]=-E_J\int\limits_0^{1/T} d\tau\bigl[
\cos(\varphi_1+\phi/2)+\cos(\varphi_2-\phi/2)
\bigr]
\label{int}
\end{equation}
accounts for the Josephson energy of the contacts, where we set $\varphi_1=\varphi(0,\tau)$, $\varphi_2=\varphi(L,\tau)$.  For simplicity, in Eq. (\ref{int})) we do not include the charging energy of the point contacts which could always be absorbed into the first term of the wire action (\ref{wire_action}).

As the wire action (\ref{wire_action}) is Gaussian it is possible to exactly integrate out the phase variable $\varphi (x)$ at all values $x$ except for $x=0,L$. Then we arrive at the reduced effective action $S_R$ which depends on the two phase variables  $\varphi_1$ and $\varphi_2$. The grand partition function $\mathcal{Z}$ reads
\begin{eqnarray}
\nonumber
\mathcal{Z}=\int D\varphi(x)\,{\rm e}^{-S_{TL}[\varphi(x)]-S_{J}[\varphi(0),\varphi(L)]}
=\int D\varphi_1 D\varphi_2\, {\rm e}^{-S_R[\varphi_1,\varphi_2]-S_{J}[\varphi_1,\varphi_2]},
\label{B}
\end{eqnarray}
where
\begin{equation}
S_R[\varphi_1,\varphi_2]=
\frac{1}{2}\tr\left[
\begin{pmatrix}
\varphi_1 & \varphi_2
\end{pmatrix}
\begin{pmatrix}
G_0(0) & G_0(L) \\
G_0(L) & G_0(0)
\end{pmatrix}^{-1}
\begin{pmatrix}
\varphi_1 \\ \varphi_2
\end{pmatrix}\right]
.\label{eq.3}
\end{equation}
Here the trace also includes integration over the imaginary time. The Green function $G_0=\langle \varphi(x,\tau) \varphi(0,0)\rangle_{S_0}$ has the form
\begin{equation}
G_0(\omega_n,x)=\frac{4e^2}{C}\int\frac{dq}{2\pi}\,\frac{{\rm e}^{iqx}}{\omega_n^2+v^2q^2}
=\frac{4\pi}{g|\omega_n|}{\rm e}^{-\left|
\frac{\omega_n x}{v}
\right|}.
\end{equation}
In order to diagonalize  the quadratic part of the action it is convenient to express $S_R$ in terms of the variables $\varphi_\pm=(\varphi_1\pm \varphi_2)/2$. Then we obtain
\begin{align}
\nonumber
S_R+S_{J}
=\frac{1}{2}\sum\limits_{a=\pm}\tr\Bigl[
\varphi_aG_{0,a}^{-1}\varphi_a
\Bigr]-2E_J\int\limits_0^{1/T} d\tau\,\cos(\varphi_+) \cos(\varphi_--\phi/2)\label{S_eff}
\end{align}
with the propagators
\begin{equation}
G_{0,\pm}(\omega_n)=\frac{2\pi}{g|\omega_n|}\left(1\pm {\rm e}^{-\left|
\frac{\omega_n L}{v}
\right|}
\right).
\end{equation}

Let us stress that the phase variables $\varphi_+$ and $\varphi_-$ account for different physics in our problem. 
The phase $\varphi_-=(\varphi(L)-\varphi(0))/2$ determines the supercurrent flowing in-between two contacts
inside the wire segment of length $L$. Hence, configurations with non-zero $\varphi_-$ have non-zero energies due the kinetic inductance of the wire and the mode corresponding to $\varphi_-$ has a mass equal to $gv/2\pi L$. The variable $\varphi_+$, in contrast, describes simultaneous shifts of both phases $\varphi(0)$ and $\varphi(L)$ by the same value without producing any phase gradient along the wire. Thus, in the absence of interactions the mode corresponding to  $\varphi_+$ is massless. At the same time, below we will observe that fluctuations of $\varphi_+$ yield renormalization of the Josephson coupling energies $E_J$ of the contacts and, as such, should also be taken into account. 

\subsection{Variational analysis and propagators}

Let us make use of the variational perturbation theory as described, e.g., in Ref. \cite{klnrt}. The main idea here is 
to improve the standard perturbation expansion by adding an extra term $\delta S$ which depends on the variational parameters to the quadratic part of the action $S_R$. In order to accomplish this goal the partition function (\ref{B}) can be identically rewritten as
\begin{align}
\mathcal{Z}&=\int D\varphi_1 D\varphi_2\,{\rm e}^{-S_{\rm tr}}{\rm e}^{-(S_{J}-\delta S)}
\label{eq.7}
\end{align}
with the trial action $S_{\rm tr}=S_R+\delta S$. The last exponent can then be conveniently expanded in powers of
$S_{J} -\delta S$. Being expanded to {\it all} orders, the partition function (\ref{eq.7}) obviously remains independent of the choice of $\delta S$ and the variational parameters.  However, such a dependence emerges as long as only a finite number of terms of this expansion is kept. Then the most accurate approximation is achieved by minimizing the result of the perturbative expansion with respect to the variational parameters.

In what follows we choose the trial action in the form
\begin{align}
S_{\rm tr}=\frac{1}{2}\tr\left[\varphi_+(G_{0+}^{-1}+m_+)\varphi_+\right]
+\frac{1}{2}\tr\left[(\varphi_--\psi)(G_{0-}^{-1}+m_-)(\varphi_--\psi)\right], \label{ansatz}
\end{align}
which corresponds to effectively performing a self-consistent harmonic approximation (SCHA). 

Here the parameters $m_\pm$ represent the interaction-generated effective masses for the modes associated with the phase variables $\varphi_\pm$. The parameter $\psi$ accounts for the average value of the combination $(\varphi(L)-\varphi(0))/2$. We note that a somewhat similar variational calculation with a massive term was proposed in the context of Brownian motion of a quantum particle in a periodic potential with linear Ohmic dissipation \cite{FisherZwerger}. The results obtained within the framework of this variational approach agree with those derived by means of more rigorous techniques \cite{SZ90}. 

Expanding the last exponent in Eq. (\ref{eq.7}) in powers of 
$S_{J} -\delta S$ and evaluating the integrals, for  the free energy $\mathcal{F}=-T\ln \mathcal{Z}$ we get
\begin{equation}
\mathcal{F}=\mathcal{F}_0+\mathcal{F}_1+ {\rm higher \;order \;terms},
\label{14}
\end{equation}
where
\begin{align}
\nonumber
\mathcal{F}_0=\frac{T}{2}\left(\tr\,\ln\, G^{-1}_+ + \tr\,\ln\, G^{-1}_-\right),\quad 
\mathcal{F}_1=\bigl\langle S_{\rm int}-\delta S\bigr\rangle_{\rm tr}.
\label{16}
\end{align}
Neglecting all higher order terms in the expansion (\ref{14}) and evaluating the average in Eq. (\ref{16})
with respect to $S_{\rm tr}$  (\ref{ansatz}), we obtain
\begin{multline}
\mathcal{F}_1=-\frac{m_+}{2} G_+(0)
-\frac{m_-}{2}G_-(0)
+\frac{1}{2}\psi\, G_{0-}^{-1}(\omega_n=0)\,\psi 
\\
-2E_J\cos(\psi-\phi/2) {\rm e}^{-\bigl(G_+(0)+G_-(0)\bigr)/2},
\end{multline}
where $G_\pm^{-1}=G_{0,\pm}^{-1}+m_\pm$ and $G_\pm (0)=T\sum\limits_{\omega_n} G_\pm(\omega_n)$ with $\omega_n=\pi T(2n+1)$ being the Matsubara frequency.

Evaluating variational derivatives of $\mathcal{F}$ with respect to $m_\pm$ and $\psi$, one readily finds
\begin{align}
\label{18}
&\frac{\delta \mathcal{F}_0}{\delta m_\pm}=G_\pm(0)/2,\\
&\frac{\delta \mathcal{F}_0}{\delta \psi}=0,\\
&\frac{\delta \mathcal{F}_1}{\delta m_\pm}=- G_\pm(0)/2
-\frac{1}{2}\frac{\delta G_\pm(0)}{\delta m_\pm}\left(
m_\pm-2E_J\cos(\psi-\phi/2){\rm e}^{-\bigl(G_+(0)+G_-(0)\bigr)/2}
\right),\\
&\frac{\delta \mathcal{F}_1}{\delta \psi}=2E_J\sin(\psi-\phi/2){\rm e}^{-\bigl(G_+(0)+G_-(0)\bigr)/2}+G_{0-}^{-1}(\omega_n=0)\psi.
\label{21}
\end{align}
Imposing the extremum conditions $\delta\mathcal{F}/\delta m_\pm=\delta\mathcal{F}/\delta \psi=0$ and making use of Eqs.
(\ref{18})-(\ref{21}) we arrive at the following set of SCHA equations:
\begin{equation}
m_+=m_-\equiv m
\label{2222}
\end{equation}
and
\begin{align}
&2E_J\cos(\psi-\phi/2){\rm e}^{-\bigl(G_+(0)+G_-(0)\bigr)/2}-m=0,\label{eq_RG}\\
&2E_J\sin(\psi-\phi/2){\rm e}^{-\bigl(G_+(0)+G_-(0)\bigr)/2}+\frac{gv}{2\pi L}\psi=0.\label{eq_mot}
\end{align}
Note that the mass parameters $m_\pm$ in Eq. (\ref{2222}) remain equal to each other since the two Josephson junctions involved in our problem are identical. Equation (\ref{eq_RG}) provides the relation between the effective mass $m$ and the fluctuation-induced renormalization of the Josephson coupling energy $E_J$. Equation (\ref{eq_mot}) represents the equation of motion for $\psi$. It coincides with the equation of motion for the phase $\varphi_-$ with $E_J$ renormalized by quantum fluctuations.

As we already pointed out, the wire effective action in the form (\ref{wire_action}) applied in the low energy limit, i.e. for $\omega, vq \ll \Delta$. Hence, a proper ultraviolet cutoff should be imposed which respects both causality and the fluctuation-dissipation relation. This goal is achieved by modifying the spectral density 
$$J_\pm(\omega)=-\frac{1}{\pi}{\rm Im}\, G^R_\pm(\omega)$$ 
making it decay at $\omega>\Delta$. The retarded Green function $G_\pm^R(\omega)$ can be obtained from its Matsubara counterpart by means of the standard analytic continuation procedure
$$G_\pm^R(\omega)=-G_\pm(i\omega_n)|_{i\omega_n\rightarrow \omega+i0}, \;\;\;\omega_n>0.$$ 
Then the Matsubara frequency summation in $G_\pm(0)$ can be performed with the aid of the contour integration in the complex plane.

Employing our regularization procedure we obtain
\begin{align}
G_\pm(0)&=T\sum\limits_{\omega_n}G_\pm(\omega_n)=\frac{2\pi}{g}T\sum\limits_{\omega_n}\left(
\frac{|\omega_n|}{1\pm{\rm e}^{-|\omega_n L/v|}}+\mu
\right)^{-1}\nonumber\\
&=\frac{1}{4\pi i}\int\limits_\mathcal{C}dz\, G_\pm(-iz)\coth\frac{z}{2T}\nonumber\\
&=\frac{i}{4\pi}\int\limits_{-\infty}^\infty d\omega \Bigl(G^R_\pm(\omega)-G^A_\pm(\omega)\Bigr)\coth\frac{\omega}{2T}\nonumber\\
&=\int\limits_0^\Delta d\omega\, J_\pm(\omega)\coth\frac{\omega}{2T},
\end{align}
where the spectral density functions $J_\pm(\omega)$ read
\begin{align}
&J_\pm(\omega)=-\frac{1}{\pi}{\rm Im}\left[\frac{2\pi}{g}
\left(
\frac{i\omega}{ \Bigl(1\pm {\rm e}^{i\omega L/v}\Bigr)}-\mu
\right)^{-1}
\right],
\end{align}
and we define $\mu=2\pi m/g$. In the limit $\mu L/v\ll 1$ these expressions reduce to 
\begin{equation}
J_+(\omega) =\frac{4}{g}\frac{\omega}{\omega^2+4\mu^2}, \quad J_-(\omega) =0.
\end{equation}

\subsection{Quantum phase transition and supercurrent}
Let us now evaluate the supercurrent $I$ flowing in the wire segment of length $L$ in-between two Josephson junctions. This current
can be defined as
\begin{align}
\nonumber
I&=-2eT\frac{1}{\mathcal{Z}}\frac{d\mathcal{Z}}{d\phi}=2e \frac{d \mathcal{F}}{d\phi}\\
&=-2eE_J\sin(\psi-\phi/2){\rm e}^{-\bigl(G_+(0)+G_-(0)\bigr)/2}\nonumber\\
&= \frac{gev}{2\pi L}\psi.\label{curr}
\end{align}
Thus, within the framework of our approach the effect of phase fluctuations is accounted for by
effective renormalization of the critical current by the factor ${\rm e}^{-\bigl(G_+(0)+G_-(0)\bigr)/2}$. 

In the zero temperature limit $T\to 0$ the solution of Eq. (\ref{eq_RG}) takes the form
\begin{equation}
\mu=\left\{
\begin{matrix}
\Delta \left(
\frac{4\pi E_J\cos(\psi-\phi/2)}{g\Delta}
\right)^{\frac{g}{g-2}}, & g>2, \\
0, & g< 2,
\end{matrix}
\right. 
\end{equation}
while the renormalized equation of motion (\ref{eq_mot}) can be rewritten as 
\begin{equation}
\frac{\mu L}{v} \tan(\psi-\phi/2)+\psi=0 \label{rewr_eq_mot}.
\end{equation}
We observe that for $g< 2$ one has $\psi=0$ and, hence, the supercurrent $I$ inside the wire 
is completely suppressed by strong quantum fluctuations of the phase. on the other hand, at bigger values of $g>2$ a non-vanishing supercurrent $I$ can flow across the wire segment between the two Josephson junctions. 

We arrive at an important conclusion: A quantum phase transition occurs at $g=2$ separating two different phases with non-superconducting ($g< 2$) and superconducting-like ($g>2$) behavior. This dissipative QPT belongs to the same universality class as the so-called Schmid phase transition in resistively shunted Josephson junctions \cite{Albert,SZ90}. It is curious that this QPT occurs at exactly the same value of the parameter $g$ where the superconducting gap singularity in the local electron density of states gets suppressed due to interaction between electrons and a dissipative bath formed by Mooij-Sch\"on plasmons. 

Let us now focus our attention on the superconducting-like phase $g>2$ and evaluate the supercurrent
$I$ affected by quantum fluctuations of the phase $\varphi$ inside the wire. For this purpose let us combine
the solution of the equation
\begin{align}
\frac{\Delta L}{v} \left(
\frac{4\pi E_J}{g\Delta}
\right)^{\frac{g}{g-2}}\sin(\psi-\phi/2)[\cos(\psi-\phi/2)]^{\frac{2}{g-2}}+\psi =0
\end{align}
with Eq. (\ref{curr}).  We observe that there exists a new length scale $L^*$ in our problem associated with the effective mass. Making use of the Ambegaokar-Baratoff formula for the Josephson coupling energy $E_J=g_N\Delta/8$ (where $g_N$ the dimensionless normal state conductance of each tunnel junction) we can express $L^*$ in the form 
\begin{equation}
L^*= \frac{v}{\Delta} \left(
\frac{2g}{\pi g_N}
\right)^{g/(g-2)}
\label{L*}
\end{equation} 
Here we are merely interested in the case of small-size tunnel junctions with few conducting channels serving as probes aiming to disturb the superconducting wire as little as possible. Accordingly, one typically has $g_N \ll g$ and $L^*$ diverges at $g \to 2$ remaining much longer than the characteristic length scale $v/\Delta$ at any value of $g > 2$. 

The length scale  (\ref{L*}) separates two different fluctuation regimes. For $L\gg L^*$ the wire kinetic inductance contribution remains small as compared to that of the Josephson junctions. Then the phase difference across the wire segment in-between the junctions does not fluctuate being equal to $\varphi(L)-\varphi(0)=\phi$. In this case we reproduce the standard mean field current-phase relation 
\begin{equation}
I(\phi)=\frac{gev}{4\pi L}\phi.
\label{mf}
\end{equation}
In the opposite limit  $L\ll L^*$ the renormalization of $E_J$ already becomes important and phase fluctuations tend to suppress the supercurrent flowing inside the wire. In this limit we arrive at the $L$-independent result
\begin{equation}
I(\phi)=\frac{gev}{2\pi L^*}\sin\left(\frac{\phi}{2}\right)\left[\cos\left(\frac{\phi}{2}\right)\right]^{\frac{2}{g-2}}.
\label{I(T=0)}
\end{equation}

Comparing the results (\ref{mf}) and (\ref{I(T=0)}) we observe that quantum fluctuations can strongly
affect both the supercurrent magnitude and the current-phase relation. The dependence $I(\phi)$ (\ref{I(T=0)}) in the presence of fluctuations becomes smoother than in Eq. (\ref{mf}) and the absolute value of the supercurrent is reduced 
by the factor $\sim L/L^*$. Additional reduction of $I$ originates from the term in the square brackets 
in Eq. (\ref{I(T=0)}):  The supercurrent gets suppressed stronger for bigger values of $\phi$ The latter effect becomes particularly significant for $g$ sufficiently close to 2. For $\phi\rightarrow \pi$ and any $g>2$ the supercurrent tends to zero as $I(\phi) \propto (\pi -\phi)^\frac{g}{g-2}$. 

\begin{figure}
\includegraphics[width=0.45\linewidth]{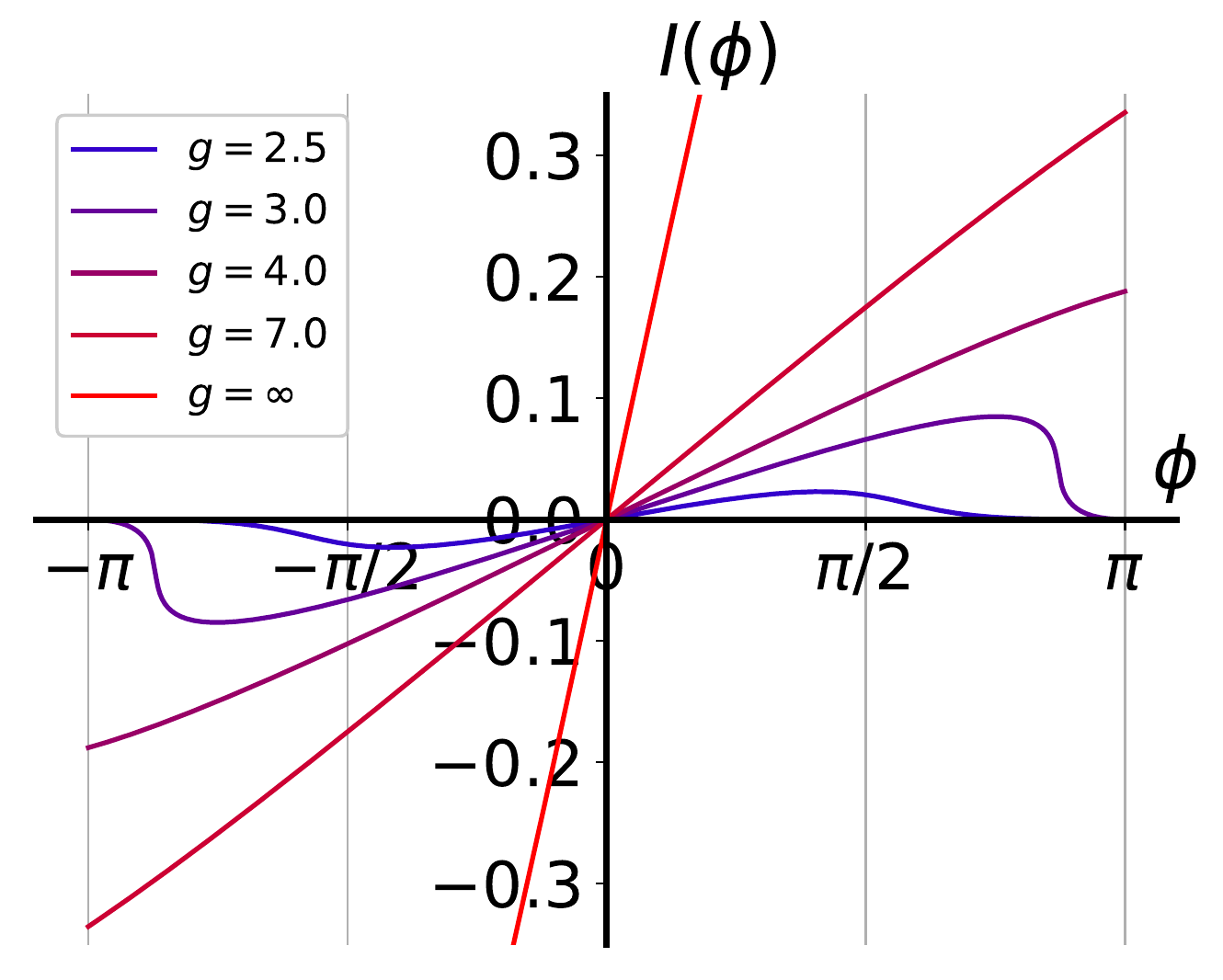}
\includegraphics[width=0.45\linewidth]{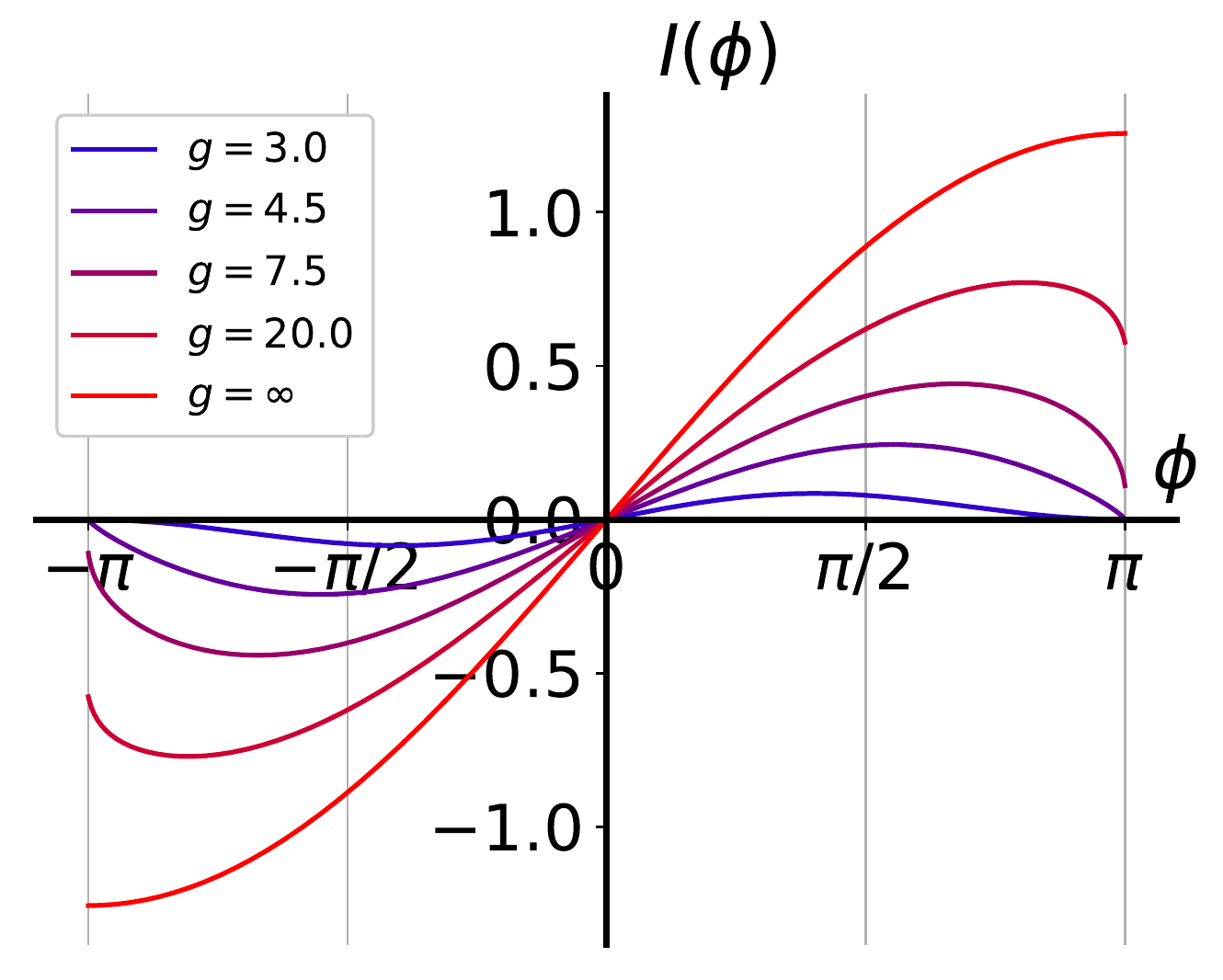}
\caption{The phase dependent supercurrent $I(\phi)$ (expressed in units of $e\Delta/(2\pi)$)  for $E_J/\Delta=0.1$ and different $g$. The upper and lower panels correspond respectively to $\Delta L/v=20$ and to $\Delta L/v=0.5$.}
\label{Fig:14}
\end{figure}

We also point out that for $L$ not much smaller than $L^*$ the supercurrent  $I( \phi \to \pi)$ behaves somewhat differently: It vanishes only for $2<g<4$, whereas at $g>4$ one has 
\begin{equation}
I( \phi \to \pi)\approx \frac{ge\Delta}{2\pi }\left(\frac{\Delta L}{v}\right)^\frac{2}{g-4}\left(\frac{\pi g_N}{2g}\right)^\frac{g}{g-4},
\end{equation}
i.e. for such values of $g$ the current-phase relation remains discontinuous at $\phi=\pi$. The dependencies $I(\phi)$ evaluated for different values of $g$ and $L$ are also displayed in Fig. \ref{Fig:14}. 

Finally, we note that the form of the current-phase relation (\ref{I(T=0)}) derived here resembles that obtained for resistively shunted Josephson junctions in the presence of quantum fluctuations of the phase \cite{PZ}.

\subsection{Effect of QPS and localization of Cooper pairs}

Let us recall that the effective action $S_{TL}$ (\ref{wire_action}) employed in the above analysis only accounts for the effect of Gaussian fluctuations of the superconducting phase and does not yet include quantum phase slips. In order to describe QPS effects inside the wire it is convenient to turn to the dual $\chi$-representation for the wire effective action $S_{\rm eff}[J(\tau)=0]$ (\ref{efacchi}) derived in Sec. V.
As we already discussed, this effective action defines the sine-Gordon model which has a QPT at $T \to 0$ and $\lambda=2$ (or $g=16$)
separating two different phases \cite{ZGOZ}. Provided $g>16$ "positive" and "negative" quantum phase slips are bound in close "neutral"  pairs which do not disrupt phase coherence at any relevant scales exceeding the superconducting coherence length $\xi$. For such values of $g$ QPS effects are irrelevant for any of the above results for the supercurrent $I(\phi)$ which remain applicable without any modifications. 

\begin{figure}
\includegraphics[width=0.7\linewidth]{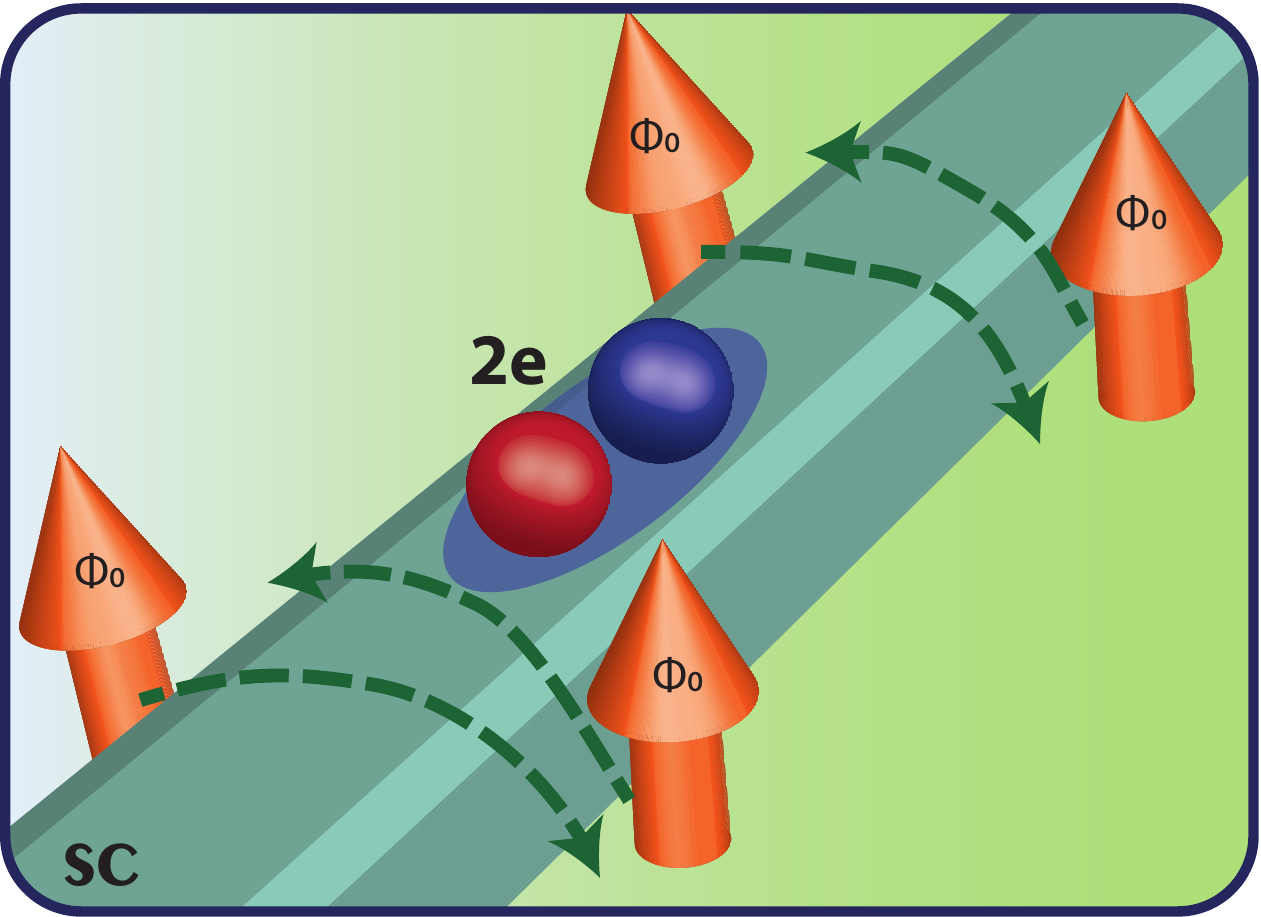}
\caption{Localization of Cooper pairs in the ground state of a uniform superconducting nanowire due to strong zero-point fluctuations of magnetic flux.}
\label{Fig:15}
\end{figure}

On the other hand,  for $g<16$ quantum phase slips are no longer bound in pairs. In this phase relevant excitations of our sine-Gordon theory are kinks and anti-kinks (as well as their bound states) with an effective gap in the spectrum \cite{LZ,GNT}
\begin{equation}
\tilde \Delta \propto \gamma_{QPS}^{\frac{1}{2-\lambda}}.
\end{equation}
The appearance of this gap for $g<16$ (or $\lambda <2$) gives rise to the correlation length $L_c \propto 1/\tilde \Delta$ which, similarly to the case of superconducting nanorings, can be evaluated by equating the renormalized QPS amplitude at this length scale $L_c\tilde \gamma_{QPS}(L_c)$ (see Eq. (\ref{ren})) to the inductive energy $\Phi_0^2/(2{\mathcal L}_{\rm kin}L_c)\sim \xi \Delta g_\xi /L_c$ with some numerical prefactor. Keeping track of all relevant prefactors, in the most relevant limit $\xi \ll v/\Delta$ we obtain
\begin{equation}
L_c \approx \left(\frac{\pi}{8\sqrt{2b}}\right)^{\frac{1}{1-\lambda/2}} \xi \exp\left(\frac{ag_\xi}{2-\lambda}\right)\left(\frac{\xi \Delta}{v}\right)^{\frac{\lambda}{2-\lambda}},\label{LQPS}
\end{equation}
cf. also Eq. (\ref{rc}). The appearance of this fundamental length scale in our problem is directly related to the phase-charge (or flux-charge) duality. It can be interpreted as a result of spontaneous tunneling of magnetic fluxons $\Phi_0$ back and forth across the wire, as it is illustrated in Fig. \ref{Fig:15}. These strong quantum fluctuations of magnetic flux wipe out phase coherence at distances $\gtrsim L_c$ and yield effective {\it localization of Cooper pairs} at such length scales. Accordingly, one may interpret the energy 
\begin{equation}
\tilde E\sim \xi \Delta g_\xi /L_c \propto \tilde \Delta
\end{equation}
as an effective Coulomb gap for the wire segment of length $\sim L_c$. Viewing our nanowire as a chain of $N\approx L/L_c$ independent segments one may conclude that its total Coulomb gap in the insulating regime could be as high as $\sim \xi \Delta g_\xi L/L_c^2$.

In the context of a setup considered here the correlation length  \eqref{LQPS} is essentially irrelevant for $g<2$ since in this case the supercurrent $I$ is totally suppressed already by smooth phase fluctuations. 

On the other hand, for $2<g<16$ the length scale (\ref{LQPS}) becomes important. At such values of $g$ there exist two correlation lengths, $L^*$ and $L_c$ defined respectively in Eqs. (\ref{L*}) and (\ref{LQPS}). The first of these two lengths diverges at one of the phase boundaries $g=2$ whereas the second one tends to infinity at another phase boundary $g=16$.

Consider the situation with $L^*<L_{c}$, in which case there exist three regimes. At $L<L^*$ the supercurrent is strongly affected only by smooth phase fluctuations and not by quantum phase slips. This regime is accounted for by Eq. (\ref{I(T=0)}). At $L^*<L<L_c$ the supercurrent is insensitive to any kind of phase fluctuations and, hence, it is given by a simple mean field formula (\ref{mf}). Finally, for $L > L_c$ the supercurrent gets exponentially suppressed by quantum phase slips and similarly to Eq. (\ref{lsc2}) we have 
\begin{eqnarray}
\nonumber
I(\phi) \sim \frac{eg_\xi\Delta \sqrt{L}}{\sqrt{\xi}}\left(\frac{v}{L \Delta}\right)^{\frac{3\lambda}{4}} 
\exp\left(-\frac{3ag_\xi}{4}-\left(\frac{L}{L_c}\right)^{1-\lambda/2}\right)\sin(\phi ).
\label{expsup}
\end{eqnarray}
In practical terms, the latter regime can be regarded as non-superconducting provided $L$ strongly exceeds $L_c$.
 
It is also possible to realize the opposite situation with $L^*>L_{c}$, in particular for values of $g$ close to 2. In this case the length $L^*$ becomes irrelevant, and one distinguishes only two regimes: $L<L_c$ and $L>L_c$. The first
one is again superconducting with the supercurrent $I(\phi)$ affected by smooth phase fluctuations according to Eq. (\ref{I(T=0)}), whereas the second regime corresponds to exponential suppression of the supercurrent due to proliferating QPS, cf.
Eq. (\ref{expsup}). No room for the mean field regime (\ref{mf}) exists at $L^*>L_{c}$.

\subsection{An alternative setup}
\begin{figure}
\includegraphics[width=0.6\linewidth]{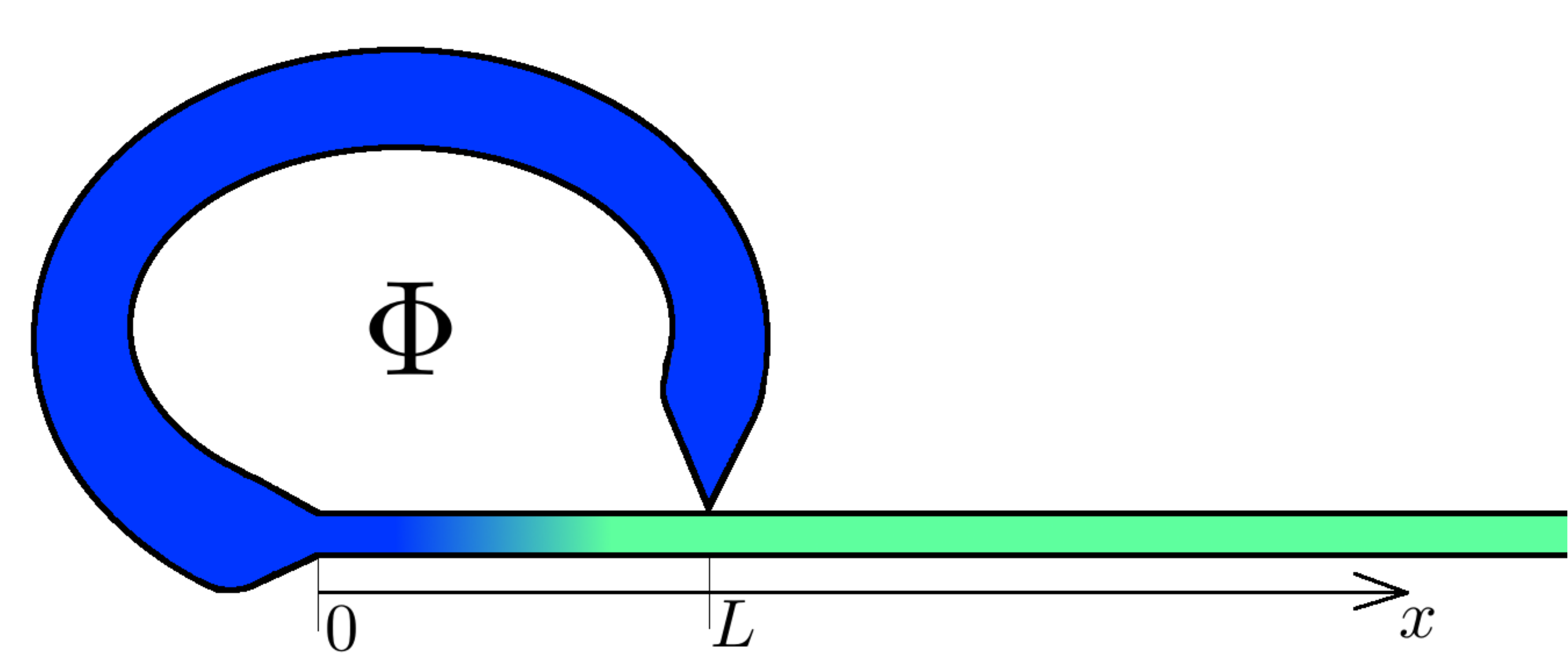}
\caption{A superconducting nanowire directly attached to a bulk open superconducting ring at its left end. The ring is also attached to the wire via a small-area tunnel junction at a distance $L$ from its left end. The open ring is pierced by an external magnetic flux $\Phi$.}
\label{Fig:16}
\end{figure}

As we already emphasized, the setup displayed in Fig. \ref{Fig:13} enables one  to pass an equilibrium supercurrent across a wire segment of an arbitrary length $L$ without restricting phase fluctuations inside the wire by any means. Below we will consider an alternative setup that allows to effectively restrict the space available for phase fluctuations by ''pinning'' the superconducting phase $\varphi$ at one point inside the wire \cite{RSZ20}. This setup is schematically shown in Fig. \ref{Fig:16}. As before, it includes a long and thin superconducting nanowire with one of its ends attached to a bulk superconducting reservoir. This reservoir has a form of an open ring whose opposite end is attached to the wire via a small-area tunnel junction at a distance $L$ along the wire. The open ring is pierced by an external magnetic flux $\Phi$ which controls the phase difference $\phi=2\pi \Phi/\Phi_0$ between its ends. Accordingly, the phase at the left end of the wire is pinned by the reservoir and is set equal to zero, i.e. $\varphi(x=0)=0$.

We will demonstrate that such topology controlled phase pinning severely enhances the ability of the wire to conduct supercurrent. This effect can be interpreted in terms of the absence of a massless mode responsible for the destruction of superconductivity at $g<2$ in the setup of Fig.  \ref{Fig:16}. Instead, the nanowire embedded in our present setup exhibits a transition between ''more'' and ''less'' superconducting phases characterized by different types of long-range behavior. Our observable of interest is again the supercurrent $I(\phi)$ flowing through the wire segment of length $L$ between the left wire end and the junction. As before, the phase $\phi$ is restricted to the interval $\phi\in (-\pi,\pi)$.

In order to proceed we will again describe our system with the aid of the effective action (\ref{system_action}), where
the wire contribution $S_{TL}$ (\ref{wire_action}) remains the same, whereas the Josephson term $S_J$ now has a somewhat simpler form
\begin{equation}
S_{J}[\varphi(L)]=-E_J\int\limits_0^{1/T} d\tau\cos\bigl(\varphi(L)-\phi\bigr)
\end{equation}
as compared to Eq. (\ref{int}).

Our further analysis will be fully analogous to that already carried out above for the setup of Fig. \ref{Fig:13}. Integrating out the phase variable $\varphi(x)$ at all points along the wire except for its value at $x=L$ we arrive at the reduced effective action
\begin{equation}
S_R+S_J=\frac{1}{2}\tr\left[\varphi\, G_0^{-1}\,\varphi\right]-E_J\int\limits_0^{1/T} d\tau \cos(\varphi-\phi)
\end{equation}
with
\begin{equation}
G_0(\omega_n)=\frac{8\pi}{g\omega_n}\tanh\left(\frac{\omega_n L}{v}\right)
\end{equation}
We observe that fluctuations of the phase variable are massive with $m_0=gv/8\pi L$. The absence of a massless mode in our setup (in contrast to that of Fig. \ref{Fig:13}) is a direct consequence of phase pinning at $x=0$ which prohibits uniform shifts of the phase inside the wire.

Again employing the SCHA-type of analysis we define the trial action
\begin{equation}
S_{\rm tr}=
\frac{1}{2}\tr\left[(\varphi-\psi)(G_{0}^{-1}+m)(\varphi-\psi)\right], \label{ansatz2}
\end{equation}
where the variational parameter $m$ accounts for the interaction-induced effective mass for the $\varphi$-mode and $\psi$ determines the average value of the phase difference.  Evaluating the free energy of the system as a function of these two parameters and minimizing it with respect to both $m$ and $\psi$, we arrive at the following SCHA equations 
\begin{align}
&E_J\cos(\psi-\phi){\rm e}^{-G(0)/2}-m=0,\label{eq_RG2}\\
&E_J\sin(\psi-\phi){\rm e}^{-G(0)/2}+\frac{gv}{8\pi L}\psi=0,\label{eq_mot2}
\end{align}
where 
\begin{equation}
G(0)=T\sum\limits_{\omega_n}\left(G_{0}^{-1}(\omega_n)+m\right)^{-1}.\label{GF}
\end{equation} 

As before, the effect of phase fluctuations reduces to effective renormalization of $E_J$ by the factor ${\rm e}^{-G(0)/2}$. The supercurrent $I$ is then found from the equation
\begin{equation}
I=\frac{gev}{4\pi L}\psi .
\label{I}
\end{equation} 

At $L\lesssim v/\Delta$ phase fluctuations are strongly suppressed and the system remains in the mean-field regime.  In the opposite limit of large $L$ the solution of Eq. (\ref{eq_RG2}) exhibits two qualitatively distinct regimes. At $g<4$ we find $|m|\ll v/L$. Therefore, the emergent mass is negligible, and the effect of fluctuations is purely Gaussian. The equation of motion (\ref{eq_mot2}) is then rewritten as
\begin{equation}
E_J\sin(\psi-\phi)\left(\frac{\Delta L}{v}\right)^{-4/g}+\frac{gv}{8\pi L}\psi=0.\label{gaussian_eq_mot}
\end{equation}
In the interesting for us limit of small $E_J$ we may readily set $\frac{8\pi E_J}{g\Delta}< 1$. In this case the sine term is renormalized to zero faster than the kinetic inductance contribution $\propto L^{-1}$ and, hence, we obtain
\begin{equation}
I(\phi)= 2eE_J \left(\frac{v}{\Delta L}\right)^{4/g} \sin\phi.\label{I_small_g}
\end{equation}
This expression demonstrates that for $g<4$ phase fluctuations (i) modify the current-phase relation making it sine-like instead of the sawtooth-like and (ii) yield a decrease of the supercurrent 
as compared to the standard Josephson formula $I(\phi)=2eE_J\sin\phi$ that applies in the limit $L\rightarrow 0$. In addition, we observe that in the presence of fluctuations the supercurrent (\ref{I_small_g}) decays faster with increasing $L$ than the standard mean field dependence $I \propto 1/L$.

Let us now turn to the case $g>4$. Resolving Eq. (\ref{eq_RG2}) in the limit $L\rightarrow\infty$ we obtain 
\begin{equation}
m=\left\{
\begin{matrix}
\left[E_J\cos(\psi-\phi)\left(\frac{8\pi}{\Delta g}\right)^\frac{4}{g}\right]^\frac{g}{g-4}, & \cos(\psi-\phi)>0, \\
-gv/8\pi L+o(1/L), & \cos(\psi-\phi)<0.
\end{matrix}
\right.\label{m}
\end{equation}
This solution remains valid only as long as $L$ exceeds the length scale  
\begin{equation}
L^*=\frac{v}{\Delta}\left(
\frac{g}{\pi g_N}
\right)^\frac{g}{g-4},\label{L**}
\end{equation}
cf. Eq. (\ref{L*}).

This length separates the regime $L>L^*$ where fluctuations yield non-Gaussian renormalization of the interaction potential from the Gaussian regime $L \ll L^*$ where $|m|\ll gv/8\pi L$. As long as $v/\Delta\ll L\ll L^*$ the current is again given by Eq. (\ref{I_small_g}). 

For $g>4$ the renormalized Josephson coupling energy decreases slower than $1/L$ and at $L\sim L^*$ it becomes of the same order as the kinetic inductance contribution. At even larger distances mass renormalization saturates to the value defined in Eq. (\ref{m}). The kinetic inductance contribution, on the contrary, decreases as $1/L$. Therefore, at $L\gg L^*$ the phase is pinned to the lowest minimum of the renormalized Josephson junction potential, i.e. we have $\psi =\phi$. In this case the current-phase relation reduces to the standard mean field form (\ref{mf}).

Finally, we note that the effects of QPS inside the superconducting nanowire are analyzed in exactly the same manner as it was already done
above in Sec. IXD, except one should now distinguish two phases -- $g < 4$ and $4<g<16$ (instead of $g < 2$ and $2<g<16$) --
and replace the expression for $L^*$ (\ref{L*}) by Eq. (\ref{L**}).

For $g<4$ there exists only one correlation length (\ref{LQPS}) in our problem. At $L\ll L_c$ QPS effects are irrelevant and the 
supercurrent suppression is merely due to smooth phase fluctuations. In this limit Eq. (\ref{I_small_g}) applies and the supercurrent decays as a power-law with increasing $L$. As soon as $L$ exceeds $L_c$ quantum phase slips come into play and the supercurrent decay becomes exponential with $L$ according to Eq. (\ref{expsup}). 

At $4<g<16$ there already exist two different correlation lengths,  $L^*$ and $L_c$. The first one diverges as $g\rightarrow 4$ while the second one tends to infinity at $g\rightarrow 16$. Depending on the relation between these two lengths, different regimes can occur. 

Consider, e.g., the limit $L^* \ll L_c$ which can always be realized for sufficiently large values of $g_\xi$. In this case, at shorter length scales $L<L^*$ only smooth phase fluctuations affect the supercurrent causing its power-law suppression with increasing $L$ and the sinusoidal current-phase relation, see Eq. (\ref{I_small_g}). At $L^* < L <L_c$ both smooth phase fluctuations and quantum phase slips are irrelevant and the supercurrent is defined by the standard mean field result $I \propto 1/L$ (\ref{mf}) describing the sawtooth-shaped current-phase relation. Finally, at $L \gg L_c$ the current is exponentially suppressed and the current-phase relation again reduces to the sine form (\ref{expsup}).

\subsection{Discussion}
The above analysis demonstrates that superconducting properties of metallic nanowires depend not only on their parameters, but also on the topology of the experimental setup and on the way the experiment is being performed. The ability of the wire to carry supercurrent also varies at different length scales being affected by different kinds of fluctuations including, on one hand, sound-like collective plasma modes forming a quantum dissipative environment for electrons inside the wire and, on the other hand, quantum phase slips. As the bath of plasma modes is (almost) Ohmic the low temperature system behavior resembles that involving a Schmid-like dissipative QPT \cite{Albert,SZ90} either at $g=2$ or at $g=4$ depending on the setup under consideration. The presence of quantum phase slips naturally leads to a BKT-type quantum phase transition at $g=16$ \cite{ZGOZ}. 

While the system displayed in Fig. \ref{Fig:13} allows for unrestricted fluctuations of the superconducting phase, the setup of Fig. \ref{Fig:16} effectively pins the phase at one of the wire ends. In the former case a gapless Ohmic mode  (associated with uniform phase shifts along the wire) appears at any $L$. Fluctuations associated with this gapless mode cause a Schmid-like QPT at $g=2$. As a result, the wire completely loses superconductivity at $g<2$, whereas the phase with $2<g<16$ is mixed, i.e. it is non-superconducting in the long length limit and superconducting at shorter scales, even though the gapless mode causes additional suppression of current in the limit $L\rightarrow 0$. Comparing this situation with the one encountered for the structure of Fig. \ref{Fig:16}, we observe that in the latter case superconductivity is severely enhanced as a result of the soft mode suppression due to phase pinning. This effect turns the QPT into a transition between ''less'' and ''more'' superconducting phases at $g=4$ . Note that a similar phase transition was also discussed in Ref. \cite{HG} in the context of superconducting nanorings interrupted by a Josephson junction.

\section{Quantum phase slips in capacitively coupled superconducting nanowires}

Let us now extend our analysis of QPS-related effects to yet another structure which consists of capacitively coupled superconducting nanowires. We will demonstrate that quantum fluctuations in one of the two wires effectively "add up" to those of another one even without any direct electric contact between them, thereby giving rise to a number of interesting effects, such as, e.g., splitting of plasmon modes and interaction-induced SIT shifting in each of the wires.

\subsection{The model}
\begin{figure}
\includegraphics[width=0.99\linewidth]{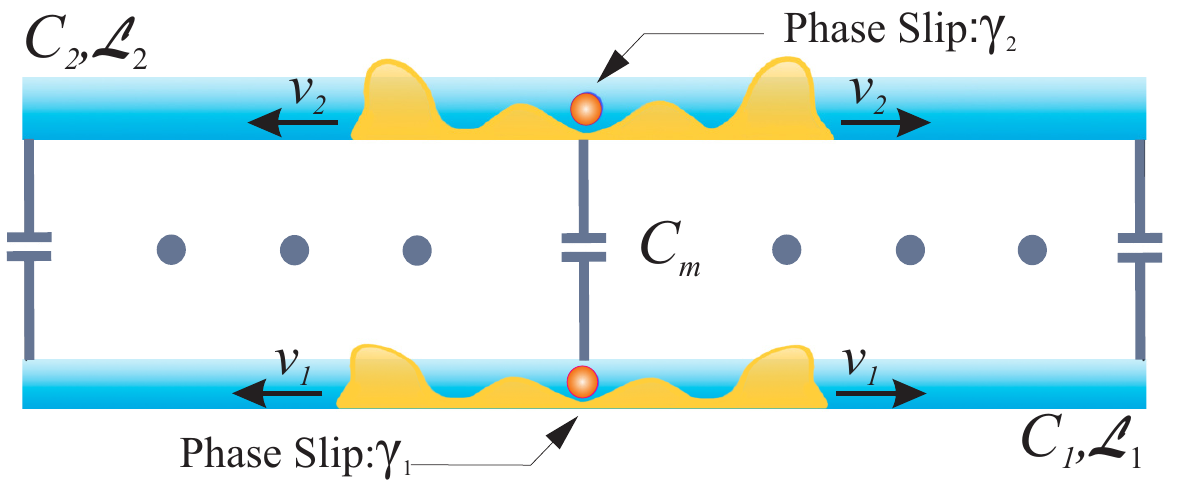}
\caption{Two capacitively coupled superconducting nanowires.}
\label{Fig:17}
\end{figure}

Consider the system of two long parallel to each other superconducting nanowires, as it is schematically shown in Fig. \ref{Fig:17}. 
As before, the wires are described by geometric capacitances $C_1$ and $C_2$ (per unit wire length) and kinetic inductances $\mathcal{L}_{1}$ and $\mathcal{L}_{2}$ (times length) effectively representing the two transmission lines. Capacitive coupling between these two nanowires is accounted for by the mutual capacitance $C_m$. Generalizing our analysis in Sec. IVA to the above structure, we arrive at the contribution to the system Hamiltonian that keeps track of both electric and magnetic energies in these coupled transmission lines. It reads  \cite{LSZ19,LSZ20}
\begin{eqnarray}
\hat{H}_{TL}=\frac{1}{2}\sum_{i,j=1,2}\int dx (\mathcal{L}^{-1}_{ij}\hat{\Phi}_{i}(x)\hat{\Phi}_{j}(x)
+(1/\Phi^{2}_{0})C^{-1}_{ij}(\partial_x \hat{\chi}_{i}(x)\partial_x \hat{\chi}_{j}(x)),
\label{eq4}
\end{eqnarray}
where $x$ is the coordinate along the wires, $\mathcal{L}_{ij}$ and $C_{ij}$ denote the matrix elements of the inductance and capacitance matrices
\begin{equation}\label{eq6}
\check{\mathcal{L}}=\left(\begin{array}{crl}
\mathcal{L}_{1} & 0\\
0 & \mathcal{L}_{2}\\\end{array}\right),   \quad \check{C}=\left(\begin{array}{crl}
C_{1} & C_{m}\\
C_{m} & C_{2}\\\end{array}\right)
\end{equation}
As before, the Hamiltonian (\ref{eq4}) is expressed in terms of the dual operators $\hat{\chi}(x)$ and $\hat{\Phi}(x)$ which 
obey the canonical commutation relation 
\begin{equation}\label{eq2}
    [\hat{\Phi}_i(x),\hat{\chi}_j(x^{\prime})]=-i\delta_{ij}\Phi_{0}\delta(x-x^{\prime}).
\end{equation} 

Provided the wires are thick enough the low energy Hamiltonian in Eq. (\ref{eq4}) is sufficient. However, for thinner wires one should also account for the effect of quantum phase slips. The corresponding contribution to the total Hamiltonian for our system can be expressed in the form (cf., Eq. (\ref{HQPS0}))
\begin{equation}\label{eq5}
\hat{H}_{QPS}=-\sum_{j=1,2}\gamma_{j} \int dx\cos(\hat{\chi}_{j}(x)),
\end{equation}
where 
\begin{equation}
\gamma_j \sim (g_{j \xi}\Delta/\xi)\exp (-ag_{j \xi}), \;\;\;a \sim 1, \;\;\;  j=1,2
\label{gaQPS}
\end{equation}
are the QPS amplitudes per unit wire length, $\Delta$ is the superconducting order parameter in each of the wires, $g_{j \xi} =R_q/R_{j \xi}$ and $R_{j \xi}$ is the normal state resistance of the $j$-th wire segment of length equal to the superconducting coherence length $\xi$.

The total Hamiltonian for the system under consideration equals to the sum of the two terms in Eqs. (\ref{eq4}) and (\ref{eq5}),
\begin{equation}\label{eq3}
\hat{H}=\hat{H}_{TL}+\hat{H}_{QPS},
\end{equation} 
representing an effective sine-Gordon model that will be treated below.

\subsection{Plasma mode splitting}

\begin{figure}
\includegraphics[width=0.32\linewidth]{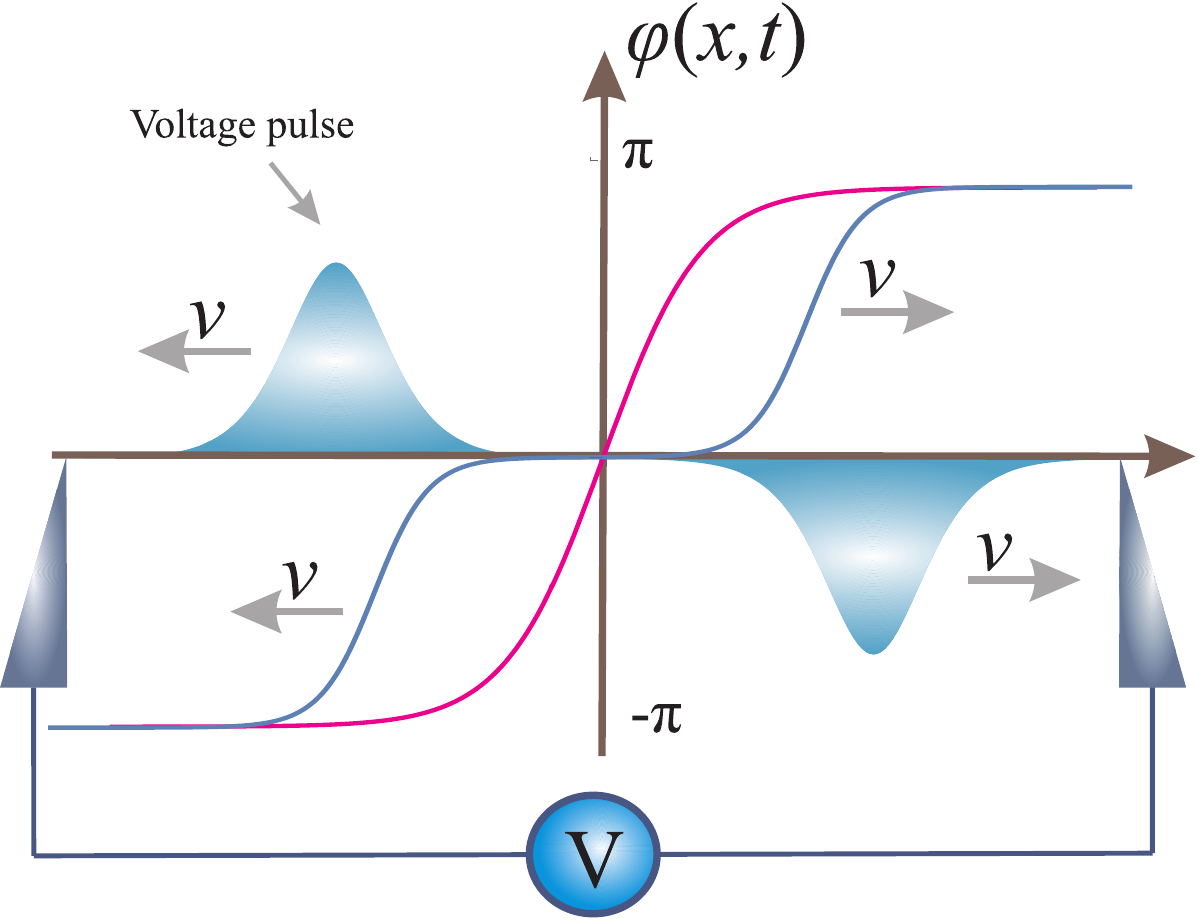}
\includegraphics[width=0.32\linewidth]{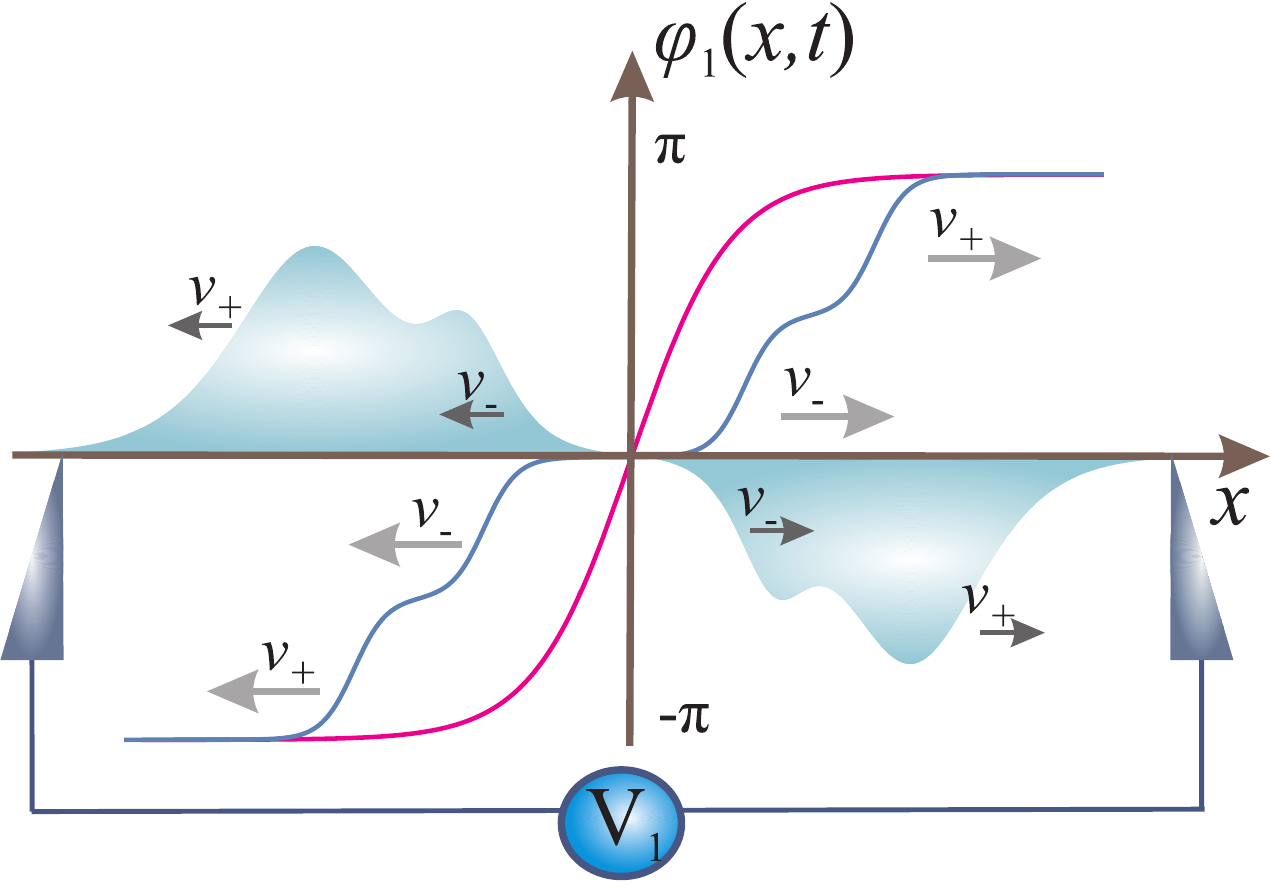}
\includegraphics[width=0.32\linewidth]{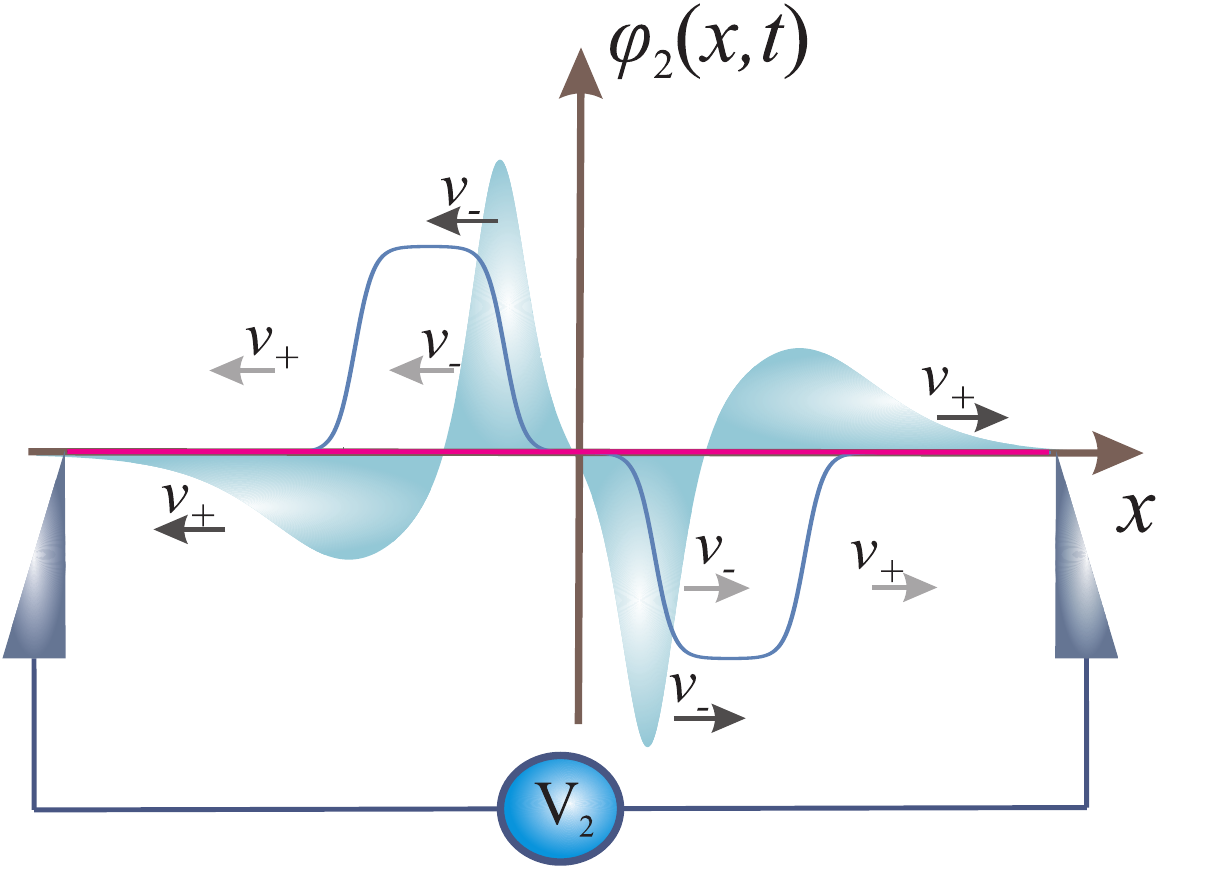}
\caption{Time-dependent phase configurations describing a QPS event at $t=0$ (red) and $t>0$ (blue) together with propagating voltage pulses generated by this QPS event in a single superconducting nanowire (left panel) and in the first of the two capacitively coupled superconducting nanowires (middle panel).In the case of the two wires each of the voltage pulses is split into two propagating with different velocities $v_{\pm}$. Right panel: ime-dependent phase configurations at $t=0$ (red) and $t>0$ (blue) together with propagating voltage pulses in the second of the two capacitively coupled superconducting nanowires generated by a QPS event in the first one.}
\label{Fig:18}
\end{figure}

Any QPS event causes redistribution of charges inside the wire and generates a pair of voltage pulses propagating simultaneously in the opposite directions along the wire with the velocity $v$, see also Fig. \ref{Fig:18} (left panel). In a single wire this process is controlled, e.g., by the wave equation (\ref{we}) for the operator  $\hat{\chi}$. In the case of two capacitively coupled wires a QPS event in one of the wires generates voltage pulses in {\it both wires}. The corresponding generalization of the wave equation (\ref{we}) for the operators $\hat{\chi}_{1,2}$ in each of the two wires follows directly from the system Hamiltonian (\ref{eq4}), and the corresponding voltage operators $\hat{V}_{1,2}$ read
\begin{equation}\label{eq7}
\hat{V}_{i}(t)=1/\Phi_{0}\sum_{j=1,2}C^{-1}_{ij}(\partial_x\hat{ \chi}_{j}(x_{1},t)-\partial_x \hat{\chi}_{j}(x_{2},t)).
\end{equation}

It will be convenient for us to go over to the phase representation and express the corresponding equation of motion for the phases $\varphi_{1,2}$ in both wires in the form (cf., e.g., Eq.(\ref{speq}))
\begin{equation}
\left(\check{1}\partial_t^2-\check{\mathcal{V}}^2\partial_x^2\right)\left(\begin{array}{crl}
\varphi_1(x,t) \\
\varphi_2(x,t)\\\end{array}\right)=0,
\label{speq2}
\end{equation}
where $\check{\mathcal{V}}= (\check{C}\check{\mathcal L})^{-1/2}$ is the velocity matrix that accounts for plasmon modes propagating along the wires. Assume now that a QPS event occurs at the initial time moment $t=0$ at the point $x=0$ inside the first wire. As we just discussed, at $t>0$ voltage pulses originating from this QPS event will propagate in both wires. Resolving Eq. (\ref{speq2}) together with the proper initial condition corresponding to a QPS event (cf., e.g., Eq. (\ref{wind})), we arrive at the following picture. 

In the first wire each of the two voltage pulses propagating in opposite directions turns out to be split into two pulses of the same sign propagating with different velocities $v_{+}$ and $v_{-}$, as it is also illustrated in Fig. \ref{Fig:18} (middle panel). Each of the voltage pulses generated in the second wire by a QPS event in the first one is also split into two, they also propagate with two different velocities $v_{+}$ and $v_{-}$, however the signs of these voltage pulses are now opposite to each other, cf. Fig. \ref{Fig:18} (right panel). The velocities of these split plasmon modes are determined by the eigenvalues of the velocity matrix $\check{\mathcal{V}}$. They read \cite{LSZdrag}
\begin{equation}
v_{\pm}=\frac{1}{2\kappa}\left[\sqrt{v_1^2+v_2^2+2v_1v_2\kappa}\pm \frac{\sqrt{(v_1^2-v_2^2)^2+\frac{4C_m^2v_1^2v_2^2}{C_1C_2}}}{\sqrt{v_1^2+v_2^2+2v_1v_2\kappa}}\right],
\label{vpm}
\end{equation}
where $\kappa=\sqrt{1-C_m^2/(C_1C_2)}$ and $v_i=1/\sqrt{C_i\mathcal{L}_i}$ ($i=1,2$) is the velocity of the Mooij-Sch\"on modes in the $i$-th wire for $C_m \to 0$. In the case of identical wires with $C_1=C_2=C$ and ${\mathcal L}_1={\mathcal L}_2={\mathcal L}$ the result (\ref{vpm})
reduces to a particularly simple form
\begin{equation}
v_{\pm}=1/\sqrt{{\mathcal L}(C \mp C_m)}.
\end{equation}
This expression indicates that one of these velocities may strongly increase provided the wires are located close enough to each other in which case the cross-capacitance $C_m$ may become of order $C$. In the absence of inter-wire interaction we have $\kappa=1$ and Eq. (\ref{vpm})
obviously yields $v_+=v_1$ and $v_-=v_2$.

\subsection{Quantum phase transitions: renormalization group analysis}

Let us now turn to the issue of a superconductor-insulator QPT in capacitively coupled superconducting nanowires
with proliferating quantum phase slips. As we just demonstrated, the presence of capacitive coupling between the nanowires may significantly affect plasmon propagation in our system and, hence, alter the interaction between quantum phase slips. In this case, SIT in each of the wires is controlled not only by the parameters of the corresponding wire but also by those of the neighboring one as well as by the mutual capacitance. Interestingly enough, superconducting nanowires with properly chosen parameters may turn insulating once they are brought sufficiently close to each other \cite{LSZ20}.  

In order to quantitatively describe QPT in coupled superconducting nanowires we will employ the renormalization group (RG) analysis.
This approach is well developed and was successfully applied to a variety of problems in condensed matter theory, such as, e.g., 
the problem of weak Coulomb blockade in tunnel \cite{SZ90,GS,PZ91,Zwerger} and non-tunnel \cite{KN,GZ04,BN} barriers between normal metals or that of a dissipative phase transition in resistively shunted Josephson junctions \cite{SZ90,Albert,Blg84,GHM}. In the case of superconducting nanowires QPT was described \cite{ZGOZ} with the aid of RG equations equivalent to those initially developed for two-dimensional superconducting films \cite{BKT} which exhibit classical BKT phase transition driven by temperature. In contrast, as we already discussed above, quantum SIT in quasi-one dimensional superconducting wires is controlled by the parameter  $\lambda$ proportional to the square root of the wire cross section $s$.

Provided the two superconducting wires  depicted in Fig. \ref{Fig:17} are decoupled from each other, i.e. for $C_m \to 0$, one should expect two independent QPT to occur in these two wires respectively at $\lambda_1=2$ and at $\lambda_2=2$, where $\lambda_{1,2} =(R_q/8)\sqrt{C_{1,2}/{\mathcal L}_{1,2}}$.  In the presence of capacitive coupling between the wires these two QPT get modified. In order to account for these modifications let us express the grand partition function of our system ${\mathcal Z}={\rm Tr}\exp (-\hat{H}/T)$ in terms of the path 
integral
\begin{equation}
{\mathcal Z}=\int D\chi_1\int D\chi_2\exp (-S[\chi_1,\chi_2]),
\label{Z}
\end{equation}
where 
\begin{eqnarray}
    S=\frac{1}{2\Phi^{2}_{0}}\sum_{i,j=1,2}\int dxd\tau\Big(\xi\Delta\mathcal{L}_{ij}\partial_{\tau}\chi_{i}\partial_{\tau}\chi_{j}
    +\frac{1}{\xi\Delta}C^{-1}_{ij}\partial_{x}\chi_{i}\partial_{x}\chi_{j}\Big)-\sum_{i=1,2}y_{i}\int dxd\tau \cos(\chi_{i})
\label{eq8}
\end{eqnarray}
is the effective action corresponding to the Hamiltonian (\ref{eq3}) and $y_{i}= \gamma_{i} \xi/\Delta$
denote the effective fugacity for the gas of quantum phase slips in the $i$-th wire. For the sake of convenience in Eq. (\ref{eq8}) we rescaled the spatial coordinate in units of $x_0$, i.e. $x \to x\xi $ and the time coordinate in units of $\tau_0$, i.e. $\tau \to \tau /\Delta$. 

As usually, let us divide the $\chi$-variables into fast and slow components $\chi_{i}=\chi_{i}^{f}+\chi_{i}^{s}$, where
\begin{eqnarray}\nonumber
    \chi^{f}_{i}(x,\tau)=\int_{\Lambda<\omega^{2}+q^{2}<\Lambda+\delta\Lambda}\frac{d\omega dq}{2\pi}\chi_{\omega,q}e^{i\omega \tau+iqx},\\\chi^{s}_{i}(x,\tau)=\int_{\omega^{2}+q^{2}<\Lambda}\frac{d\omega dq}{2\pi}\chi_{\omega,q}e^{i\omega \tau+iqx}.
\nonumber
\end{eqnarray}

Assuming $\delta\Lambda/\Lambda \ll 1$, expanding in $\chi_{i}^{f}$ and integrating these fast variables out we employ the perturbation theory in $y_{1,2}$ and observe that in order to account for the leading order corrections it suffices to evaluate the following matrix Green function at coincident points
\begin{equation}
\check{G}^{f}(0,0)=\Phi^{2}_{0}\int \frac{d\omega dq}{(2\pi)^2}\Big(\xi\Delta\check{\mathcal L}\omega^2+\frac{1}{\xi\Delta}\check{C}^{-1}q^2\Big)^{-1}
=2(\delta \Lambda/\Lambda)\check{\lambda},
\label{G00}
\end{equation}
where $\check{\lambda}=(R_q/8)\check{\mathcal{V}}\check{C}$. The matrix $\check{\lambda}$ reads
\begin{equation}
    \check{
\lambda}=\frac{1}{\sqrt{\frac{1}{v^{2}_{1}}+\frac{1}{v^{2}_{2}}+\frac{2\kappa}{v_{1}v_{2}}}}\left(\begin{array}{crl}
\lambda_1\left(\frac{1}{v_1}+\frac{\kappa}{v_{2}}\right)& R_qC_{m}/8\\
R_qC_{m}/8& \lambda_2\left(\frac{1}{v_2}+\frac{\kappa}{v_{1}}\right)\\\end{array}\right).
\end{equation}
Following the standard analysis \cite{BKT} and proceeding to bigger and bigger scales $\Lambda$, we eventually arrive at the following RG equations for the QPS fugacities $y_1$ and $y_2$: 
\begin{equation}\label{eq11}
    \frac{dy_{i}}{d\log \Lambda}=(2-\lambda_{ii})y_{i},\quad i=1,2,
\end{equation}
where $\lambda_{11}$ and $\lambda_{22}$ are diagonal elements of the matrix $\check{\lambda}$. Note that here we restrict our RG analysis to the lowest  order in $y_{1,2}$, in which case other parameters of our problem remain unrenormalized.  

It follows immediately from Eqs. (\ref{eq11}) that our system exhibits two BKT-like QPT at $\lambda_{11}=2$ and $\lambda_{22}=2$. For the first wire the corresponding phase transition point is fixed by the condition \cite{LSZ20}
\begin{equation}
\lambda_1=2\frac{\sqrt{1+\frac{v^2_1}{v^2_2}+2\kappa\frac{v_1}{v_2}}}{1+\kappa\frac{v_1}{v_2}}.
\label{QPT}
\end{equation}
The same condition for the second wire is derived from Eq. (\ref{QPT}) simply by interchanging the indices $1 \leftrightarrow 2$.

These results allow to conclude that in the presence of capacitive coupling quantum fluctuations in one of these wires tend to decrease superconducting properties of the other one. As a result, SIT in both wires occurs at larger values of $\lambda_{1,2}$ than in the absence of such coupling.  

Equation (\ref{QPT}) demonstrates that the magnitude of this effect depends on the ratio of the plasmon velocities in the two wires $v_1/v_2$ and on the strength of the capacitive coupling controlled by $C_m$. Provided the wire cross sections $s_1$ and $s_2$ differ strongly the plasmon velocities $v_i \propto \sqrt{s_i}$ also differ considerably. Provided, e.g., the first wire is much thinner than the second one we have $v_1 \ll v_2$ and, hence, the QPT condition (\ref{QPT}) in the first wire remains almost unaffected for any capacitive coupling strength. If, on the contrary, the first wire is much thicker than the second one, then one has $v_1 \gg v_2$ and the condition (\ref{QPT}) reduces to $\lambda_1\simeq 2/\sqrt{1-C_m^2/(C_1C_2)}$ implying that the critical value $\lambda_1$ can exceed 2 considerably for sufficiently large values $C_m$. 

\begin{figure}
\includegraphics[width=\linewidth]{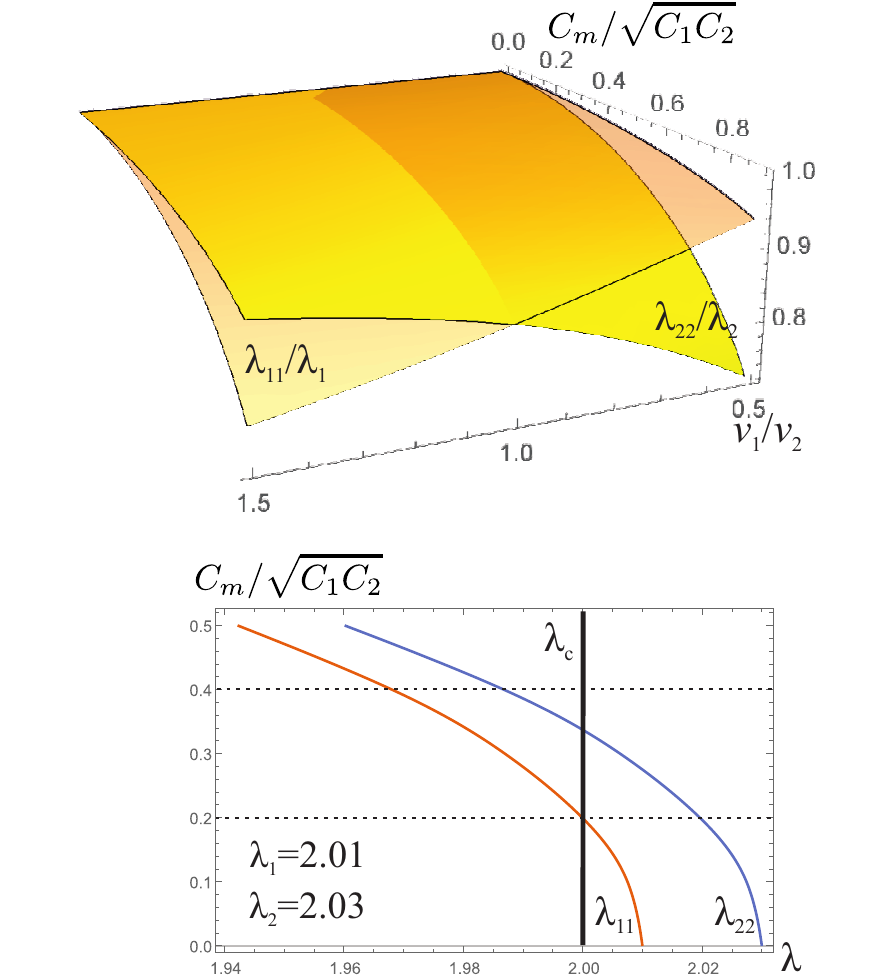}
\caption{Top: Critical surfaces corresponding to SIT at $\lambda_{11}=2$ and $\lambda_{22}$=2.
Bottom: Phase diagram for two capacitively coupled superconducting nanowires with $\lambda_{1}=2.01$ and $\lambda_{2}=2.03$.
Both curves $\lambda_{11}(C_m)$ and $\lambda_{22}(C_m)$ decrease and cross the critical line $\lambda_c=2$ with increasing mutual capacitance $C_m$.}
\label{Fig:19}
\end{figure}

It is fairly obvious that capacitive coupling depends on the distance between the wires. While at large distances this coupling is negligible, as the wires get closer to each other the value $C_m$ increases and, hence, their mutual influence increases as well. Let us choose the wire parameters in such a way that for $C_m =0$ both $\lambda_1$ and $\lambda_2$ slightly exceed 2, i.e. the wires remain in the superconducting phase being relatively close to SIT. Moving the wires closer to each other we "turn on" capacitive coupling between them and, hence, decrease both values $\lambda_1$ and $\lambda_2$ below 2. As a result, two superconducting wires become insulating as soon as they are brought sufficiently close to each other. This remarkable physical phenomenon is illustrated by the phase diagram in Fig. \ref{Fig:19}.

We can also add that transport properties can be investigated in exactly the same manner 
as in the case of a single nanowire. Generalization of the technique \cite{ZGOZ} is straightforward and yields
\begin{equation}\label{eq17}
    R_{i}(T)\propto \gamma^{2}_{i} T^{2\lambda_{ii}-3}, \quad i=1,2,
\end{equation}
where $R_{i}(T)$ is the linear resistance of the $i$-th wire. This result remains applicable either for $\lambda_{ii} >2$ or for any $\lambda_{ii}$ at sufficiently high temperatures. 

The effects discussed here can be observed in a variety of structures involving superconducting nanowires. For instance, superconducting nanowires in the form of a meander are quite frequently employed in experiments, see, e.g., Ref. \cite{Gre12}. Different segments of such a wire remain parallel to each other being close enough to develop electromagnetic coupling. One can expect, therefore, that the wire of such a geometry should be "less superconducting" than the same wire that has the form of a straight line.

\begin{figure}
\includegraphics[width=\linewidth]{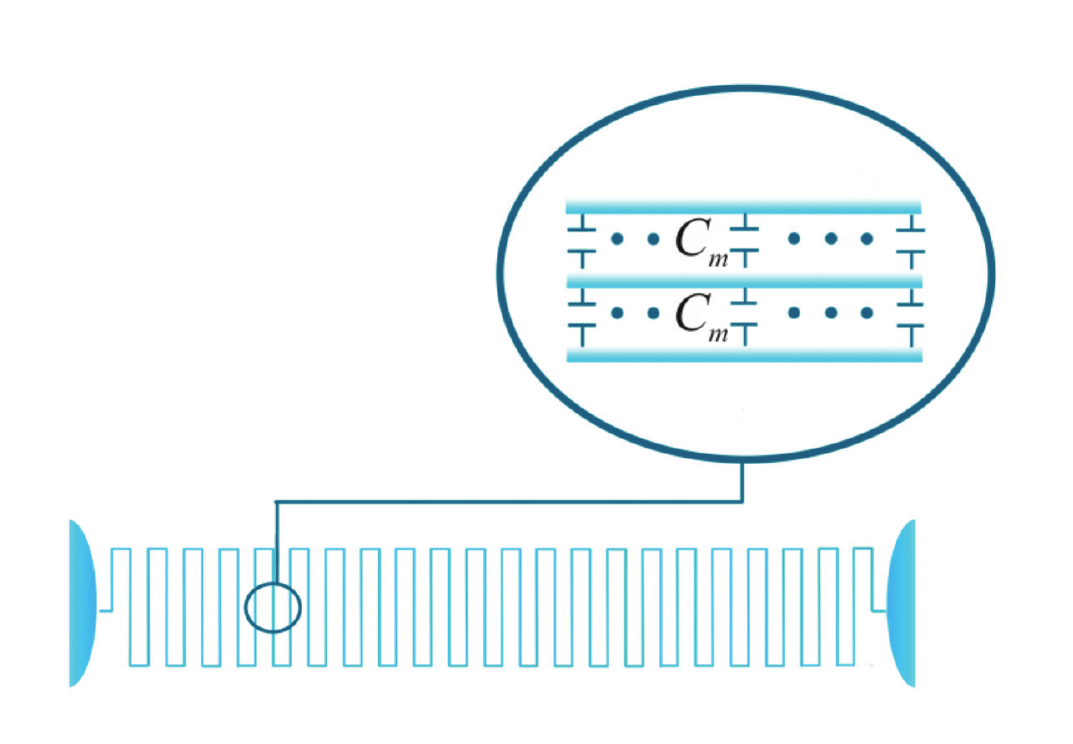}
\caption{A superconducting nanowire in the form of a meander.}
\label{Fig:20}
\end{figure}

Let us first mimic the behavior of the wire depicted in Fig. \ref{Fig:20} by considering three identical parallel to each other 
capacitively coupled superconducting nanowires. For simplicity we will assume the nearest neighbor interaction, i.e. the second (central) nanowire is coupled to both the first and the third nanowires via the mutual capacitance $C_m$ whereas the latter two are decoupled from each other.  Quantum properties of this system are described by the same effective action (\ref{eq8}) where the inductance and capacitance 
matrices now read
\begin{equation}
\check{\mathcal{L}} =\left(\begin{array}{crl}
\mathcal{L} & 0 & 0\\
0 &\mathcal{L}& 0\\
0& 0& \mathcal{L}\\\end{array}\right), \quad 
\check{C}=\left(\begin{array}{crl}
C & C_{m} & 0\\
C_{m}& C & C_{m}\\
0& C_{m}& C\\\end{array}\right),
\end{equation}
and the summation runs over the indices $i,j=1,2,3$. Proceeding along the same lines as above we again arrive at 
Eq. (\ref{G00}), where the diagonal elements of the matrix $\check{\lambda}$ now read \cite{LSZ20}
\begin{eqnarray}
\label{22}
\lambda_{22}=\frac{\lambda}{2}\left(\sqrt{1-\frac{\sqrt{2}C_{m}}{C}} + \sqrt{1+\frac{\sqrt{2}C_{m}}{C}}\right),
\end{eqnarray}
$\lambda_{11}=\lambda_{33}= \lambda/2+ \lambda_{22}/4$ and $\lambda =g/8$ with $g$ defined in Eq. (\ref{gaZ}). We again recover the RG equations (\ref{eq11}), now with $i=1,2,3$. Being combined with Eq. (\ref{22}), these RG equations demonstrate that in the presence of capacitive coupling between nanowires SET in the central nanowire occurs at $\lambda_{22}=2$ implying $\lambda > 2$ similarly to the case of two coupled nanowires.

The RG equation (\ref{eq11}) with $i=2$ combined with Eq. (\ref{22}) also accounts for QPT in the wire having the form of a meander displayed in Fig. \ref{Fig:20}. In this case, within the approximation of the nearest neighbor capacitive interaction between the wire segments QPT occurs at
\begin{equation}
\lambda= \frac{4}{\sqrt{1-\sqrt{2}\frac{C_{m}}{C}} + \sqrt{1+\sqrt{2}\frac{C_{m}}{C}}},
\end{equation}
i.e. the critical value of the parameter $\lambda$ exceeds 2 for any $C_m\neq 0$. The approximation of the nearest neighbor interaction appears to be well justified in the limit $C_m \ll C$. For stronger interactions with $C_m \sim C$ this approximation becomes insufficient for a quantitative analysis. Qualitatively, our key observations should remain applicable also in this case: A nanowire in the form of a straight line with $\lambda$ slightly exceeding the critical value 2 should demonstrate superconducting-like behavior with $R(T) \propto T^{2\lambda -3}$ \cite{ZGOZ} whereas the wire with exactly the same parameters may turn insulating provided it has the form of a meander with capacitive coupling between its segments.

\section{Concluding remarks}

In this review we have made an effort to cover a number of recent developments related to quantum properties of superconducting nanowires
putting an emphasis on fundamental aspects of the theory of such systems. Unlike in the case of bulk superconductors, the low temperature physics of quasi-one-dimensional nanowires and nanorings is essentially determined by quantum fluctuations which are, in turn, controlled by two different parameters, the dimensionless normal state conductance of the wire segment $g_\xi$  (\ref{gxi}) and the dimensionless wire admittance $g$ (\ref{gaZ}). Both these parameters decrease with decreasing wire cross section making quantum fluctuations progressively more pronounced.

Provided the parameter $g_\xi$ remains very large, the low temperature behavior of the system is determined by small (Gaussian) quantum fluctuations of the phase of the order parameter which -- for not very large values of $g$ -- significantly affect the electron density of states in superconducting nanowires and cause supercurrent noise in superconducting nanorings. As soon as the dimensionless conductance $g_\xi$ becomes not too large quantum phase slips come into play. As a result, current-biased superconducting nanowires acquire a non-zero resistance and exhibit shot noise of the voltage. These phenomena can be conveniently interpreted in terms of tunneling of quantum fluxons 
(i.e. the flux quanta $\Phi_0$) across the nanowire. In the context of phase-charge duality, these fluxons can be treated as effective quantum "particles" exactly dual to Cooper pairs with charge $2e$. Such "particles" obey complicated full counting statistics which, however, reduces to Poissonian one in the zero frequency limit.

Quantum phase slips may strongly affect both the supercurrent and its fluctuations in superconducting nanorings. This effect can be particularly pronounced if QPS remain unbound, i.e. for $g < 16$ or, equivalently, for $\lambda < 2$. In this case and provided the ring radius exceeds the critical value $R_c$ (\ref{rc}), even at $T \to 0$ strong quantum fluctuations essentially "dephase" and suppress supercurrent that could flow across the ring. 

The same non-perturbative length scale $L_c$ (\ref{LQPS}) emerges in superconducting nanowires. For $g < 16$ and $T \to 0$ such nanowires do show an insulating behavior at scales exceeding typical size of a "superconducting domain" $L_c$, whereas at shorter length scales they may exhibit superconducting properties albeit possibly affected by quantum fluctuations of the phase. Hence, $L_c$ can be interpreted as {\it localization length for Cooper pairs}. 

Direct experimental evidence for this kind of behavior and for the presence of this localization length was very recently observed in nominally uniform titanium nanowires \cite{NP}. Moreover, reanalyzing similar data  reported previously \cite{BT,Lau,Bezr8} for a large number of $MoGe$ nanowires we conclude that these data are also consistent with the above physical picture involving the correlation length $L_c$ (\ref{LQPS}), i.e. the superconducting $MoGe$ samples \cite{BT,Lau,Bezr8} obey the condition $L \lesssim L_c$, whereas the non-superconducting ones typically have the length $L$ exceeding $L_c$. It is also important to emphasize that the nanowires employed in all these experiments did not contain any grains or dielectric barriers. Hence, similarly to normal metallic structures \cite{book} these observations can be interpreted as a manifestation of {\it weak Coulomb blockade} of Cooper pairs that may occur even in the absence of any tunnel barriers.

Further non-trivial properties of superconducting nanowires may be sensitive to specific topology of the experimental setup under consideration. A number of interesting phenomena associated with quantum phase slips also occurs in capacitively coupled superconducting nanowires.

Finally, it is worthwhile to point out that non-trivial quantum properties of superconducting nanowires and nanorings open up plenty of possibilities for their applications in nanoelectronics, metrology and quantum information technology. Various devices, such as, e.g., single-charge transistor \cite{HZ} and charge quantum interference device \cite{Zhenya} have already been demonstrated. Superconducting nanowires can also be employed as central elements for QPS flux qubits \cite{MH} as well as for creating a QPS-based standard of electric current  \cite{Wang} and single photon detectors \cite{Gre12}. We are confident that intensive investigations of intriguing fundamental properties of quasi-one-dimensional superconducting structures as well as their technological applications will continue in the near future.

\vspace{0.5cm}

\centerline{\bf Acknowledgements}
We would like to thank K.Yu. Arutyunov, A. Radkevich and A. Latyshev  for stimulating discussions and collaboration on a number of issues touched upon in this review. Our work was supported by RFBR grant 19-12-50260.


\begin{thebibliography}{99}
\bibitem{FN} For this very reason the notion "coherent quantum phase slips" sometimes used in the literature is to a large extent tautological. No such objects as "incoherent quantum phase slips" (or, by duality, "incoherent Cooper pairs") exist in Nature. At the same time, quantum tunneling of both fluxons and Copper pairs can, of course, be made incoherent provided, e.g., enough extra dissipation is added to the system.
\end{thebibliography}
\end{document}